\documentclass[final,5p,times,twocolumn]{elsarticle}

\usepackage{graphicx}
\usepackage{amssymb}
\usepackage{amsthm}

\biboptions{comma,square,numbers,sort&compress}

\usepackage[utf8]{inputenc}
\usepackage[T1]{fontenc}
\usepackage{lmodern}
\usepackage{microtype}
\usepackage[english]{babel}
\usepackage[hyperindex=true,colorlinks=true]{hyperref}
\usepackage{amsmath}
\usepackage{amsfonts}
\usepackage[force]{feynmp-auto}

\newtheorem*{proposition}{Proposition}

\newenvironment{definition}[1][Definition]{\begin{trivlist}
\item[\hskip \labelsep {\bfseries #1}]}{\end{trivlist}}

\journal{Computer Physics Communications}

 {
   \begin{list}{\bf\arabic{enumi} --}
    {\usecounter{enumi} \setlength{\parsep}{0pt}
     \setlength{\itemsep}{3pt} \settowidth{\labelwidth}{10 --}
     \sloppy
    }}{\end{list}}

 {
   \begin{list}{\bf\Alph{enumii}/}
    {\usecounter{enumii} \setlength{\parsep}{0pt}
     \setlength{\itemsep}{3pt} \settowidth{\labelwidth}{A/}
     \sloppy
    }}{\end{list}}

 {
   \begin{list}{\bf\alph{enumii}/}
    {\usecounter{enumii} \setlength{\parsep}{0pt}
     \setlength{\itemsep}{3pt} \settowidth{\labelwidth}{a/}
     \sloppy
    }}{\end{list}}

 {
   \begin{list}{\bf\roman{enumiii}/}
    {\usecounter{enumiii} \setlength{\parsep}{0pt}
     \setlength{\itemsep}{3pt} \settowidth{\labelwidth}{a/}
     \sloppy
    }}{\end{list}}

 {
   \begin{list}{$\bullet$}
    {\setlength{\parsep}{3pt}
     \setlength{\itemsep}{3pt} \settowidth{\labelwidth}{a/}
     \sloppy
    }}{\end{list}}

\makeatletter     
\def\overbracket#1{\mathop{\vbox{\ialign{##\crcr\noalign{\kern2\p@}
\downbracketfill\crcr\noalign{\kern6\p@\nointerlineskip}
$\hfil\displaystyle{#1}\hfil$\crcr}}}\limits}
\def\downbracketfill{$\m@th
\makesm@sh{\kern4\p@\llap{\vrule\@height.7\p@\@depth4.\p@\@width.7\p@}}%
\leaders\vrule\@height.7\p@\hfill
\makesm@sh{\rlap{\vrule\@height.7\p@\@depth4.\p@\@width.7\p@}\kern4\p@}$}
\def\underbracket#1{\mathop{\vtop{\ialign{##\crcr
$\hfil\displaystyle{#1}\hfil$\crcr\noalign{\kern3\p@\nointerlineskip}
\upbracketfill\crcr\noalign{\kern3\p@}}}}\limits}
\def\overparenthesis#1{\mathop{\vbox{\ialign{##\crcr\noalign{\kern3\p@}
\downparenthfill\crcr\noalign{\kern3\p@\nointerlineskip}
$\hfil\displaystyle{#1}\hfil$\crcr}}}\limits}
\def\underparenthesis#1{\mathop{\vtop{\ialign{##\crcr
$\hfil\displaystyle{#1}\hfil$\crcr\noalign{\kern3\p@\nointerlineskip}
\upparenthfill\crcr\noalign{\kern3\p@}}}}\limits}
\def\downparenthfill{$\m@th\braceld\leaders\vrule\hfill\bracerd$}
\def\upparenthfill{$\m@th\bracelu\leaders\vrule\hfill\braceru$}
\def\upbracketfill{$\m@th\makesm@sh{\llap{\vrule\@height3\p@\@width.7\p@}}%
\leaders\vrule\@height.7\p@\hfill
\makesm@sh{\rlap{\vrule\@height3\p@\@width.7\p@}}$}
\makeatother
\newcommand{\ra}   {\rangle}

\newcommand{\ket}[1]{{|#1\rangle}}
\newcommand{\bracket}[2]{{\langle #1 |  #2 \rangle}}
\newcommand{\bracketp}[3]{{\langle #1 |  #2 | #3 \rangle}}

\newcommand{\be}{\begin{equation}}
\newcommand{\ee}{\end{equation}}
\newcommand{\bd}{\begin{displaymath}}
\newcommand{\ed}{\end{displaymath}}
\newcommand{\bqr}{\begin{eqnarray}}
\newcommand{\eqr}{\end{eqnarray}}
\newcommand{\bqrr}{\begin{eqnarray*}}
\newcommand{\eqrr}{\end{eqnarray*}}
\newcommand{\bt}{\begin{tabular}}
\newcommand{\et}{\end{tabular}}
\newcommand{\bc}{\begin{center}}
\newcommand{\ec}{\end{center}}

\begin{document}

\allowdisplaybreaks

\begin{frontmatter}


\title{\texttt{ADG}: Automated generation and evaluation of many-body diagrams \\
I. Bogoliubov many-body perturbation theory}

\author[a,b]{P.~Arthuis}
\ead{pierre.arthuis@cea.fr}
\author[a,c]{T.~Duguet}
\ead{thomas.duguet@cea.fr}
\author[d]{A.~Tichai}
\ead{alexander.tichai@cea.fr}
\author[e]{R.-D.~Lasseri}
\ead{lasseri@ipno.in2p3.fr}
\author[b]{J.-P.~Ebran}
\ead{jean-paul.ebran@cea.fr}

\address[a]{IRFU, CEA, Université Paris - Saclay, F-91191 Gif-sur-Yvette, France}
\address[b]{CEA, DAM, DIF, F-91297 Arpajon, France}
\address[c]{KU Leuven, Instituut voor Kern- en Stralingsfysica, 3001 Leuven, Belgium}
\address[d]{ESNT, IRFU, CEA, Université Paris - Saclay, F-91191 Gif-sur-Yvette, France}
\address[e]{Institut de Physique Nucléaire, CNRS/IN2P3, Univ.\@ Paris-Sud, Université Paris-Saclay, F-91406 Orsay, France}

\begin{abstract}
We describe the first version (v1.0.0) of the code \textbf{\texttt{ADG}} that automatically (1) generates all valid Bogoliubov many-body perturbation theory (BMBPT) diagrams and (2) evaluates their algebraic expression to be implemented for numerical applications. This is achieved at any perturbative order $p$ for a Hamiltonian containing both two-body (four-legs) and three-body (six-legs) interactions (vertices). The automated generation of BMBPT diagrams of order $p$ relies on elements of graph theory, i.e., it is achieved by  producing all oriented adjacency matrices of size $(p+1) \times (p+1)$ satisfying topological Feynman's rules. The automated evaluation of BMBPT diagrams of order $p$ relies both on the application of algebraic Feynman's rules and on the identification of a powerful diagrammatic rule providing the result of the remaining $p$-tuple time integral. The diagrammatic rule in question constitutes a novel finding allowing for the straight summation of large classes of time-ordered diagrams at play in the time-independent formulation of BMBPT. Correspondingly, the traditional resolvent rule employed to compute time-ordered diagrams happens to be a particular case of the general rule presently identified. The code \textbf{\texttt{ADG}} is written in \emph{Python2.7} and uses the graph manipulation package \emph{NetworkX}. The code is also able to generate and evaluate Hartree-Fock-MBPT (HF-MBPT) diagrams and is made flexible enough to be expanded throughout the years to tackle the diagrammatics at play in various many-body formalisms that already exist or are yet to be formulated. 

\end{abstract}

\begin{keyword}
many-body theory \sep ab initio \sep perturbation theory \sep Feynman diagrams
\PACS 21.60.De
\end{keyword}

\end{frontmatter}



{\bf PROGRAM SUMMARY}

\begin{small}
\noindent
{\em Program Title:} \texttt{ADG} \\
{\em Licensing provisions:} GPLv3                                   \\
{\em Programming language:} Python2.7 \\
{\em Nature of problem:} \\
  As formal and numerical developments in many-body-perturbation-theory-based \emph{ab initio} methods make higher orders reachable, producing and evaluating all the diagrams become rapidly undoable on a handmade basis as both their number and complexity grows quickly, making it prone to mistakes and oversights.
  \\
{\em Solution method:}\\
  BMBPT diagrams are encoded as square matrices known as oriented adjacency matrices in graph theory, and then turned into graph objects using the \emph{NetworkX} package. Checks on the diagrams and evaluation of their time-integrated expression is then done on a purely diagrammatic basis. HF-MBPT diagrams are produced and evaluated as well using the same principle.
   \\

\end{small}

\section{Introduction}
\label{s:intro}

Diagrams have long been used in combination with formalisms, e.g., many-body perturbation theory (MBPT)~\cite{goldstone57a,hugenholtz57a,Shavitt_Bartlett_2009,Ti16,Hu16,Ti17}, self-consistent Green's function (SCGF) theory~\cite{Dickhoff:2004xx,CiBa13,Carbone:2013eqa,Soma:2011aj,SoCi13}, coupled-cluster (CC) theory~\cite{KoDe04,BaRo07,HaPa10,PiGo09,BiLa14,Si15} etc, designed to solve the many-body Schr\"odinger equation, be it in nuclear physics, quantum chemistry, atomic physics or solid-state physics. Many-body diagrams belong to a series of tools introduced  to compute the expectation value of products of (many) operators in a vacuum state in an incrementally faster, more flexible and less error-prone way. The first step in this series relied on the introduction of the second quantization formalism that makes algebraic manipulations much more efficient than within the first quantization formalism. The next step consisted in the elaboration of Wick's theorem~\cite{wick50a}, which is nothing but a procedure to capture the result in a condensed and systematic fashion. Still, the combinatorics associated with the application of Wick's theorem becomes quickly cumbersome whenever a long string of creation and annihilation operators is involved. Furthermore, many terms generated via the application of Wick's theorem happen to give identical contributions to the end results. Many-body diagrams were introduced next to provide a pictorial representation of the various contributions and, even more importantly, to capture at once all identical contributions generated via the straight application of Wick's theorem, thus reducing the combinatorics tremendously. The procedure results (i) in a set of \emph{topological} rules to generate all valid diagrams and (ii) in a set of \emph{algebraic} rules to evaluate their expressions, including a prefactor accounting for all identical contributions. 

While diagrams have proven to be extremely useful, their number grows tremendously when applying, e.g., MBPT beyond the lowest orders, thus leading to yet another combinatorial challenge. This translates into the difficulty to both \emph{generate} all allowed diagrams at a given order without missing any and to \emph{evaluate} their expression in a quick and error-safe way. Consequently, yet another tool must be introduced to tackle this difficulty. As a matter of fact, there have been several attempts to generate MBPT diagrams automatically, e.g. see Refs.~\cite{Paldus1973a,Wong1973,Kaldor1976,Csepes1988,Lyons:1994ew,Stevenson2003}. However, it is of primer interest to also evaluate their algebraic expressions automatically~\cite{Csepes1988,Lyons:1994ew} in view of performing their numerical implementation. 

It happens that the past decade has witnessed the development and/or the application of new formalisms to tackle the nuclear many-body problem~\cite{Tsukiyama:2010rj,Soma:2011aj,HeBo13,Hergert:2014iaa,Duguet:2014jja,Bogner:2014baa,Jansen:2014qxa,Duguet:2015yle,Gebrerufael:2016xih,Tichai:2017rqe,Art18b}, some of which rely on original, i.e., more general, diagrammatics~\cite{Duguet:2014jja,Duguet:2015yle}. This profusion of methods, along with the rapid progress of computational power allowing for high-order implementations, welcomes the development of a versatile code capable of both generating and evaluating diagrams. 

Many-body diagrams come in various forms and flavors. First, the diagrammatic framework depends on the nature of the reference state at play in the formalism. Second, most many-body methods can be designed within a time-dependent or a time-independent formalism, eventually leading to the same result\footnote{Dealing with static properties of an isolated system, the end results are obviously independent of time.}. While a time-independent formalism naturally translates into time-ordered diagrams, a time-dependent formulation can be represented by a time-unordered diagrammatic, i.e., by diagrams containing an explicit integration over time variables, thus capturing different time orderings of the vertices at once.

In the present publication, we focus our attention on Bogoliubov many-body perturbation theory (BMBPT) that has been recently formulated~\cite{Duguet:2015yle,Art18b} to tackle (near) degenerate Fermi systems, e.g. open-shell nuclei displaying a superfluid character. Even more recently, second- and third-order BMBPT calculations have been performed in mid- and heavy-mass nuclei and show great promises~\cite{Tichai:2018mll}. This many-body method perturbatively expands the exact solution of the Schrödinger equation around a so-called Bogoliubov reference state, i.e., a general product state breaking $U(1)$ global-gauge symmetry associated with the conservation of particle number in the system. As such, BMBPT generalizes standard Rayleigh-Schrödinger MBPT based on a single-reference Slater determinant, which is thus recovered as a particular subcase. The BMBPT diagrammatic is itself a first step towards the more general diagrammatic at play in the so-called particle-number projected Bogoliubov many-body perturbation theory (PNP-BMBPT)~\cite{Duguet:2015yle}. As a matter of fact, our goal is to develop a numerical tool that is flexible enough to be expanded throughout the years to tackle the diagrammatics at play in various many-body formalisms (already existing or yet to be formulated). 

Having originally formulated BMBPT on the basis of a time-dependent formalism~\cite{Duguet:2015yle,Art18b}, the diagrammatic of present interest relies on the use of time-unordered Feynman diagrams. The first benefit is that Feynman diagrams are naturally employed within SCGF theory that we wish to address as a next step. Indeed, there is an interest in generating and evaluating the diagrams at play in Gorkov-SCGF~\cite{Soma:2011aj,SoCi13} within the so-called ADC(3) truncation scheme~\cite{Schirmer83}. The second gain relates to a combinatorial advantage. Indeed, a Feynman diagram captures at once all time-orderings of involved operator vertices, which possibly corresponds to summing many time-ordered diagrams in a time-independent formalism. The disadvantage is that Feynman's diagrammatic rules provide the algebraic expression of the diagram \emph{prior} to integrating over time variables such that the multiple time integral remains to be performed\footnote{Contrarily, the diagrammatic rules associated with time-ordered diagrams directly provide the final, i.e., time-integrated, expressions to be implemented in a numerical application at the price of dealing with a (much) larger number of diagrams. See Sec.~\ref{resolventrule} for an in-depth discussion on this point.}. This feature constitutes a challenge for the automated evaluation of diagrams. The present finding of a rule to compute the end result of these time integrals on the sole given of the Feynman diagram, independently of its topology and perturbative order, constitutes a formal breakthrough regarding the evaluation of many-body diagrams. 

In this paper, we present our strategy to automatically generate and evaluate  all Feynman diagrams appearing at an arbitrary  order in BMBPT. The algorithms developped to achieve this goal have been implemented in a numerical code called \textbf{\texttt{ADG}} for \emph{Automatic Diagram Generator}, written in \emph{Python2.7} and using the graph manipulation package \emph{NetworkX}~\cite{SciPyProceedings_11}, which will be detailed as well. The program is also able to produce standard MBPT diagrams using similar techniques as those detailed below. Given that MBPT is indeed a subcase of BMBPT, we have chosen not to detail this feature of the code in the body of the paper. Extensions of the program to tackle other diagrammatic flavors are to be considered in the future.

The paper is organized as follows. Section~\ref{s:bmbpt_diags} recalls the basics of BMBPT and of the associated diagrammatic, underlining the difficulties to be overcome in order to achieve an automated generation and evaluation of diagrams of arbitrary orders. Building on this, Secs.~\ref{secautomaticgeneration} and~\ref{secautomaticevaluation} detail the method developped to reach such an objective. Section~\ref{resolventrule} discusses how the present finding sheds some light on the connection between time-unordered and time-ordered diagrammatics. While Sec.~\ref{programuse} details how the \textbf{\texttt{ADG}} code operates, conclusions are given in Sec.~\ref{conclusions}. Two appendices follow to provide details regarding the formalism and the structure of the program.

\section{BMBPT diagrammatics}
\label{s:bmbpt_diags}

\subsection{Basics ingredients}
\label{subs:basicsingredients}

Bogoliubov MBPT consists of expanding the exact A-body ground-state energy in perturbation around a Bogoliubov vacuum that breaks $U(1)$ global gauge symmetry associated with particle-number conservation. Breaking $U(1)$ symmetry allows one to deal with Cooper pair's instability associated with the superfluid character of open-shell nuclei. Doing so, the degeneracy of a Slater determinant with respect to particle-hole excitations is lifted via the use of a more general Bogoliubov state and commuted into a degeneracy with respect to symmetry transformations of the group. As a consequence, the ill-defined (i.e., singular) MBPT expansion of exact quantities around a Slater determinant is replaced by a well-behaved one.

The set up of the formalism starts with the introduction of the Bogoliubov reference state
\begin{align}
\ket{\Phi} \equiv \mathcal{C} \prod_k \beta_k \vert 0 \ra\, ,
\end{align}
where $\mathcal{C}$ is a complex normalization constant and $\vert 0 \ra$ denotes the physical vacuum. The Bogoliubov state is a vacuum for the set of quasi-particle operators obtained from operators associated with a basis of the one-body Hilbert space via a unitary linear transformation of the form~\cite{ring80a}
\begin{subequations}
\begin{align}
\beta_k &\equiv \sum_p U^*_{pk} c_p + V_{pk} c^\dagger_p\, , \\
\beta_k^\dagger &\equiv \sum_p U^*_{pk} c^\dagger_p + V_{pk} c_p \, ,
\end{align}
\end{subequations}
i.e., $\beta_k \ket{\Phi} = 0$ for all $k$. One possiblity to actually specify the Bogoliubov reference state $\ket{\Phi}$ is to require that it solves the Hartree-Fock-Bogoliubov (HFB) variational problem. This fixes the transformation matrices $(U,V)$~\cite{ring80a} and delivers the set of quasi-particle energies $\{E_k > 0\}$ defining the unperturbed part of the Hamiltonian later on (see Eqs.~\eqref{split2}-\eqref{onebodypiece}). We do not impose this choice here such that the reference state and the associated unperturbed Hamiltonian can be defined more generally.

The Bogoliubov reference state is not an eigenstate of the particle number operator $A$. The same is true of the perturbatively corrected state generated from it, unless the perturbative expansion is resummed to all orders. Consequently, one must at least enforce that the expectation value of $A$ matches the actual number of particles A of the targeted system. Correspondingly, the Hamiltonian $H$ is to be replaced by the grand potential $\Omega \equiv H - \lambda A$ in the set up of the many-body formalism~\cite{Duguet:2015yle,Art18b}, where $\lambda$ denotes the chemical potential.

One is interested in evaluating a ground-state observable $\text{O}^{\text{A}}_0$ whose associated operator $O$ commutes with $\Omega$, i.e., $H$, $A$ or $\Omega$ itself. The operator $O$ typically contains one-body, two-body and three-body terms\footnote{State-of-the-art nuclear Hamiltonians are indeed modeled in terms of one-, two- and three-body operators. Higher-body operators can be employed as well. From the formal point of view, it poses no fundamental difficulty but further complexifies the diagrammatic and its bookeeping. As for the automated generation of diagrams, it poses no fundamental difficulty but requires to handle the memory needed to deal with the increased combinatorial.}. The operator in the Schrödinger representation is expressed in an arbitrary basis of the one-body Hilbert space as
\begin{align}
O &\equiv o^{[2]} + o^{[4]} + o^{[6]} \label{e:ham}  \\
&\equiv  o^{11} +  o^{22} +  o^{33}  \nonumber \\
&\equiv  \frac{1}{(1!)^2} \sum _{p_1p_2} o^{11}_{p_1p_2} c^{\dagger}_{p_1} c_{p_2} \nonumber \\
&\phantom{=}+\frac{1}{(2!)^2} \sum _{p_1p_2p_3p_4} o^{22}_{p_1p_2p_3p_4}  c^{\dagger}_{p_1} c^{\dagger}_{p_2} c_{p_4} c_{p_3} \nonumber \\
&\phantom{=}+\frac{1}{(3!)^2} \sum _{p_1p_2p_3p_4p_5p_6} o^{33}_{p_1p_2p_3p_4p_5p_6}  c^{\dagger}_{p_1} c^{\dagger}_{p_2} c^{\dagger}_{p_3} c_{p_6} c_{p_5} c_{p_4}  \, . \nonumber
\end{align} 
Each term $o^{kk}$ of the particle-number conserving operator $O$ is obviously characterized by the equal number $k$ of particle creation and annihilation operators. The class $o^{[2k]}$ is nothing but the term $o^{kk}$ of $k$-body character. Matrix elements are fully antisymmetric, i.e.
\begin{equation}
o^{kk}_{p_1 \ldots p_{k} p_{k+1} \ldots p_{2k}} = (-1)^{\sigma(P)} o^{kk}_{P(p_1 \ldots p_{k} | p_{k+1} \ldots p_{2k})} \, ,
\end{equation}
where $\sigma(P)$ refers to the signature of the permutation $P$.  The notation $P(\ldots | \ldots)$ denotes a separation into the $k$ particle-creation operators and the $k$ particle-annihilation operators such that permutations are only considered between members of the same group. 

The next step consists of normal ordering $O$ with respect to the Bogoliubov vacuum $\ket{\Phi}$, thus obtaining
\begin{align}
\label{e:oqpas}
O &\equiv O^{[0]} + O^{[2]} + O^{[4]} + O^{[6]} \\
&\equiv O^{00} + \Big[O^{11} + \{O^{20} + O^{02}\}\Big]  \nonumber\\
&\phantom{=} + \Big[O^{22} + \{O^{31} + O^{13}\} + \{O^{40} + O^{04}\}\Big] \nonumber\\
   & \phantom{=} + \Big[O^{33} + \{O^{42} + O^{24}\} + \{O^{51} + O^{15}\} + \{O^{60} + O^{06}\}\Big] \nonumber \\
&= O^{00} \nonumber\\
& \phantom{=} + \frac{1}{(1!)^2}\sum_{k_1 k_2} O^{11}_{k_1 k_2}\beta^{\dagger}_{k_1} \beta_{k_2} \nonumber\\
& \phantom{=} + \frac{1}{2!}\sum_{k_1 k_2} \Big \{O^{20}_{k_1 k_2} \beta^{\dagger}_{k_1}
 \beta^{\dagger}_{k_2} + O^{02}_{k_1 k_2}   \beta_{k_2} \beta_{k_1} \Big \} \nonumber\\
& \phantom{=} + \frac{1}{(2!)^{2}} \sum_{k_1 k_2 k_3 k_4} \hspace{-5pt}O^{22}_{k_1 k_2 k_3 k_4}
   \beta^{\dagger}_{k_1} \beta^{\dagger}_{k_2} \beta_{k_4}\beta_{k_3} \nonumber\\
   & \phantom{=} + \frac{1}{3!1!}\sum_{k_1 k_2 k_3 k_4}\hspace{-5pt}\Big \{ O^{31}_{k_1 k_2 k_3 k_4}
   \beta^{\dagger}_{k_1}\beta^{\dagger}_{k_2}\beta^{\dagger}_{k_3}\beta_{k_4} \nonumber\\
   & \phantom{= + \frac{1}{1!3!}\sum_{k_1 k_2 k_3 k_4}} + O^{13}_{k_1 k_2 k_3 k_4} \beta^{\dagger}_{k_1} \beta_{k_4} \beta_{k_3} \beta_{k_2}  \Big \} \nonumber\\
  & \phantom{=} +  \frac{1}{4!} \sum_{k_1 k_2 k_3 k_4}\hspace{-5pt}\Big \{ O^{40}_{k_1 k_2 k_3 k_4}
   \beta^{\dagger}_{k_1}\beta^{\dagger}_{k_2}\beta^{\dagger}_{k_3}\beta^{\dagger}_{k_4}  \nonumber\\
   & \phantom{= +  \frac{1}{4!} \sum_{k_1 k_2 k_3 k_4}} + O^{04}_{k_1 k_2 k_3 k_4}  \beta_{k_4} \beta_{k_3} \beta_{k_2} \beta_{k_1}  \Big \}  \nonumber\\
   & \phantom{=} + \frac{1}{(3!)^2} \sum_{k_1 k_2 k_3 k_4 k_5 k_6} \hspace{-5pt}
   O^{33}_{k_1 k_2 k_3 k_4 k_5 k_6}
   \beta^{\dagger}_{k_1}\beta^{\dagger}_{k_2}\beta^{\dagger}_{k_3}\beta_{k_6}\beta_{k_5}\beta_{k_4} \nonumber\\
   & \phantom{=} + \frac{1}{2! \, 4!} \sum_{k_1 k_2 k_3 k_4 k_5 k_6} \hspace{-5pt}\Big \{
   O^{42}_{k_1 k_2 k_3 k_4 k_5 k_6}\beta^{\dagger}_{k_1}\beta^{\dagger}_{k_2}\beta^{\dagger}_{k_3}
   \beta^{\dagger}_{k_4}\beta_{k_6}\beta_{k_5} \nonumber\\
   & \phantom{= + \frac{1}{2! \, 4!} \sum_{k_1 k_2 k_3 k_4 k_5 k_6}} +  O^{24}_{k_1 k_2 k_3 k_4 k_5 k_6}
   \beta^{\dagger}_{k_1}\beta^{\dagger}_{k_2}\beta_{k_6}\beta_{k_5}\beta_{k_4}\beta_{k_3} \Big \} \nonumber\\
   & \phantom{=} + \frac{1}{5!1!} \sum_{k_1 k_2 k_3 k_4 k_5 k_6} \hspace{-5pt}\Big \{
   O^{51}_{k_1 k_2 k_3 k_4 k_5 k_6}\beta^{\dagger}_{k_1}\beta^{\dagger}_{k_2}\beta^{\dagger}_{k_3}
   \beta^{\dagger}_{k_4}\beta^{\dagger}_{k_5}\beta_{k_6} \nonumber\\
   & \phantom{= + \frac{1}{1!5!} \sum_{k_1 k_2 k_3 k_4 k_5 k_6}}  +  O^{15}_{k_1 k_2 k_3 k_4 k_5 k_6}
   \beta^{\dagger}_{k_1}\beta_{k_6}\beta_{k_5}\beta_{k_4}\beta_{k_3}\beta_{k_2} \Big \} \nonumber\\
   & \phantom{=} + \frac{1}{6!} \sum_{k_1 k_2 k_3 k_4 k_5 k_6} \hspace{-5pt} \Big \{
   O^{60}_{k_1 k_2 k_3 k_4 k_5 k_6}\beta^{\dagger}_{k_1}\beta^{\dagger}_{k_2}\beta^{\dagger}_{k_3}
   \beta^{\dagger}_{k_4}\beta^{\dagger}_{k_5}\beta^{\dagger}_{k_6} \nonumber\\
   & \phantom{= + \frac{1}{6!} \sum_{k_1 k_2 k_3 k_4 k_5 k_6}} +  O^{06}_{k_1 k_2 k_3 k_4 k_5 k_6}
   \beta_{k_6}\beta_{k_5}\beta_{k_4}\beta_{k_3}\beta_{k_2}\beta_{k_1} \Big \} \ , \nonumber\\
   \end{align}
where the expressions of the matrix elements of each operator $O^{ij}$ in terms of those of the operators $o^{kk}$ and of the $(U,V)$ matrices can be found in Ref.~\cite{Si15}. Each term $O^{ij}$ is characterized by its number $i$ ($j$) of quasiparticle creation (annihilation) operators. Because $O$ has been normal-ordered  with respect to $| \Phi \rangle$, all quasiparticle creation operators (if any) are located to the left of all quasiparticle annihilation operators (if any).  The class $O^{[2k]}$ groups all the terms $O^{ij}$ of \emph{effective} $k$-body character, i.e.,\@ with $i+j=2k$. Matrix elements are fully antisymmetric, i.e.,
\begin{equation}
O^{ij}_{k_1 \ldots k_{i} k_{i+1} \ldots k_{i+j}} = (-1)^{\sigma(P)} O^{ij}_{P(k_1 \ldots k_i | k_{i+1} \ldots k_{i+j})}  \, .
\end{equation}
More details and properties can be found in Refs.~\cite{Si15,Duguet:2015yle}.

State-of-the-art many-body calculations are typically performed within the normal-ordered two-body approximation (NO2B)~\cite{Roth:2011vt}, i.e., neglecting the residual three-body part $O^{[6]}$ in the above equation. In the present work, however, the diagrammatic is worked out in presence of the effective three-body part, i.e.,\@ in presence of six-legs vertices (see below), which significantly increases the number of possible diagrams at a given order and the complexity of their topology. Correspondingly, the code can eventually be run with or without including the effective three-body part of the operators at play.

\subsection{Time-dependent perturbation theory}
\label{subs:basicsBMBPT}

The grand potential is split into an unperturbed part $\Omega_{0}$ and a residual part $\Omega_1$
\begin{equation}
\label{split1}
\Omega = \Omega_{0} + \Omega_{1} \ ,
\end{equation}
such that
\begin{subequations}
\label{split2}
\begin{align}
\Omega_{0} &\equiv \Omega^{00}+\bar{\Omega}^{11} \ , \\
\Omega_{1} &\equiv \Omega^{20} + \breve{\Omega}^{11} + \Omega^{02} \nonumber \\
  &\phantom{\equiv } + \Omega^{40} + \Omega^{31} + \Omega^{22} +  \Omega^{13} + \Omega^{04} \nonumber \\
  &\phantom{\equiv } + \Omega^{60}+ \Omega^{51}+ \Omega^{42}+ \Omega^{33}+ \Omega^{24}+ \Omega^{15}+ \Omega^{06} \nonumber \ ,
 \label{e:perturbation}
\end{align}
\end{subequations}
with $\breve{\Omega}^{11}\equiv\Omega^{11}- \bar{\Omega}^{11}$ and where the one-body part of $\Omega_{0}$ is diagonal, i.e.,
\begin{equation}
\bar{\Omega}^{11} \equiv \sum_{k} E_k \beta^{\dagger}_k \beta_k \, , \label{onebodypiece}
\end{equation}
with $E_k > 0$ for all $k$. For a given number of interacting fermions, the key is to choose $\Omega_0$ with a low-enough symmetry for its ground state $| \Phi \rangle$ to be non-degenerate with respect to elementary excitations. For open-shell superfluid nuclei, this leads to choosing an operator $\Omega_0$ that breaks particle-number conservation, i.e., while $\Omega$ commutes with $U(1)$ transformations, we are interested in the case where $\Omega_0$, and thus $\Omega_1$, do not. Introducing many-body states generated via an even number of quasi-particle excitations of the vacuum
\begin{equation}
| \Phi^{k_1 k_2\ldots} \rangle \equiv \beta^{\dagger}_{k_1} \, \beta^{\dagger}_{k_2} \,  \ldots  |  \Phi \rangle \, , 
\end{equation}
the unperturbed grand potential $\Omega_{0}$ is fully characterized by its complete set of orthonormal eigenstates in Fock space
\begin{subequations}
\begin{align}
\Omega_{0}\, |  \Phi \rangle &= \Omega^{00} \, |  \Phi \rangle \, , \\
\Omega_{0}\, |  \Phi^{k_1 k_2\ldots} \rangle &= \left[\Omega^{00} + E_{k_1 k_2 \ldots}\right] |  \Phi^{k_1 k_2\ldots} \rangle  \label{phi} \, ,
\end{align}
\end{subequations}
where the strict positivity of unperturbed excitations $E_{k_1 k_2 \ldots} \equiv E_{k_1} + E_{k_2} +\ldots $ characterizes the lifting of the particle-hole degeneracy authorized by the spontaneous breaking of $U(1)$ symmetry in open-shell nuclei at the mean-field level.

In the particular case where $\ket{\Phi}$ solves the HFB variational problem, one has that $\Omega^{20}=\breve{\Omega}^{11}=\Omega^{02}=0$ such that $\Omega_1$ reduces to $\Omega^{[4]}+\Omega^{[6]}$. This choice defines the \emph{canonical} version of BMBPT and reduces significantly the number of non-zero diagrams to be considered. However, we do not make this \textit{a priori} hypothesis such that the reference state $\ket{\Phi}$ and the corresponding unperturbed grand potential $\Omega_{0}$ can be defined more generally, eventually leading to the appearance of \emph{non-canonical} diagrams involving $\Omega^{20}$, $\breve{\Omega}^{11}$ and $\Omega^{02}$ vertices. 

On the basis of the above splitting of $\Omega$, one introduces the interaction representation of operators in the quasi-particle basis, e.g.,
\begin{align}
O^{31}(\tau) &\equiv e^{+\tau \Omega_{0}} O^{31} e^{-\tau \Omega_{0}}  \\
&=\frac{1}{3!}\sum_{k_1 k_2 k_3 k_4}  O^{31}_{k_1 k_2 k_3 k_4}
   \beta^{\dagger}_{k_1}(\tau)\beta^{\dagger}_{k_2}(\tau)\beta^{\dagger}_{k_3}(\tau)\beta_{k_4}(\tau) \, ,\nonumber
\end{align} 
where
\begin{subequations}
\label{aalphatau}
\begin{align}
\beta_{k} (\tau)  &\equiv e^{+\tau \Omega_{0}} \, \beta_{k} \, e^{-\tau \Omega_{0}}=e^{-\tau E_k} \, \beta_{k} \label{aalphatau2} \ , \\
\beta_{k}^{\dagger}(\tau)  &\equiv e^{+\tau \Omega_{0}} \, \beta_{k}^{\dagger} \, e^{-\tau \Omega_{0}}=e^{+\tau E_{k}} \, \beta_{k}^{\dagger} \ . \label{aalphatau1}
\end{align}
\end{subequations}

Defining the operator evolution in imaginary time and expanding it in powers of $\Omega_1$~\cite{blaizot86}
\begin{align}
\mathcal{U}(\tau) &\equiv e^{-\tau \Omega} \nonumber \\
&= e^{-\tau \Omega_{0}} \, \textmd{T}e^{-\int_{0}^{\tau}\!\mathrm{d}\tau \Omega_{1}(\tau) }  \, ,
 \label{evol1}
\end{align}
where $\textmd{T}$ denotes the time-ordering operator\footnote{The time-ordering operator orders a product of operators in decreasing order according to their time labels (i.e., larger times to the left) and multiplies the result with the signature of the permutation used to achieve the corresponding reordering.}, the ground-state observable of interest is accessed through~\cite{Duguet:2015yle,Art18b}
\begin{align}
\text{O}^{\text{A}}_0 &\equiv  \lim\limits_{\tau \to \infty} \frac{\langle \Phi |{\cal U}(\tau) O | \Phi \rangle}{\langle \Phi |{\cal U}(\tau) | \Phi \rangle} \nonumber  \\
&= \lim\limits_{\tau \to \infty}  \langle \Phi | \textmd{T}e^{-\int_{0}^{\tau}dt \Omega_{1}\left(t\right)} O |  \Phi \rangle_{c}   \nonumber \\
&= \langle \Phi | O |  \Phi \rangle \nonumber \\
& \,\,\,\,-\frac{1}{1!}\int_{0}^{+\infty}d\tau_1 \langle \Phi | \textmd{T}\left[  \Omega_{1}\left(  \tau_1\right)O(0)\right]|  \Phi \rangle_{c}  \nonumber \\
& \,\,\,\,+\frac{1}{2!}\int_{0}^{+\infty} \!d\tau_{1} d\tau_{2} \langle \Phi |\textmd{T}\left[  \Omega_{1}\left(  \tau_{1}\right)  \Omega_{1}\left(  \tau_{2}\right) O(0)\right] |  \Phi \rangle_{c} \nonumber \\
& \,\,\,\,-...  \ , \label{observableO1}
\end{align}
where the lower index $c$ refers to the restriction to connected diagrams, thus, yielding a size-extensive many-body framework that properly scales with system size. The time-independent operator $O$ could be inserted at no cost within the time-ordering by providing it with a fictitious and harmless time dependence $t=0$. Indeed, all $\Omega_{1}\left(  \tau_k\right)$ operators appear to the left of $O$ and occur at a larger time given that their corresponding time variables are positive.

Invoking perturbation theory consists of truncating the Taylor expansion of the time-evolution operator in Eq.~\eqref{observableO1}. Gathering all terms up to order $p$, the observable $\text{O}^{\text{A}}_0$ sums matrix elements of products of up to $p+1$ time-dependent operators\footnote{The expansion starts at order $p=0$ that corresponds to the term containing no $\Omega_{1}$ operator and no time integral in Eq.~\eqref{observableO1}.}. The running time variables are integrated over from $0$ to $\tau \rightarrow +\infty$ whereas the time label attributed to the operator $O$ itself remains fixed at $t=0$, i.e.,\@ contributions of order $p$ contain a $p$-tuple time integral that needs to be performed to generate the end result under the required form. 

Each matrix element in Eq.~\eqref{observableO1} is computed via the application of time-dependent Wick's theorem~\cite{blaizot86} with respect to the Bogoliubov reference state. This results in the sum of all possible products of elementary contractions. In BMBPT, the two elementary contractions, i.e.,\@ unperturbed time-dependent propagators, at play are
\begin{subequations}
\label{e:propagators}
\begin{align}
G^{+- (0)}_{k_1k_2}(\tau_1, \tau_2) &\equiv \frac{\bracketp{ \Phi }{  \textmd{T}[\beta^{\dagger}_{k_1}(\tau_1) \beta_{k_2}(\tau_2)] }{ \Phi }}{ \bracket{\Phi}{\Phi} } \nonumber \\
&= - e^{-(\tau_2-\tau_1)E_{k_1}} \theta(\tau_2-\tau_1) \delta_{k_1k_2} \ , \label{e:propagatorsGpm} \\
G^{-+ (0)}_{k_1k_2}(\tau_1, \tau_2) &\equiv \frac{\bracketp{ \Phi }{  \textmd{T}[\beta_{k_1}(\tau_1) \beta^{\dagger}_{k_2}(\tau_2)] }{ \Phi }}{\bracket{\Phi}{\Phi}} \nonumber \\
&= + e^{-(\tau_1-\tau_2)E_{k_1}} \theta(\tau_1-\tau_2) \delta_{k_1k_2} \ , \label{e:propagatorsGmp}
\end{align}
\end{subequations}
which are in fact just one by virtue of the antisymmetry relation
\begin{equation}
G^{+- (0)}_{k_1k_2}(\tau_1, \tau_2) = -G^{-+ (0)}_{k_2k_1}(\tau_2, \tau_1)  \ .
\label{e:prop_antisym}
\end{equation}
Equal-time, i.e., $\tau_1= \tau_2$, unperturbed propagators deserve special attention. Equal-time propagators can solely arise from contracting two quasi-particle operators belonging to the same normal-ordered operator displaying creation operators to the left of annihilation ones. It necessarily leads to selecting a normal-ordered contraction that is identically zero. As a result, no equal-time propagator, and, thus, no contraction of an interaction vertex onto itself, can occur. 

In an even more general context than BMBPT, e.g., PNP-BMBPT~\cite{Duguet:2015yle}, one or two more non-zero propagators can appear by contracting two creation or two annihilation operators together. This makes the application of Wick's theorem and the diagrammatic that derives from it more general and involved. We keep the implementation of such an extension for a future version of the code.

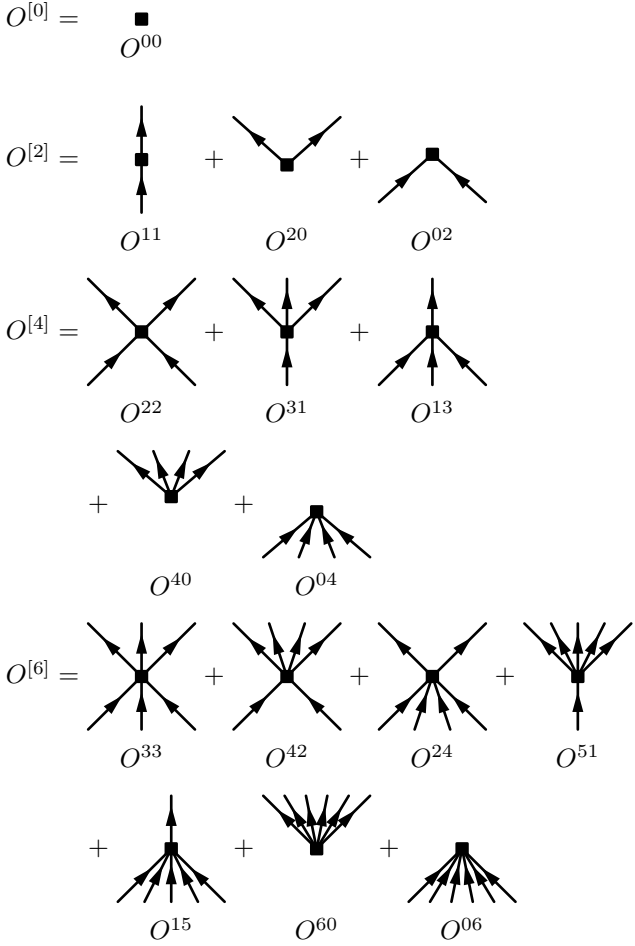
\begin{figure}[t!]
$O^{[0]} =$
	\parbox{40pt}{\begin{fmffile}{O00}
	\begin{fmfgraph*}(40,40)
	\fmftop{t1} \fmfbottom{b1}
	\fmf{phantom}{b1,i1}
	\fmf{phantom}{i1,t1}
	\fmfv{label=$O^{00}$,label.angle=-90,d.shape=square,d.filled=full,d.size=2thick}{i1}
	\end{fmfgraph*}
	\end{fmffile}}

\vspace{\baselineskip}

$O^{[2]} =$
	\parbox{40pt}{\begin{fmffile}{O11}
	\begin{fmfgraph*}(40,40)
	\fmfcmd{style_def half_prop expr p =
    draw_plain p;
    shrink(.7);
        cfill (marrow (p, .5))
    endshrink;
	enddef;}
	\fmftop{t1} \fmfbottom{b1}
	\fmf{half_prop}{b1,i1}
	\fmf{half_prop}{i1,t1}
	\fmfv{d.shape=square,d.filled=full,d.size=2thick}{i1}
	\fmflabel{$O^{11}$}{b1}
	\end{fmfgraph*}
	\end{fmffile}}
+
	\parbox{40pt}{\begin{fmffile}{O20}
	\begin{fmfgraph*}(40,40)
	\fmfcmd{style_def half_prop expr p =
    draw_plain p;
    shrink(.7);
        cfill (marrow (p, .5))
    endshrink;
	enddef;}
	\fmftop{t1,t2} \fmfbottom{b1}
	\fmfstraight
	\fmf{half_prop}{i1,t2}
	\fmf{half_prop}{i1,t1}
	\fmf{phantom,tension=2}{b1,i1}
	\fmfv{d.shape=square,d.filled=full,d.size=2thick}{i1}
	\fmflabel{$O^{20}$}{b1}
	\end{fmfgraph*}
	\end{fmffile}}
+
	\parbox{40pt}{\begin{fmffile}{O02}
	\begin{fmfgraph*}(40,40)
	\fmfcmd{style_def half_prop expr p =
    draw_plain p;
    shrink(.7);
        cfill (marrow (p, .5))
    endshrink;
	enddef;}
	\fmftop{t1} \fmfbottom{b1,b2,b3}
	\fmfstraight
	\fmf{half_prop}{b1,i1}
	\fmf{half_prop}{b3,i1}
	\fmf{phantom,tension=2}{i1,t1}
	\fmfv{d.shape=square,d.filled=full,d.size=2thick}{i1}
	\fmflabel{$O^{02}$}{b2}
	\end{fmfgraph*}
	\end{fmffile}}

\vspace{2\baselineskip}

$O^{[4]} =$
	\parbox{40pt}{\begin{fmffile}{O22}
	\begin{fmfgraph*}(40,40)
	\fmfcmd{style_def half_prop expr p =
    draw_plain p;
    shrink(.7);
        cfill (marrow (p, .5))
    endshrink;
	enddef;}
	\fmfstraight
	\fmftop{t1,t2} \fmfbottom{b1,b2,b3}
	\fmf{half_prop}{b1,i1}
	\fmf{half_prop}{b3,i1}
	\fmf{half_prop}{i1,t1}
	\fmf{half_prop}{i1,t2}
	\fmfv{d.shape=square,d.filled=full,d.size=2thick}{i1}
	\fmflabel{$O^{22}$}{b2}
	\end{fmfgraph*}
	\end{fmffile}}
+
	\parbox{40pt}{\begin{fmffile}{O31}
	\begin{fmfgraph*}(40,40)
	\fmfcmd{style_def half_prop expr p =
    draw_plain p;
    shrink(.7);
        cfill (marrow (p, .5))
    endshrink;
	enddef;}
	\fmfstraight
	\fmftop{t1,t2,t3} \fmfbottom{b1}
	\fmf{half_prop}{i1,t2}
	\fmf{half_prop}{i1,t1}
	\fmf{half_prop}{i1,t3}
	\fmf{half_prop,tension=3}{b1,i1}
	\fmfv{d.shape=square,d.filled=full,d.size=2thick}{i1}
	\fmflabel{$O^{31}$}{b1}
	\end{fmfgraph*}
	\end{fmffile}}
+
	\parbox{40pt}{\begin{fmffile}{O13}
	\begin{fmfgraph*}(40,40)
	\fmfcmd{style_def half_prop expr p =
    draw_plain p;
    shrink(.7);
        cfill (marrow (p, .5))
    endshrink;
	enddef;}
	\fmfstraight
	\fmftop{t1} \fmfbottom{b1,b2,b3}
	\fmf{half_prop}{b1,i1}
	\fmf{half_prop}{b2,i1}
	\fmf{half_prop}{b3,i1}
	\fmf{half_prop,tension=3}{i1,t1}
	\fmfv{d.shape=square,d.filled=full,d.size=2thick}{i1}
	\fmflabel{$O^{13}$}{b2}
	\end{fmfgraph*}
	\end{fmffile}}

\vspace{2\baselineskip}	

$\phantom{O^{[4]} =}$
+
	\parbox{40pt}{\begin{fmffile}{O40}
	\begin{fmfgraph*}(40,40)
	\fmfcmd{style_def half_prop expr p =
    draw_plain p;
    shrink(.7);
        cfill (marrow (p, .5))
    endshrink;
	enddef;}
	\fmfstraight
	\fmftop{t1,t2,t3,t4} \fmfbottom{b1}
	\fmf{half_prop}{i1,t1}
	\fmf{half_prop}{i1,t2}
	\fmf{half_prop}{i1,t3}
	\fmf{half_prop}{i1,t4}
	\fmf{phantom,tension=3}{i1,b1}
	\fmfv{d.shape=square,d.filled=full,d.size=2thick}{i1}
	\fmflabel{$O^{40}$}{b1}
	\end{fmfgraph*}
	\end{fmffile}}
+
	\parbox{40pt}{\begin{fmffile}{O04}
	\begin{fmfgraph*}(40,40)
	\fmfcmd{style_def half_prop expr p =
    draw_plain p;
    shrink(.7);
        cfill (marrow (p, .5))
    endshrink;
	enddef;}
	\fmfstraight
	\fmftop{t1} \fmfbottom{b1,b2,b3,b4}
	\fmf{half_prop}{b1,i1}
	\fmf{half_prop}{b2,i1}
	\fmf{half_prop}{b3,i1}
	\fmf{half_prop}{b4,i1}
	\fmf{phantom,tension=3}{i1,t1}
	\fmfv{d.shape=square,d.filled=full,d.size=2thick}{i1}
	\fmffreeze
	\fmf{phantom,label=$O^{04}$}{b2,b3}
	\end{fmfgraph*}
	\end{fmffile}}

\vspace{2\baselineskip}

$O^{[6]} =$
	\parbox{40pt}{\begin{fmffile}{O33}
	\begin{fmfgraph*}(40,40)
	\fmfcmd{style_def half_prop expr p =
    draw_plain p;
    shrink(.7);
        cfill (marrow (p, .5))
    endshrink;
	enddef;}
	\fmfstraight
	\fmftop{t1,t2,t3} \fmfbottom{b1,b2,b3}
	\fmf{half_prop}{b1,i1}
	\fmf{half_prop}{b2,i1}
	\fmf{half_prop}{b3,i1}
	\fmf{half_prop}{i1,t1}
	\fmf{half_prop}{i1,t2}
	\fmf{half_prop}{i1,t3}
	\fmfv{d.shape=square,d.filled=full,d.size=2thick}{i1}
	\fmflabel{$O^{33}$}{b2}
	\end{fmfgraph*}
	\end{fmffile}}
+
	\parbox{40pt}{\begin{fmffile}{O42}
	\begin{fmfgraph*}(40,40)
	\fmfcmd{style_def half_prop expr p =
    draw_plain p;
    shrink(.7);
        cfill (marrow (p, .5))
    endshrink;
	enddef;}
	\fmfstraight
	\fmftop{t1,t2,t3,t4} \fmfbottom{b1,b2,b3}
	\fmf{half_prop}{i1,t2}
	\fmf{half_prop}{i1,t1}
	\fmf{half_prop}{i1,t3}
	\fmf{half_prop}{i1,t4}
	\fmf{half_prop,tension=2}{b3,i1}
	\fmf{half_prop,tension=2}{b1,i1}
	\fmfv{d.shape=square,d.filled=full,d.size=2thick}{i1}
	\fmflabel{$O^{42}$}{b2}
	\end{fmfgraph*}
	\end{fmffile}}
+
	\parbox{40pt}{\begin{fmffile}{O24}
	\begin{fmfgraph*}(40,40)
	\fmfcmd{style_def half_prop expr p =
    draw_plain p;
    shrink(.7);
        cfill (marrow (p, .5))
    endshrink;
	enddef;}
	\fmfstraight
	\fmftop{t1,t2} \fmfbottom{b1,b2,b3,b4}
	\fmf{half_prop}{b1,i1}
	\fmf{half_prop}{b2,i1}
	\fmf{half_prop}{b3,i1}
	\fmf{half_prop}{b4,i1}
	\fmf{half_prop,tension=2}{i1,t2}
	\fmf{half_prop,tension=2}{i1,t1}
	\fmfv{d.shape=square,d.filled=full,d.size=2thick}{i1}
	\fmffreeze
	\fmf{phantom,label=$O^{24}$}{b2,b3}
	\end{fmfgraph*}
	\end{fmffile}}
+
	\parbox{40pt}{\begin{fmffile}{O51}
	\begin{fmfgraph*}(40,40)
	\fmfcmd{style_def half_prop expr p =
    draw_plain p;
    shrink(.7);
        cfill (marrow (p, .5))
    endshrink;
	enddef;}
	\fmfstraight
	\fmftop{t1,t2,t3,t4,t5} \fmfbottom{b1}
	\fmf{half_prop}{i1,t2}
	\fmf{half_prop}{i1,t1}
	\fmf{half_prop}{i1,t3}
	\fmf{half_prop}{i1,t4}
	\fmf{half_prop}{i1,t5}
	\fmf{half_prop,tension=5}{b1,i1}
	\fmfv{d.shape=square,d.filled=full,d.size=2thick}{i1}
	\fmflabel{$O^{51}$}{b1}
	\end{fmfgraph*}
	\end{fmffile}}
	
\vspace{2\baselineskip}	

$\phantom{O^{[6]} =}$
+
	\parbox{40pt}{\begin{fmffile}{O15}
	\begin{fmfgraph*}(40,40)
	\fmfcmd{style_def half_prop expr p =
    draw_plain p;
    shrink(.7);
        cfill (marrow (p, .5))
    endshrink;
	enddef;}
	\fmfstraight
	\fmftop{t1} \fmfbottom{b1,b2,b3,b4,b5}
	\fmf{half_prop}{b1,i1}
	\fmf{half_prop}{b2,i1}
	\fmf{half_prop}{b3,i1}
	\fmf{half_prop}{b4,i1}
	\fmf{half_prop}{b5,i1}
	\fmf{half_prop,tension=5}{i1,t1}
	\fmfv{d.shape=square,d.filled=full,d.size=2thick}{i1}
	\fmflabel{$O^{15}$}{b3}
	\end{fmfgraph*}
	\end{fmffile}}
+
	\parbox{40pt}{\begin{fmffile}{O60}
	\begin{fmfgraph*}(40,40)
	\fmfcmd{style_def half_prop expr p =
    draw_plain p;
    shrink(.7);
        cfill (marrow (p, .5))
    endshrink;
	enddef;}
	\fmfstraight
	\fmftop{t1,t2,t3,t4,t5,t6} \fmfbottom{b1}
	\fmf{half_prop}{i1,t1}
	\fmf{half_prop}{i1,t2}
	\fmf{half_prop}{i1,t3}
	\fmf{half_prop}{i1,t4}
	\fmf{half_prop}{i1,t5}
	\fmf{half_prop}{i1,t6}
	\fmf{phantom,tension=6}{i1,b1}
	\fmfv{d.shape=square,d.filled=full,d.size=2thick}{i1}
	\fmflabel{$O^{60}$}{b1}
	\end{fmfgraph*}
	\end{fmffile}}
+
	\parbox{40pt}{\begin{fmffile}{O06}
	\begin{fmfgraph*}(40,40)
	\fmfcmd{style_def half_prop expr p =
    draw_plain p;
    shrink(.7);
        cfill (marrow (p, .5))
    endshrink;
	enddef;}
	\fmfstraight
	\fmftop{t1} \fmfbottom{b1,b2,b3,b4,b5,b6}
	\fmf{half_prop}{b1,i1}
	\fmf{half_prop}{b2,i1}
	\fmf{half_prop}{b3,i1}
	\fmf{half_prop}{b4,i1}
	\fmf{half_prop}{b5,i1}
	\fmf{half_prop}{b6,i1}
	\fmf{phantom,tension=6}{i1,t1}
	\fmfv{d.shape=square,d.filled=full,d.size=2thick}{i1}
	\fmffreeze
	\fmf{phantom,label=$O^{06}$}{b3,b4}
	\end{fmfgraph*}
	\end{fmffile}}

\vspace{\baselineskip}

\caption{
\label{f:verticesO}
Canonical diagrammatic representation of normal-ordered contributions to the operator $O$ in the Schr\"odinger representation.}
\end{figure}

\subsection{Diagrammatic representation}
\label{subs:prod_bmbpt_diags}

As discussed in the introduction, the pedestrian application of Wick's theorem becomes quickly cumbersome as the order $p$ increases. Furthermore, it leads to computing independently many contributions that are in fact identical. By identifying the corresponding pattern, one can design a diagrammatic representation of the various contributions and evaluate their algebraic expressions such that a single diagram captures all identical contributions  at once. In order to achieve this goal, one must first introduce the diagrammatic representation of the building blocks. 

The operator $O$ expressed in the quasi-particle basis is displayed in the Schrödinger representation in Fig.~\ref{f:verticesO} as a sum of Hugenholtz vertices denoting its various normal-ordered contributions $O^{ij}$. The antisymmetrized matrix element $O^{ij}_{k_1 \ldots k_i k_{i+1} \ldots k_{i+j}}$ must be assigned to the corresponding square vertex, where $i$ ($j$) denotes the number of lines traveling out (into) of the vertex and representing quasiparticle creation (annihilation) operators. The operator $O(\tau)$ in the interaction representation possesses the same diagrammatic except that a time $\tau$ is attributed to each of the vertices, i.e., to each of the lines coming in or out of them.

\begin{figure}[t!]
\begin{center}
\parbox{40pt}{\begin{fmffile}{OrientRule}
	\begin{fmfgraph*}(40,40)
	\fmfcmd{style_def half_prop expr p =
    draw_plain p;
    shrink(.7);
        cfill (marrow (p, .5))
    endshrink;
	enddef;}
	\fmfstraight
	\fmftop{t1,t2} \fmfbottom{b1,b2}
	\fmfv{label=$+O^{22}_{k_1 k_2 k_3 k_4}$,label.angle=20,label.dist=15pt}{i1}
	\fmf{half_prop}{b1,i1}
	\fmf{half_prop}{b2,i1}
	\fmf{half_prop}{i1,t1}
	\fmf{half_prop}{i1,t2}
	\fmfv{d.shape=square,d.filled=full,d.size=2thick}{i1}
	\fmfv{l=$k_1$,l.d=0.05w}{t1}
	\fmfv{l=$k_2$,l.d=0.05w}{t2}
	\fmfv{l=$k_3$,l.d=0.03w}{b1}
	\fmfv{l=$k_4$,l.d=0.05w}{b2}
	\fmffreeze
	\fmfleft{l1}
	\fmfright{r1}
	\fmf{plain,fore=red}{l1,i1}
	\fmf{dashes,fore=red}{i1,r1}
	\end{fmfgraph*}
	\end{fmffile}}
\hspace{40pt} = \hspace{5pt}
	\parbox{40pt}{\begin{fmffile}{OrientRule_1}
	\begin{fmfgraph*}(40,40)
	\fmfcmd{style_def half_prop expr p =
    draw_plain p;
    shrink(.7);
        cfill (marrow (p, .5))
    endshrink;
	enddef;}
	\fmftop{t1} \fmfbottom{b1,b2,b3,b4}
	\fmfv{label=$+O^{22}_{k_1 k_2 k_3 k_4}$,label.angle=90}{i1}
	\fmf{half_prop}{b1,i1}
	\fmf{half_prop}{b2,i1}
	\fmf{half_prop}{i1,b3}
	\fmf{half_prop}{i1,b4}
	\fmf{phantom,tension=3}{i1,t1}
	\fmfv{d.shape=square,d.filled=full,d.size=2thick}{i1}
	\fmfv{l=$k_3$,l.d=0.03w}{b1}
	\fmfv{l=$k_4$,l.d=0.05w}{b2}
	\fmfv{l=$k_2$,l.d=0.05w}{b3}
	\fmfv{l=$k_1$,l.d=0.05w}{b4}
	\end{fmfgraph*}
	\end{fmffile}}
\hspace{5pt} = \hspace{10pt}
	\parbox{40pt}{\begin{fmffile}{OrientRule_2}
	\begin{fmfgraph*}(40,40)
	\fmfcmd{style_def half_prop expr p =
    draw_plain p;
    shrink(.7);
        cfill (marrow (p, .5))
    endshrink;
	enddef;}
	\fmftop{t1} \fmfbottom{b1,b2,b3,b4}
	\fmfv{label=$-O^{22}_{k_1 k_2 k_3 k_4}$,label.angle=90}{i1}
	\fmf{half_prop}{b1,i1}
	\fmf{half_prop}{i1,b2}
	\fmf{half_prop}{b3,i1}
	\fmf{half_prop}{i1,b4}
	\fmf{phantom,tension=3}{i1,t1}
	\fmfv{d.shape=square,d.filled=full,d.size=2thick}{i1}
	\fmfv{l=$k_3$,l.d=0.03w}{b1}
	\fmfv{l=$k_2$,l.d=0.05w}{b2}
	\fmfv{l=$k_4$,l.d=0.05w}{b3}
	\fmfv{l=$k_1$,l.d=0.05w}{b4}
	\end{fmfgraph*}
	\end{fmffile}}
\end{center}
\caption{
\label{variousvertices3}
Rules to apply when departing from the canonical diagrammatic representation of a normal-ordered operator. Oriented lines can be rotated through the dashed line but not through the full line.}
\end{figure}
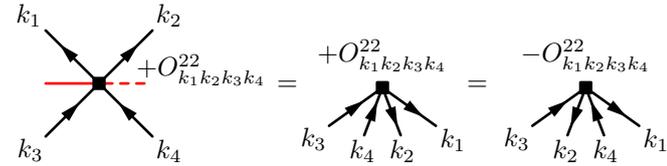

In the canonical representation used in Fig.~\ref{f:verticesO}, all oriented lines go up, i.e., lines representing quasiparticle creation (annihilation) operators appear above (below) the vertex. Accordingly, indices $k_1 \ldots k_i$ must be assigned consecutively from the leftmost to the rightmost line above the vertex, while $k_{i+1} \ldots k_{i+j}$ must be similarly assigned consecutively for lines below the vertex. In the diagrammatic representation of the observable  $\text{O}^{\text{A}}_0$, it is possible for a line to propagate downwards. This can be obtained unambiguously starting from the canonical representation of Fig.~\ref{f:verticesO}  at the price of adding a specific rule. As illustrated in Fig.~\ref{variousvertices3} for the diagram representing $O^{22}$, lines must only be rotated through the right of the diagram, i.e., going through the dashed line, while it is forbidden to rotate them through the full line. Additionally, a minus sign must be added to the amplitude $O^{ij}_{k_1 \ldots k_i k_{i+1} \ldots k_{i+j}}$ associated with the canonical diagram each time two lines cross as illustrated in Fig.~\ref{variousvertices3}.

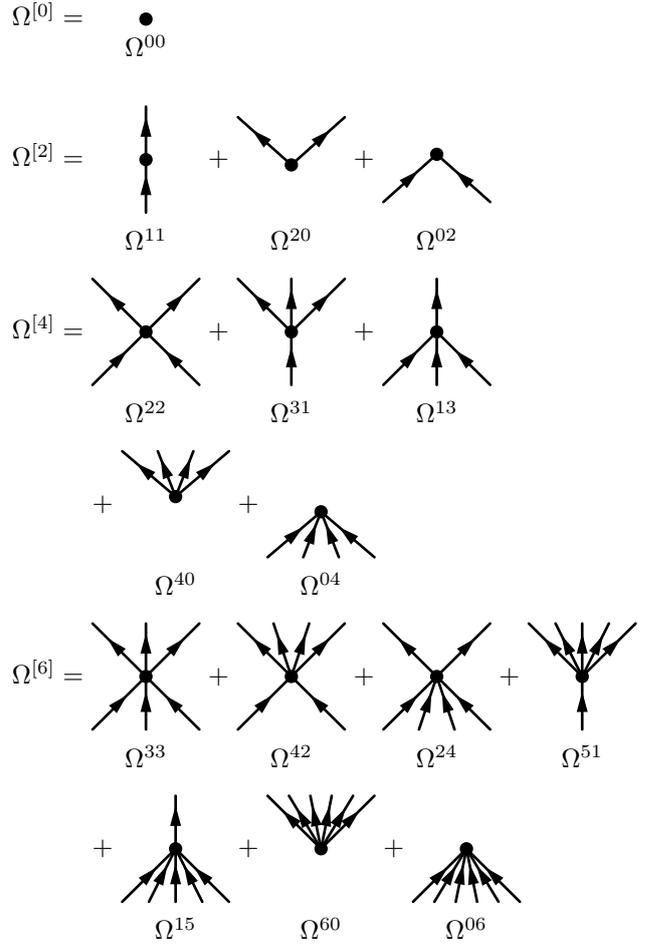
\begin{figure}[t!]
$\Omega^{[0]} =$
	\parbox{40pt}{\begin{fmffile}{Omega00}
	\begin{fmfgraph*}(40,40)
	\fmftop{t1} \fmfbottom{b1}
	\fmf{phantom}{b1,i1}
	\fmf{phantom}{i1,t1}
	\fmfv{label=$\Omega^{00}$,label.angle=-90,d.shape=circle,d.filled=full,d.size=2thick}{i1}
	\end{fmfgraph*}
	\end{fmffile}}

\vspace{\baselineskip}

$\Omega^{[2]} =$
	\parbox{40pt}{\begin{fmffile}{Omega11}
	\begin{fmfgraph*}(40,40)
	\fmfcmd{style_def half_prop expr p =
    draw_plain p;
    shrink(.7);
        cfill (marrow (p, .5))
    endshrink;
	enddef;}
	\fmftop{t1} \fmfbottom{b1}
	\fmf{half_prop}{b1,i1}
	\fmf{half_prop}{i1,t1}
	\fmfv{d.shape=circle,d.filled=full,d.size=2thick}{i1}
	\fmflabel{$\Omega^{11}$}{b1}
	\end{fmfgraph*}
	\end{fmffile}}
+
	\parbox{40pt}{\begin{fmffile}{Omega20}
	\begin{fmfgraph*}(40,40)
	\fmfcmd{style_def half_prop expr p =
    draw_plain p;
    shrink(.7);
        cfill (marrow (p, .5))
    endshrink;
	enddef;}
	\fmftop{t1,t2} \fmfbottom{b1}
	\fmfstraight
	\fmf{half_prop}{i1,t2}
	\fmf{half_prop}{i1,t1}
	\fmf{phantom,tension=2}{b1,i1}
	\fmfv{d.shape=circle,d.filled=full,d.size=2thick}{i1}
	\fmflabel{$\Omega^{20}$}{b1}
	\end{fmfgraph*}
	\end{fmffile}}
+
	\parbox{40pt}{\begin{fmffile}{Omega02}
	\begin{fmfgraph*}(40,40)
	\fmfcmd{style_def half_prop expr p =
    draw_plain p;
    shrink(.7);
        cfill (marrow (p, .5))
    endshrink;
	enddef;}
	\fmftop{t1} \fmfbottom{b1,b2,b3}
	\fmfstraight
	\fmf{half_prop}{b1,i1}
	\fmf{half_prop}{b3,i1}
	\fmf{phantom,tension=2}{i1,t1}
	\fmfv{d.shape=circle,d.filled=full,d.size=2thick}{i1}
	\fmflabel{$\Omega^{02}$}{b2}
	\end{fmfgraph*}
	\end{fmffile}}

\vspace{2\baselineskip}

$\Omega^{[4]} =$
	\parbox{40pt}{\begin{fmffile}{Omega22}
	\begin{fmfgraph*}(40,40)
	\fmfcmd{style_def half_prop expr p =
    draw_plain p;
    shrink(.7);
        cfill (marrow (p, .5))
    endshrink;
	enddef;}
	\fmfstraight
	\fmftop{t1,t2} \fmfbottom{b1,b2,b3}
	\fmf{half_prop}{b1,i1}
	\fmf{half_prop}{b3,i1}
	\fmf{half_prop}{i1,t1}
	\fmf{half_prop}{i1,t2}
	\fmfv{d.shape=circle,d.filled=full,d.size=2thick}{i1}
	\fmflabel{$\Omega^{22}$}{b2}
	\end{fmfgraph*}
	\end{fmffile}}
+
	\parbox{40pt}{\begin{fmffile}{Omega31}
	\begin{fmfgraph*}(40,40)
	\fmfcmd{style_def half_prop expr p =
    draw_plain p;
    shrink(.7);
        cfill (marrow (p, .5))
    endshrink;
	enddef;}
	\fmfstraight
	\fmftop{t1,t2,t3} \fmfbottom{b1}
	\fmf{half_prop}{i1,t2}
	\fmf{half_prop}{i1,t1}
	\fmf{half_prop}{i1,t3}
	\fmf{half_prop,tension=3}{b1,i1}
	\fmfv{d.shape=circle,d.filled=full,d.size=2thick}{i1}
	\fmflabel{$\Omega^{31}$}{b1}
	\end{fmfgraph*}
	\end{fmffile}}
+
	\parbox{40pt}{\begin{fmffile}{Omega13}
	\begin{fmfgraph*}(40,40)
	\fmfcmd{style_def half_prop expr p =
    draw_plain p;
    shrink(.7);
        cfill (marrow (p, .5))
    endshrink;
	enddef;}
	\fmfstraight
	\fmftop{t1} \fmfbottom{b1,b2,b3}
	\fmf{half_prop}{b1,i1}
	\fmf{half_prop}{b2,i1}
	\fmf{half_prop}{b3,i1}
	\fmf{half_prop,tension=3}{i1,t1}
	\fmfv{d.shape=circle,d.filled=full,d.size=2thick}{i1}
	\fmflabel{$\Omega^{13}$}{b2}
	\end{fmfgraph*}
	\end{fmffile}}
	
\vspace{2\baselineskip}	

$\phantom{\Omega^{[4]} =}$
+
	\parbox{40pt}{\begin{fmffile}{Omega40}
	\begin{fmfgraph*}(40,40)
	\fmfcmd{style_def half_prop expr p =
    draw_plain p;
    shrink(.7);
        cfill (marrow (p, .5))
    endshrink;
	enddef;}
	\fmfstraight
	\fmftop{t1,t2,t3,t4} \fmfbottom{b1}
	\fmf{half_prop}{i1,t1}
	\fmf{half_prop}{i1,t2}
	\fmf{half_prop}{i1,t3}
	\fmf{half_prop}{i1,t4}
	\fmf{phantom,tension=3}{i1,b1}
	\fmfv{d.shape=circle,d.filled=full,d.size=2thick}{i1}
	\fmflabel{$\Omega^{40}$}{b1}
	\end{fmfgraph*}
	\end{fmffile}}
+
	\parbox{40pt}{\begin{fmffile}{Omega04}
	\begin{fmfgraph*}(40,40)
	\fmfcmd{style_def half_prop expr p =
    draw_plain p;
    shrink(.7);
        cfill (marrow (p, .5))
    endshrink;
	enddef;}
	\fmfstraight
	\fmftop{t1} \fmfbottom{b1,b2,b3,b4}
	\fmf{half_prop}{b1,i1}
	\fmf{half_prop}{b2,i1}
	\fmf{half_prop}{b3,i1}
	\fmf{half_prop}{b4,i1}
	\fmf{phantom,tension=3}{i1,t1}
	\fmfv{d.shape=circle,d.filled=full,d.size=2thick}{i1}
	\fmffreeze
	\fmf{phantom,label=$\Omega^{04}$}{b2,b3}
	\end{fmfgraph*}
	\end{fmffile}}

\vspace{2\baselineskip}

$\Omega^{[6]} =$
	\parbox{40pt}{\begin{fmffile}{Omega33}
	\begin{fmfgraph*}(40,40)
	\fmfcmd{style_def half_prop expr p =
    draw_plain p;
    shrink(.7);
        cfill (marrow (p, .5))
    endshrink;
	enddef;}
	\fmfstraight
	\fmftop{t1,t2,t3} \fmfbottom{b1,b2,b3}
	\fmf{half_prop}{b1,i1}
	\fmf{half_prop}{b2,i1}
	\fmf{half_prop}{b3,i1}
	\fmf{half_prop}{i1,t1}
	\fmf{half_prop}{i1,t2}
	\fmf{half_prop}{i1,t3}
	\fmfv{d.shape=circle,d.filled=full,d.size=2thick}{i1}
	\fmflabel{$\Omega^{33}$}{b2}
	\end{fmfgraph*}
	\end{fmffile}}
+
	\parbox{40pt}{\begin{fmffile}{Omega42}
	\begin{fmfgraph*}(40,40)
	\fmfcmd{style_def half_prop expr p =
    draw_plain p;
    shrink(.7);
        cfill (marrow (p, .5))
    endshrink;
	enddef;}
	\fmfstraight
	\fmftop{t1,t2,t3,t4} \fmfbottom{b1,b2,b3}
	\fmf{half_prop}{i1,t2}
	\fmf{half_prop}{i1,t1}
	\fmf{half_prop}{i1,t3}
	\fmf{half_prop}{i1,t4}
	\fmf{half_prop,tension=2}{b3,i1}
	\fmf{half_prop,tension=2}{b1,i1}
	\fmfv{d.shape=circle,d.filled=full,d.size=2thick}{i1}
	\fmflabel{$\Omega^{42}$}{b2}
	\end{fmfgraph*}
	\end{fmffile}}
+
	\parbox{40pt}{\begin{fmffile}{Omega24}
	\begin{fmfgraph*}(40,40)
	\fmfcmd{style_def half_prop expr p =
    draw_plain p;
    shrink(.7);
        cfill (marrow (p, .5))
    endshrink;
	enddef;}
	\fmfstraight
	\fmftop{t1,t2} \fmfbottom{b1,b2,b3,b4}
	\fmf{half_prop}{b1,i1}
	\fmf{half_prop}{b2,i1}
	\fmf{half_prop}{b3,i1}
	\fmf{half_prop}{b4,i1}
	\fmf{half_prop,tension=2}{i1,t2}
	\fmf{half_prop,tension=2}{i1,t1}
	\fmfv{d.shape=circle,d.filled=full,d.size=2thick}{i1}
	\fmffreeze
	\fmf{phantom,label=$\Omega^{24}$}{b2,b3}
	\end{fmfgraph*}
	\end{fmffile}}
+
	\parbox{40pt}{\begin{fmffile}{Omega51}
	\begin{fmfgraph*}(40,40)
	\fmfcmd{style_def half_prop expr p =
    draw_plain p;
    shrink(.7);
        cfill (marrow (p, .5))
    endshrink;
	enddef;}
	\fmfstraight
	\fmftop{t1,t2,t3,t4,t5} \fmfbottom{b1}
	\fmf{half_prop}{i1,t2}
	\fmf{half_prop}{i1,t1}
	\fmf{half_prop}{i1,t3}
	\fmf{half_prop}{i1,t4}
	\fmf{half_prop}{i1,t5}
	\fmf{half_prop,tension=5}{b1,i1}
	\fmfv{d.shape=circle,d.filled=full,d.size=2thick}{i1}
	\fmflabel{$\Omega^{51}$}{b1}
	\end{fmfgraph*}
	\end{fmffile}}
	
\vspace{2\baselineskip}	

$\phantom{\Omega^{[6]} =}$
+
	\parbox{40pt}{\begin{fmffile}{Omega15}
	\begin{fmfgraph*}(40,40)
	\fmfcmd{style_def half_prop expr p =
    draw_plain p;
    shrink(.7);
        cfill (marrow (p, .5))
    endshrink;
	enddef;}
	\fmfstraight
	\fmftop{t1} \fmfbottom{b1,b2,b3,b4,b5}
	\fmf{half_prop}{b1,i1}
	\fmf{half_prop}{b2,i1}
	\fmf{half_prop}{b3,i1}
	\fmf{half_prop}{b4,i1}
	\fmf{half_prop}{b5,i1}
	\fmf{half_prop,tension=5}{i1,t1}
	\fmfv{d.shape=circle,d.filled=full,d.size=2thick}{i1}
	\fmflabel{$\Omega^{15}$}{b3}
	\end{fmfgraph*}
	\end{fmffile}}
+
	\parbox{40pt}{\begin{fmffile}{Omega60}
	\begin{fmfgraph*}(40,40)
	\fmfcmd{style_def half_prop expr p =
    draw_plain p;
    shrink(.7);
        cfill (marrow (p, .5))
    endshrink;
	enddef;}
	\fmfstraight
	\fmftop{t1,t2,t3,t4,t5,t6} \fmfbottom{b1}
	\fmf{half_prop}{i1,t1}
	\fmf{half_prop}{i1,t2}
	\fmf{half_prop}{i1,t3}
	\fmf{half_prop}{i1,t4}
	\fmf{half_prop}{i1,t5}
	\fmf{half_prop}{i1,t6}
	\fmf{phantom,tension=6}{i1,b1}
	\fmfv{d.shape=circle,d.filled=full,d.size=2thick}{i1}
	\fmflabel{$\Omega^{60}$}{b1}
	\end{fmfgraph*}
	\end{fmffile}}
+
	\parbox{40pt}{\begin{fmffile}{Omega06}
	\begin{fmfgraph*}(40,40)
	\fmfcmd{style_def half_prop expr p =
    draw_plain p;
    shrink(.7);
        cfill (marrow (p, .5))
    endshrink;
	enddef;}
	\fmfstraight
	\fmftop{t1} \fmfbottom{b1,b2,b3,b4,b5,b6}
	\fmf{half_prop}{b1,i1}
	\fmf{half_prop}{b2,i1}
	\fmf{half_prop}{b3,i1}
	\fmf{half_prop}{b4,i1}
	\fmf{half_prop}{b5,i1}
	\fmf{half_prop}{b6,i1}
	\fmf{phantom,tension=6}{i1,t1}
	\fmfv{d.shape=circle,d.filled=full,d.size=2thick}{i1}
	\fmffreeze
	\fmf{phantom,label=$\Omega^{06}$}{b3,b4}
	\end{fmfgraph*}
	\end{fmffile}}

\vspace{\baselineskip}

\caption{
\label{f:verticesOmega}
Canonical diagrammatic representation of normal-ordered contributions to the grand potential operator $\Omega$ in the Schr\"odinger representation.}
\end{figure}

Since the grand canonical potential $\Omega$ is involved in the evaluation of any observable $\text{O}^{\text{A}}_0$, its own diagrammatic representation is needed and displayed in Fig.~\ref{f:verticesOmega}. The only difference with  Fig.~\ref{f:verticesO} relates to the use of dots rather than square symbols to represent the vertices. The same is easily done for other operators of interest, i.e.,\@ $H$ and $A$. It is to be noted that $\Omega_{1}$ has the same diagrammatic representation as $\Omega$ except that $\Omega^{00}$ must be omitted and $\Omega^{11}$ replaced by $\breve{\Omega}^{11}$, which requires to use a different symbol for that particular vertex\footnote{We omit to use a different symbol for $\breve{\Omega}^{11}$ in the following although it must be clear that the vertex with one line coming in and one line coming out does represent $\breve{\Omega}^{11}$ whenever it originates from the perturbative expansion of the evolution operator. This may be confusing whenever $O=\Omega$ since in this case there can also be a vertex $\Omega^{11}$ at fixed time $t=0$.}.

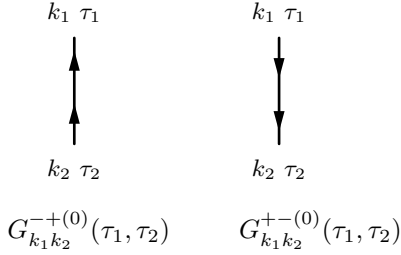
\begin{figure}[t!]
\vspace{1\baselineskip}

\begin{center}
\parbox{60pt}{\begin{fmffile}{Gplumotest}
\begin{fmfgraph*}(40,40)
\fmfcmd{style_def prop_pm expr p =
    draw_plain p;
    shrink(.7);
        cfill (marrow (p, .25));
        cfill (marrow (p, .75))
    endshrink;
enddef;
}
\fmftop{t1} \fmfbottom{b1}
\fmflabel{\small $k_1 \ \tau_1$}{t1}
\fmflabel{\small $k_2 \ \tau_2$}{b1}
\fmf{prop_pm}{b1,t1}
\end{fmfgraph*}
\end{fmffile}}
\hbox{\quad}
\parbox{60pt}{\begin{fmffile}{Gmoplutest}
\begin{fmfgraph*}(40,40)
\fmfcmd{style_def prop_mp expr p =
    draw_plain p;
    shrink(.7);
        cfill (marrow (reverse p, .25));
        cfill (marrow (reverse p, .75))
    endshrink;
enddef;
}
\fmftop{t1} \fmfbottom{b1}
\fmflabel{\small $k_1 \ \tau_1$}{t1}
\fmflabel{\small $k_2 \ \tau_2$}{b1}
\fmf{prop_mp}{b1,t1}
\end{fmfgraph*}
\end{fmffile}}

\vspace{2\baselineskip}

\parbox{60pt}{$G^{-+ (0)}_{k_1k_2}(\tau_1, \tau_2)$} \hbox{\quad\quad}
\parbox{60pt}{$G^{+- (0)}_{k_1k_2}(\tau_1, \tau_2)$}
\end{center}
\caption{
\label{f:prop}
Diagrammatic representation of the two unperturbed one-body propagators. The convention is that the left-to-right reading of a matrix element corresponds to the up-down reading of the diagram. Correspondingly, time goes upward in the diagrams.}
\end{figure}

As Wick's theorem contracts pairs of quasi-particle operators together, the lines entering the diagrammatic representation of operators are eventually connected in the computation of the observable $\text{O}^{\text{A}}_0$, thus, forming elementary contractions. Consequently, the two unperturbed propagators at play also need to be represented diagrammatically, which is done in Fig.~\ref{f:prop}. Here, the convention is that the left-to-right reading of a matrix element corresponds to the up-down reading of the diagram. Furthermore, by virtue of Eq.~\eqref{e:prop_antisym}, the reading of, e.g., a $G^{-+ (0)}$ propagator can be reinterpreted as $G^{+- (0)}$ such that a single type of propagator is necessary in the end.

\subsection{Diagram generation}
\label{subs:diag_gene}

Once the building blocks have been introduced, BMBPT Feynman diagrams representing the contributions to  $\text{O}^{\text{A}}_0$ are generated by assembling them according to a set of topological rules~\cite{Duguet:2015yle,Art18b}
\begin{enumerate}
\item A Feynman diagram of order $p$ consists of $p$ vertices $\Omega^{i_kj_k}(\tau_k)$, $i_k+j_k=2,4$ or $6$, along with one vertex $O^{mn}(0)$, $m+n=0,2,4$ or $6$, that are connected by fermionic quasi-particle lines, i.e., via non-zero propagators $G^{+- (0)}$ or $G^{-+ (0)}$. 
\item Each vertex is labeled by a time variable while each line is labeled by two time labels associated with the two vertices the line is attached to. 
\item Generating all contributions to Eq.~\eqref{observableO1} requires to form all possible diagrams, i.e., contract quasi-particle lines attached to the vertices in all possible ways while fulfilling the following restrictions.
\begin{enumerate}
\item Forbid equal-time propagators starting and ending at the same vertex as they are zero, i.e., no contraction of a vertex onto itself is to be considered.
\item Restrict the set to \emph{connected} diagrams, i.e., omit diagrams containing parts that are not connected to each other by either propagators or vertices. This implies in particular that the vertex $O^{00}$ with no line can only appear at order $p=0$\footnote{It is the only vertex appearing at order $0$ given that the vacuum expectation value of all the other terms is zero by virtue of their normal-ordered character. This is a particular occurence of the rule stipulating that no contraction of a vertex onto itself is to be considered.}.
\item The generic operator $O$ at fixed time $0$ is necessarily at the bottom of the diagram. Its contributing vertices $O^{mn}(0)$ can only have propagators going out. Indeed, a line going in would necessarily be associated with a propagator $G^{+- (0)}$  carrying a step function contradicting the fact that all the running times are positive (see Eq.~\eqref{e:propagatorsGpm}). Consequently, contributing vertices are restricted to $O^{m0}(0)$, $m=0,2,4$ or $6$.
\item Because of the time-ordering relations carried by the propagators (see Eq.~\eqref{e:propagators}), lines linking a set of vertices must not form an oriented loop. For a set of two given vertices $\Omega^{i_kj_k}(\tau_k)$ and $\Omega^{i_{k'}j_{k'}}(\tau_{k'})$, it means that lines must propagate in the same direction.
\item Restrict the set to \emph{vacuum-to-vacuum} diagrams forming a set of closed loops with no external, i.e., unpaired, lines. This condition strongly constrains which normal-ordered parts $\Omega^{i_kj_k}(\tau_k)$ and $O^{m0}(0)$ of the $p+1$ involved operators can be combined, i.e., the condition 
\begin{equation}
n_a \equiv \sum_{k=1}^{p}(j_k-i_k) -m= 0 \, , \nonumber
\end{equation}
must be fulfilled.
\item Restrict the set to \emph{topologically distinct} time-unlabelled diagrams, i.e., time-unlabelled diagrams that cannot be obtained from one another via a mere displacement, i.e., translation, of the vertices.
\end{enumerate}
\end{enumerate}

\subsection{Diagram evaluation}
\label{subs:diag_eval}

\subsubsection{Feynman expression}
\label{subs:feyndiag_eval}

The way to translate BMBPT Feynman diagrams into their mathematical expressions follows the set of algebraic rules
\begin{enumerate}
\item Each of the $p+1$ vertices contributes a factor, e.g., $\Omega^{ij}_{k_1 \ldots k_i k_{i+1} \ldots k_{i+j}}$ with the sign convention detailed in Sec.~\ref{subs:prod_bmbpt_diags}. 
\item Each of the 
\begin{equation}
n_b\equiv \left(\sum_{k=1}^{p}\left(j_k+i_k\right) +m\right)/2 \, , \nonumber
\end{equation}
lines contributes a factor $G^{gg' (0)}_{k_1k_2}(\tau_k, \tau_{k'})$, where $g=\pm=-g'$ characterize the type of elementary propagator the line corresponds to in agreement with the convention of Fig.~\ref{f:prop}. According to Eq.~\eqref{e:propagators}, each of the $n_b$ propagators carries an exponential function and a step function of the time labels associated with the two vertices it connects. 
\item A normal line can be interpreted as $G^{-+ (0)}$ or $G^{+- (0)}$ depending on the ascendant or descendant reading of the diagram. Similarly, the ordering of quasi-particle and time labels of a propagator depends on the ascendant or descendant reading of the diagram. \emph{All} the lines involved in a given diagram must be interpreted in the \emph{same} way, i.e., sticking to an ascendant or descendant way of reading the diagram all throughout. By default the diagrams are intended to be read in a descendant fashion, which corresponds to reading the matrix element it originates from in a left-right fashion\footnote{Reading them in an ascendant one is possible but requires an additional factor $(-1)^{n_p}$, with $n_p$ the number of propagators in the diagram.}.
\item All quasi-particle labels must be summed over while all running time variables must be integrated over from $0$ to $\tau\rightarrow +\infty$.
\item A sign factor $(-1)^{p+n_c}$, where $p$ denotes the order of the diagram and $n_c$ denotes the number of crossing lines in the diagram, must be considered. The overall sign results from multiplying this factor with the sign associated with each matrix element.
\item Each diagram comes with a numerical prefactor obtained from the following combination
\begin{enumerate}
\item A factor $1/(n_e)!$ must be considered for \emph{each} group of $n_e$ equivalent lines. Equivalent lines begin and end at the same vertices.
\item A symmetry factor $1/n_s$ must be considered in connection with exchanging the time labels of the vertices in all possible ways, counting the identity as one. The factor $n_s$ corresponds to the number of ways exchanging the time labels provides a time-labelled diagram that is topologically equivalent to the original one.
\end{enumerate} 
\end{enumerate} 

\begin{figure}[t!]
\begin{center}
\parbox{120pt}{\begin{fmffile}{diag10_labelled}
\begin{fmfgraph*}(120,120)
\fmfcmd{style_def prop_pm expr p =
    draw_plain p;
    shrink(.7);
        cfill (marrow (p, .25));
        cfill (marrow (p, .75))
    endshrink;
	enddef;}
\fmftop{v3}\fmfbottom{v0}
\fmf{phantom}{v0,v1}
\fmfv{d.shape=square,d.filled=full,d.size=3thick,l=$0$}{v0}
\fmf{phantom}{v1,v2}
\fmfv{d.shape=circle,d.filled=full,d.size=3thick,l=$\tau_1$}{v1}
\fmf{phantom}{v2,v3}
\fmfv{d.shape=circle,d.filled=full,d.size=3thick,l=$\tau_2$}{v2}
\fmfv{d.shape=circle,d.filled=full,d.size=3thick,l=$\tau_3$}{v3}
\fmffreeze
\fmf{prop_pm,left=0.6,tag=1}{v0,v2}
\fmf{prop_pm,right=0.6,tag=2}{v0,v2}
\fmf{prop_pm,left=0.6,tag=3}{v0,v3}
\fmf{prop_pm,right=0.6,tag=4}{v0,v3}
\fmf{prop_pm,left=0.5,tag=5}{v1,v2}
\fmf{prop_pm,right=0.5,tag=6}{v1,v2}
\fmf{prop_pm,left=0.6,tag=7}{v1,v3}
\fmf{prop_pm,right=0.6,tag=8}{v1,v3}
\fmfposition
      \fmfipath{p[]}
      \fmfiset{p1}{vpath1(__v0,__v2)}
      \fmfiset{p2}{vpath2(__v0,__v2)}
      \fmfiset{p3}{vpath3(__v0,__v3)}
      \fmfiset{p4}{vpath4(__v0,__v3)}
      \fmfiset{p5}{vpath5(__v1,__v2)}
      \fmfiset{p6}{vpath6(__v1,__v2)}
      \fmfiset{p7}{vpath7(__v1,__v3)}
      \fmfiset{p8}{vpath8(__v1,__v3)}
      \fmfiv{label=$k_1$,l.dist=.05w,l.a=-20}{point length(p1)/3 of p1}
      \fmfiv{label=$k_2$,l.dist=.05w,l.a=-160}{point length(p2)/3 of p2}
      \fmfiv{label=$k_3$,l.dist=.03w}{point length(p3)/2 of p3}
      \fmfiv{label=$k_4$,l.dist=.03w}{point length(p4)/2 of p4}
      \fmfiv{label=$k_5$,l.dist=.03w,l.a=-30}{point length(p5)/2 of p5}
      \fmfiv{label=$k_6$,l.dist=.03w,l.a=150}{point length(p6)/2 of p6}
      \fmfiv{label=$k_7$,l.dist=.05w,l.a=20}{point 3length(p7)/4 of p7}
      \fmfiv{label=$k_8$,l.dist=.05w,l.a=160}{point 3length(p8)/4 of p8}
      \fmfiv{label=$O^{40}$,l.dist=.08w,l.a=-20}{point 0 of p1}
      \fmfiv{label=$\Omega^{40}$,l.dist=.03w,l.a=-20}{point 0 of p8}
      \fmfiv{label=$\Omega^{04}$,l.dist=.07w,l.a=20}{point length(p5) of p5}
      \fmfiv{label=$\Omega^{04}$,l.dist=.05w,l.a=30}{point length(p4) of p4}
\end{fmfgraph*}
\end{fmffile}}
\end{center}
\caption{A third-order Feynman BMBPT diagram.}
\label{f:ex_bmbpt_diag}
\end{figure}
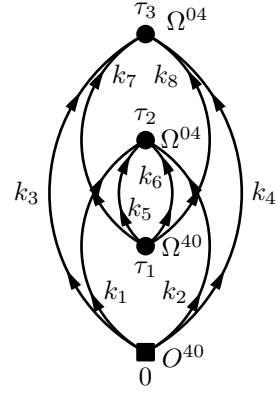

In order to illutrate the typical expression of a Feynman BMBPT diagram, let us write it for the third-order diagram displayed\footnote{It is not customary to keep two arrows visible on an oriented propagator. While it is indeed redundant to make the two arrows visible in BMBPT diagrams, we do so to anticipate the diagrammatic at play in PNP-BMBPT~\cite{Duguet:2015yle} that also invokes {\it anomalous} quasi-particle propagators, i.e., propagators carrying two arrows in opposite directions.} in Fig.~\ref{f:ex_bmbpt_diag}, i.e., 
\begin{align}
D &= \lim\limits_{\tau \to \infty}\frac{(-1)^3 }{2(2!)^4}\sum_{k_i}O^{40}_{k_{1}k_{2}k_{3}k_{4}} \Omega^{40}_{k_{5}k_{6}k_{7}k_{8}} \Omega^{04}_{k_{5}k_{6}k_{1}k_{2}} \Omega^{04}_{k_{7}k_{8}k_{3}k_{4}} \nonumber \\
&\phantom{=} \times \int_{0}^{\tau}\mathrm{d}\tau_1\mathrm{d}\tau_2\mathrm{d}\tau_3 \, \theta(\tau_2-\tau_1) \theta(\tau_3-\tau_1) \label{e:ex_bmbpt_feynman}\\
&\phantom{=\times \int_{0}^{\tau}} \times e^{-\tau_1 \epsilon_{}^{k_{5}k_{6}k_{7}k_{8}}}e^{-\tau_2 \epsilon_{k_{1}k_{2}k_{5}k_{6}}^{}}e^{-\tau_3 \epsilon_{k_{3}k_{4}k_{7}k_{8}}^{}}  \, ,\nonumber 
\end{align}
where the notation 
\begin{equation}
\epsilon^{k_{a}k_{b}\dots}_{k_{i}k_{j}\dots} \equiv  E_{k_{i}} +  E_{k_{j}} + \ldots - E_{k_{a}} -  E_{k_{b}} - \ldots
\end{equation}
was introduced. The sign, the combinatorial factors and the four matrix elements directly reflect Feynman's algebraic rules listed above and are easy to interpret. The final form of the integrand originates from expliciting the $n_b = 8$ propagators $G^{-+(0)}$ and displays a typical structure that needs to be scrutinized for the following. 
\begin{itemize}
\item While the vertex $O^{40}$ is at fixed time $0$, the $\Omega^{40}$ vertex is at running time $\tau_1$ and the two $\Omega^{04}$ vertices are at running times $\tau_2$ and $\tau_3$. The two step functions characterize the time ordering between $\Omega^{40}$ and each of the two $\Omega^{04}$ vertices it is directly connected to via propagators. Contrarily, the two $\Omega^{04}$ vertices are not connected via propagators and do not belong to a linear sequence of connected vertices such that their time labels are not ordered with respect to one another. Eventually, the fact that the three running variables are positive is directly encoded into the boundary of the triple integral.
\item Grouping appropriately the exponential functions coming from the $8$ propagators, the integrand displays one exponential factor per running time, i.e., per $\Omega^{i_kj_k}(\tau_k)$ vertex. The relevant energy factor $\epsilon^{k_{a}k_{b}\dots}_{k_{i}k_{j}\dots} $ multiplying the variable $\tau_k$ in this exponential function denotes the sum/difference of quasi-particle energies associated with the lines entering/leaving the corresponding vertex.
\end{itemize}
The two above points characterizing (a) the steps functions associated with the links between vertices and (b) the exponential function associated with each vertex can be viewed as an optimal rephrasing of the algebraic rule 3 stipulated above.

\subsubsection{Time-integrated expression}
\label{subs:feynman_to_goldstone}

The expression obtained via the application of Feynman's algebraic rules does not yet constitute the form needed for the numerical implementation of the formalism. While the sign, the combinatorial factor and the matrix elements will remain untouched, the $p$-tuple time integral must be performed in order to obtain the needed expression. Generically denoting as  $a_k$ the energy factor multiplying the time label $\tau_k$ in the integrand, the integral  associated with our example of Fig.~\ref{f:ex_bmbpt_diag} can be worked out in the following fashion
\begin{align}
T &= \lim\limits_{\tau \to \infty} \int_0^{\tau} d\tau_1 d\tau_2 d\tau_3 \theta(\tau_2 - \tau_1) \theta(\tau_3 - \tau_1) e^{-a_1\tau_1-a_2\tau_2-a_3\tau_3} \nonumber \\
&= \lim\limits_{\tau \to \infty} \int_0^{\tau} d\tau_1 \, e^{-a_1\tau_1} \int_{\tau_1}^{\tau} d\tau_2 \, e^{-a_2\tau_2} \int_{\tau_1}^{\tau} d\tau_3 \, e^{-a_3\tau_3} \nonumber \\
&= \lim\limits_{\tau \to \infty} \int_0^{\tau} d\tau_1 \, e^{-a_1\tau_1} \frac{e^{-a_2\tau} - e^{-a_2\tau_1}}{a_2} \frac{e^{-a_3\tau} - e^{-a_3\tau_1}}{a_3} \nonumber \\
&= \frac{1}{(a_1+a_2+a_3)a_2a_3} \ . \label{D}
\end{align}
Two important lessons can be learned from this particular example. 
\begin{itemize}
\item Exploiting the time-ordering relations imposed by the step functions, one performs the $p$ integrals following a specific sequence, i.e., one starts with the greatest time label whose corresponding integral is written in the right-most position before proceeding to integrals over smaller times, i.e., moving in steps towards the leftmost integral. In doing so, we see that the integrals over $\tau_2$ and $\tau_3$ are in fact independent as the two time variables are not ordered with respect to each other and only entertain a causal relation with respect to a common variable, i.e., $\tau_1$, corresponding to an \emph{earlier} time. The integral over $\tau_1$ does depend on the result of the integrals over $\tau_2$ and $\tau_3$ and is thus performed in last. 
\item While the energy variables entering the time integrand are $a_1$, $a_2$ and $a_3$, the end result takes the form of a fraction whose non-trivial factors appear in the denominator and are specific combinations of these original energy variables. Expressing $a_1$, $a_2$ and $a_3$ back in terms of quasi-particle energies, these combinations read as
\begin{align}
a_1+a_2+a_3 &= \epsilon_{k_{1}k_{2}k_{3}k_{4}}^{} \, ,  \nonumber \\
a_2 &= \epsilon_{k_{1}k_{2}k_{5}k_{6}}^{} \, , \label{energylabels}\\
a_3 &= \epsilon_{k_{3}k_{4}k_{7}k_{8}}^{}\, ,  \nonumber
\end{align}
and thus correspond to positive sums of quasi-particle energies. We will identify later on what these specific combinations  of quasi-particle energies actually correspond to.
\end{itemize}
Combining Eqs.~\eqref{D} and~\eqref{energylabels} and inserting the result back into Eq.~\eqref{e:ex_bmbpt_feynman} provides the time-integrated expression of the diagram under the needed form
\begin{equation*}
D = \frac{(-1)^3 }{2(2!)^4}\sum_{k_i}\frac{O^{40}_{k_{1}k_{2}k_{3}k_{4}} \Omega^{40}_{k_{5}k_{6}k_{7}k_{8}} \Omega^{04}_{k_{5}k_{6}k_{1}k_{2}} \Omega^{04}_{k_{7}k_{8}k_{3}k_{4}} }{\epsilon_{k_{1}k_{2}k_{3}k_{4}}^{}\ \epsilon_{k_{1}k_{2}k_{5}k_{6}}^{}\ \epsilon_{k_{3}k_{4}k_{7}k_{8}}^{}\ }   \ .
\end{equation*}

\begin{figure}[t!]
\begin{center}
\parbox{30pt}{\begin{fmffile}{PO0_1}
    \begin{fmfgraph*}(30,30)
    \fmftop{t1} \fmfbottom{b1}
    \fmf{phantom,tag=1}{b1,i1}
    \fmf{phantom}{i1,t1}
    \fmfv{d.shape=square,d.filled=full,d.size=3thick}{i1}
    \fmflabel{$0$}{i1}
    \fmfposition
    \fmfipath{p[]}
    \fmfiset{p1}{vpath1(__b1,__i1)}
    \fmfiv{label=$ O^{00}$,l.angle=-60}{point length(p1) of p1}
    \end{fmfgraph*}
    \end{fmffile}}

\vspace{\baselineskip}
PO0.1
\vspace{2\baselineskip}

\parbox{40pt}{\begin{fmffile}{PO1_1}
    \begin{fmfgraph*}(40,40)
    \fmfcmd{style_def prop_pm expr p =
    draw_plain p;
    shrink(.7);
        cfill (marrow (p, .25));
        cfill (marrow (p, .75))
    endshrink;
	enddef;}
    \fmftop{t1} \fmfbottom{b1}
    \fmf{prop_pm,left=0.5,tag=1}{b1,t1}
    \fmf{prop_pm,right=0.5,tag=2}{b1,t1}
    \fmfv{d.shape=square,d.filled=full,d.size=3thick}{b1}
    \fmfv{d.shape=circle,d.filled=full,d.size=3thick}{t1}
    \fmflabel{$0$}{b1}
    \fmflabel{$\tau_1$}{t1}
    \fmfposition
    \fmfipath{p[]}
    \fmfiset{p1}{vpath1(__b1,__t1)}
    \fmfiset{p2}{vpath2(__b1,__t1)}
    \fmfiv{label=$\ O^{20}$,l.angle=-60}{point 0 of p2}
    \fmfiv{label=$\ \Omega^{02}$,l.angle=60}{point length(p2) of p2}
    \end{fmfgraph*}
    \end{fmffile}}
\hbox{\quad\quad}
\parbox{40pt}{\begin{fmffile}{PO1_2}
    \begin{fmfgraph*}(40,40)
    \fmfcmd{style_def prop_pm expr p =
    draw_plain p;
    shrink(.7);
        cfill (marrow (p, .25));
        cfill (marrow (p, .75))
    endshrink;
	enddef;}
    \fmftop{t1} \fmfbottom{b1}
    \fmf{prop_pm,left=0.75,tag=1}{b1,t1}
    \fmf{prop_pm,right=0.75,tag=2}{b1,t1}
    \fmf{prop_pm,left=0.25,tag=3}{b1,t1}
    \fmf{prop_pm,right=0.25,tag=4}{b1,t1}
    \fmfv{d.shape=square,d.filled=full,d.size=3thick}{b1}
    \fmfv{d.shape=circle,d.filled=full,d.size=3thick}{t1}
    \fmflabel{$0$}{b1}
    \fmflabel{$\tau_1$}{t1}
    \fmfposition
    \fmfipath{p[]}
    \fmfiset{p1}{vpath1(__b1,__t1)}
    \fmfiset{p2}{vpath2(__b1,__t1)}
    \fmfiset{p3}{vpath3(__b1,__t1)}
    \fmfiset{p4}{vpath4(__b1,__t1)}
    \fmfiv{label=$\ O^{40}$,l.angle=-60}{point 0 of p4}
    \fmfiv{label=$\ \Omega^{04}$,l.angle=60}{point length(p4) of p4}
    \end{fmfgraph*}
    \end{fmffile}}

\vspace{2\baselineskip}
PO1.1 \quad\quad \quad PO1.2
\vspace{2\baselineskip}

\parbox{60pt}{\begin{fmffile}{PO2_1}
    \begin{fmfgraph*}(60,60)
    \fmfcmd{style_def prop_pm expr p =
    draw_plain p;
    shrink(.7);
        cfill (marrow (p, .25));
        cfill (marrow (p, .75))
    endshrink;
	enddef;}
    \fmftop{t1} \fmfbottom{b1}
    \fmf{prop_pm,tag=1}{b1,i1}
    \fmf{prop_pm,tag=2}{i1,t1}
    \fmffreeze
    \fmf{prop_pm,right=0.5,tag=3}{b1,t1}
    \fmfv{d.shape=square,d.filled=full,d.size=3thick,l=$O^{20}$,l.angle=180}{b1}
    \fmfv{d.shape=circle,d.filled=full,d.size=3thick,l=$\breve{\Omega}^{11}$,l.angle=180}{i1}
    \fmfv{d.shape=circle,d.filled=full,d.size=3thick,l=$\Omega^{02}$,l.angle=180}{t1}
    \fmfposition
    \fmfipath{p[]}
    \fmfiset{p1}{vpath1(__b1,__i1)}
    \fmfiset{p2}{vpath2(__i1,__t1)}
    \fmfiset{p3}{vpath3(__b1,__t1)}
    \fmfiv{label=$0$,l.angle=-90,l.dist=0.15w}{point 0 of p1}
    \fmfiv{label=$\tau_1$,l.angle=0,l.dist=0.08w}{point length(p1) of p1}
    \fmfiv{label=$\tau_2$,l.angle=90,l.dist=0.15w}{point length(p2) of p2}
    \end{fmfgraph*}
    \end{fmffile}}
\hbox{\quad}
\parbox{60pt}{\begin{fmffile}{PO2_2}
    \begin{fmfgraph*}(60,60)
    \fmfcmd{style_def prop_pm expr p =
    draw_plain p;
    shrink(.7);
        cfill (marrow (p, .25));
        cfill (marrow (p, .75))
    endshrink;
	enddef;}
    \fmftop{t1} \fmfbottom{b1}
    \fmf{prop_pm,left=0.5,tag=1}{b1,i1}
    \fmf{prop_pm,right=0.5,tag=2}{b1,i1}
    \fmf{prop_pm,left=0.5,tag=3}{i1,t1}
    \fmf{prop_pm,right=0.5,tag=4}{i1,t1}
    \fmf{phantom,tag=5}{i1,t1}
    \fmf{phantom,tag=6}{b1,i1}
    \fmfv{d.shape=square,d.filled=full,d.size=3thick,l=$O^{20}$,l.angle=180}{b1}
    \fmfv{d.shape=circle,d.filled=full,d.size=3thick,l=$\Omega^{22}$,l.angle=180}{i1}
    \fmfv{d.shape=circle,d.filled=full,d.size=3thick,l=$\Omega^{02}$,l.angle=180}{t1}
    \fmfposition
    \fmfipath{p[]}
    \fmfiset{p1}{vpath1(__b1,__i1)}
    \fmfiset{p2}{vpath2(__b1,__i1)}
    \fmfiset{p3}{vpath3(__i1,__t1)}
    \fmfiset{p4}{vpath4(__i1,__t1)}
    \fmfiset{p5}{vpath5(__i1,__t1)}
    \fmfiset{p6}{vpath6(__b1,__i1)}
    \fmfiv{label=$0$,l.angle=-90,l.dist=0.15w}{point 0 of p6}
    \fmfiv{label=$\tau_1$,l.angle=0,l.dist=0.1w}{point length(p6) of p6}
    \fmfiv{label=$\tau_2$,l.angle=90,l.dist=0.15w}{point length(p5) of p5}
    \end{fmfgraph*}
    \end{fmffile}}
\hbox{\quad}
\parbox{60pt}{\begin{fmffile}{PO2_3}
    \begin{fmfgraph*}(60,60)
    \fmfcmd{style_def prop_pm expr p =
    draw_plain p;
    shrink(.7);
        cfill (marrow (p, .25));
        cfill (marrow (p, .75))
    endshrink;
	enddef;}
	\fmfcmd{style_def prop_mp expr p =
    draw_plain p;
    shrink(.7);
        cfill (marrow (reverse p, .25));
        cfill (marrow (reverse p, .75))
    endshrink;
	enddef;}
    \fmftop{t1} \fmfbottom{b1}
    \fmf{prop_pm,left=0.5,tag=1}{b1,i1}
    \fmf{prop_pm,right=0.5,tag=2}{b1,i1}
    \fmf{prop_mp,left=0.5,tag=3}{i1,t1}
    \fmf{prop_mp,right=0.5,tag=4}{i1,t1}
    \fmf{phantom,tag=5}{i1,t1}
    \fmf{phantom,tag=6}{b1,i1}
    \fmfv{d.shape=square,d.filled=full,d.size=3thick,l=$O^{20}$,l.angle=180}{b1}
    \fmfv{d.shape=circle,d.filled=full,d.size=3thick,l=$\Omega^{04}$,l.angle=180}{i1}
    \fmfv{d.shape=circle,d.filled=full,d.size=3thick,l=$\Omega^{20}$,l.angle=180}{t1}
    \fmfposition
    \fmfipath{p[]}
    \fmfiset{p1}{vpath1(__b1,__i1)}
    \fmfiset{p2}{vpath2(__b1,__i1)}
    \fmfiset{p3}{vpath3(__i1,__t1)}
    \fmfiset{p4}{vpath4(__i1,__t1)}
    \fmfiset{p5}{vpath5(__i1,__t1)}
    \fmfiset{p6}{vpath6(__b1,__i1)}
    \fmfiv{label=$0$,l.angle=-90,l.dist=0.15w}{point 0 of p6}
    \fmfiv{label=$\tau_1$,l.angle=0,l.dist=0.1w}{point length(p6) of p6}
    \fmfiv{label=$\tau_2$,l.angle=90,l.dist=0.15w}{point length(p5) of p5}
    \end{fmfgraph*}
    \end{fmffile}}

\vspace{2\baselineskip}
PO2.1 \quad\quad\quad\quad\quad PO2.2 \quad\quad\quad\quad PO2.3
\vspace{2\baselineskip}

\parbox{60pt}{\begin{fmffile}{PO2_4}
    \begin{fmfgraph*}(60,60)
    \fmfcmd{style_def prop_pm expr p =
    draw_plain p;
    shrink(.7);
        cfill (marrow (p, .25));
        cfill (marrow (p, .75))
    endshrink;
	enddef;}
    \fmftop{t1} \fmfbottom{b1}
    \fmf{phantom,tag=5}{i1,t1}
    \fmf{phantom,tag=6}{i1,b1}
    \fmffreeze
    \fmf{prop_pm,left=0.5,tag=1}{b1,i1}
    \fmf{prop_pm,right=0.5,tag=2}{b1,i1}
    \fmf{prop_pm,left=0.6,tag=3}{b1,t1}
    \fmf{prop_pm,right=0.6,tag=4}{b1,t1}
    \fmfv{d.shape=square,d.filled=full,d.size=3thick,l=$O^{40}$,l.angle=180}{b1}
    \fmfv{d.shape=circle,d.filled=full,d.size=3thick,l=$\Omega^{02}$,l.angle=90}{i1}
    \fmfv{d.shape=circle,d.filled=full,d.size=3thick,l=$\Omega^{02}$,l.angle=180}{t1}
    \fmfposition
    \fmfipath{p[]}
    \fmfiset{p1}{vpath1(__b1,__i1)}
    \fmfiset{p2}{vpath2(__b1,__i1)}
    \fmfiset{p3}{vpath3(__b1,__t1)}
    \fmfiset{p4}{vpath4(__b1,__t1)}
    \fmfiset{p5}{vpath5(__i1,__t1)}
    \fmfiset{p6}{vpath6(__i1,__b1)}
    \fmfiv{label=$0$,l.angle=-90,l.dist=0.15w}{point length(p6) of p6}
    \fmfiv{label=$\tau_1$,l.angle=0,l.dist=0.1w}{point 0 of p6}
    \fmfiv{label=$\tau_2$,l.angle=90,l.dist=0.15w}{point length(p5) of p5}
    \end{fmfgraph*}
    \end{fmffile}}
\hbox{\quad}
\parbox{60pt}{\begin{fmffile}{PO2_5}
    \begin{fmfgraph*}(60,60)
    \fmfcmd{style_def prop_pm expr p =
    draw_plain p;
    shrink(.7);
        cfill (marrow (p, .25));
        cfill (marrow (p, .75))
    endshrink;
	enddef;}
    \fmftop{t1} \fmfbottom{b1}
    \fmf{prop_pm,tag=1}{b1,i1}
    \fmf{prop_pm,tag=2}{i1,t1}
    \fmffreeze
    \fmf{prop_pm,left=0.5,tag=3}{b1,i1}
    \fmf{prop_pm,right=0.5,tag=4}{b1,i1}
    \fmf{prop_pm,right=0.6,tag=5}{b1,t1}
    \fmfv{d.shape=square,d.filled=full,d.size=3thick,l=$O^{40}$,l.angle=180}{b1}
    \fmfv{d.shape=circle,d.filled=full,d.size=3thick,l=$\Omega^{13}$,l.angle=180}{i1}
    \fmfv{d.shape=circle,d.filled=full,d.size=3thick,l=$\Omega^{02}$,l.angle=180}{t1}
    \fmfposition
    \fmfipath{p[]}
    \fmfiset{p1}{vpath1(__b1,__i1)}
    \fmfiset{p2}{vpath2(__i1,__t1)}
    \fmfiset{p3}{vpath3(__b1,__i1)}
    \fmfiset{p4}{vpath4(__b1,__i1)}
    \fmfiset{p5}{vpath5(__b1,__t1)}
    \fmfiv{label=$0$,l.angle=-90,l.dist=0.15w}{point 0 of p1}
    \fmfiv{label=$\tau_1$,l.angle=0,l.dist=0.1w}{point 0 of p2}
    \fmfiv{label=$\tau_2$,l.angle=90,l.dist=0.15w}{point length(p2) of p2}
    \end{fmfgraph*}
    \end{fmffile}}
\hbox{\quad}
\parbox{60pt}{\begin{fmffile}{PO2_6}
    \begin{fmfgraph*}(60,60)
    \fmfcmd{style_def prop_pm expr p =
    draw_plain p;
    shrink(.7);
        cfill (marrow (p, .25));
        cfill (marrow (p, .75))
    endshrink;
	enddef;}
    \fmftop{t1} \fmfbottom{b1}
    \fmf{prop_pm,tag=1}{b1,i1}
    \fmf{prop_pm,tag=2}{t1,i1}
    \fmffreeze
    \fmf{prop_pm,left=0.5,tag=3}{b1,i1}
    \fmf{prop_pm,right=0.5,tag=4}{b1,i1}
    \fmf{prop_pm,right=0.6,tag=5}{b1,t1}
    \fmfv{d.shape=square,d.filled=full,d.size=3thick,l=$O^{40}$,l.angle=180}{b1}
    \fmfv{d.shape=circle,d.filled=full,d.size=3thick,l=$\Omega^{04}$,l.angle=180}{i1}
    \fmfv{d.shape=circle,d.filled=full,d.size=3thick,l=$\breve{\Omega}^{11}$,l.angle=180}{t1}
    \fmfposition
    \fmfipath{p[]}
    \fmfiset{p1}{vpath1(__b1,__i1)}
    \fmfiset{p2}{vpath2(__t1,__i1)}
    \fmfiset{p3}{vpath3(__b1,__i1)}
    \fmfiset{p4}{vpath4(__b1,__i1)}
    \fmfiset{p5}{vpath5(__b1,__t1)}
    \fmfiv{label=$0$,l.angle=-90,l.dist=0.15w}{point 0 of p1}
    \fmfiv{label=$\tau_1$,l.angle=0,l.dist=0.1w}{point length(p2) of p2}
    \fmfiv{label=$\tau_2$,l.angle=90,l.dist=0.15w}{point 0 of p2}
    \end{fmfgraph*}
    \end{fmffile}}

\vspace{2\baselineskip}
PO2.4 \quad\quad\quad \quad\quad PO2.5 \quad\quad\quad \quad\quad PO2.6
\vspace{2\baselineskip}

\parbox{60pt}{\begin{fmffile}{PO2_7}
    \begin{fmfgraph*}(60,60)
    \fmfcmd{style_def prop_pm expr p =
    draw_plain p;
    shrink(.7);
        cfill (marrow (p, .25));
        cfill (marrow (p, .75))
    endshrink;
	enddef;}
    \fmftop{t1} \fmfbottom{b1}
    \fmf{prop_pm,tag=1}{b1,i1}
    \fmf{prop_pm,tag=2}{i1,t1}
    \fmffreeze
    \fmf{prop_pm,left=0.5,tag=3}{i1,t1}
    \fmf{prop_pm,right=0.5,tag=4}{i1,t1}
    \fmf{prop_pm,right=0.6,tag=5}{b1,t1}
    \fmfv{d.shape=square,d.filled=full,d.size=3thick,l=$O^{20}$,l.angle=180}{b1}
    \fmfv{d.shape=circle,d.filled=full,d.size=3thick,l=$\Omega^{31}$,l.angle=180}{i1}
    \fmfv{d.shape=circle,d.filled=full,d.size=3thick,l=$\Omega^{04}$,l.angle=180}{t1}
    \fmfposition
    \fmfipath{p[]}
    \fmfiset{p1}{vpath1(__b1,__i1)}
    \fmfiset{p2}{vpath2(__i1,__t1)}
    \fmfiset{p3}{vpath3(__i1,__t1)}
    \fmfiset{p4}{vpath4(__i1,__t1)}
    \fmfiset{p5}{vpath5(__b1,__t1)}
    \fmfiv{label=$0$,l.angle=-90,l.dist=0.15w}{point 0 of p1}
    \fmfiv{label=$\tau_1$,l.angle=-20,l.dist=0.12w}{point 0 of p2}
    \fmfiv{label=$\tau_2$,l.angle=90,l.dist=0.15w}{point length(p2) of p2}
    \end{fmfgraph*}
    \end{fmffile}}
\hbox{\quad}
\parbox{60pt}{\begin{fmffile}{PO2_8}
    \begin{fmfgraph*}(60,60)
    \fmfcmd{style_def prop_pm expr p =
    draw_plain p;
    shrink(.7);
        cfill (marrow (p, .25));
        cfill (marrow (p, .75))
    endshrink;
	enddef;}
    \fmftop{t1} \fmfbottom{b1}
    \fmf{prop_pm,left=0.5,tag=1}{b1,i1}
    \fmf{prop_pm,right=0.5,tag=2}{b1,i1}
    \fmf{prop_pm,left=0.5,tag=3}{i1,t1}
    \fmf{prop_pm,right=0.5,tag=4}{i1,t1}
    \fmffreeze
    \fmf{prop_pm,left=0.6,tag=5}{b1,t1}
    \fmf{prop_pm,right=0.6,tag=6}{b1,t1}
    \fmf{phantom,tag=7}{i1,t1}
    \fmf{phantom,tag=8}{b1,i1}
    \fmfv{d.shape=square,d.filled=full,d.size=3thick,l=$O^{40}$,l.angle=180}{b1}
    \fmfv{d.shape=circle,d.filled=full,d.size=3thick,l=$\Omega^{22}$,l.angle=180,l.dist=20pt}{i1}
    \fmfv{d.shape=circle,d.filled=full,d.size=3thick,l=$\Omega^{04}$,l.angle=180}{t1}
    \fmfposition
    \fmfipath{p[]}
    \fmfiset{p1}{vpath1(__b1,__i1)}
    \fmfiset{p2}{vpath2(__b1,__i1)}
    \fmfiset{p3}{vpath3(__i1,__t1)}
    \fmfiset{p4}{vpath4(__i1,__t1)}
    \fmfiset{p5}{vpath5(__b1,__t1)}
    \fmfiset{p6}{vpath6(__b1,__t1)}
    \fmfiset{p7}{vpath7(__i1,__t1)}
    \fmfiset{p8}{vpath8(__b1,__i1)}
    \fmfiv{label=$0$,l.angle=-90,l.dist=0.15w}{point 0 of p8}
    \fmfiv{label=$\tau_1$,l.angle=-20,l.dist=0.12w}{point 0 of p7}
    \fmfiv{label=$\tau_2$,l.angle=90,l.dist=0.15w}{point length(p7) of p7}
    \end{fmfgraph*}
    \end{fmffile}}

\vspace{2\baselineskip}
PO2.7  \quad\quad\quad \quad\quad  PO2.8

\end{center}
\caption{Zero-, first- and second-order Feynman BMBPT diagrams generated from operator vertices containing four legs at most, i.e., with \texttt{deg\_max}~$=4$.}
\label{diagBMBPT012}
\end{figure}
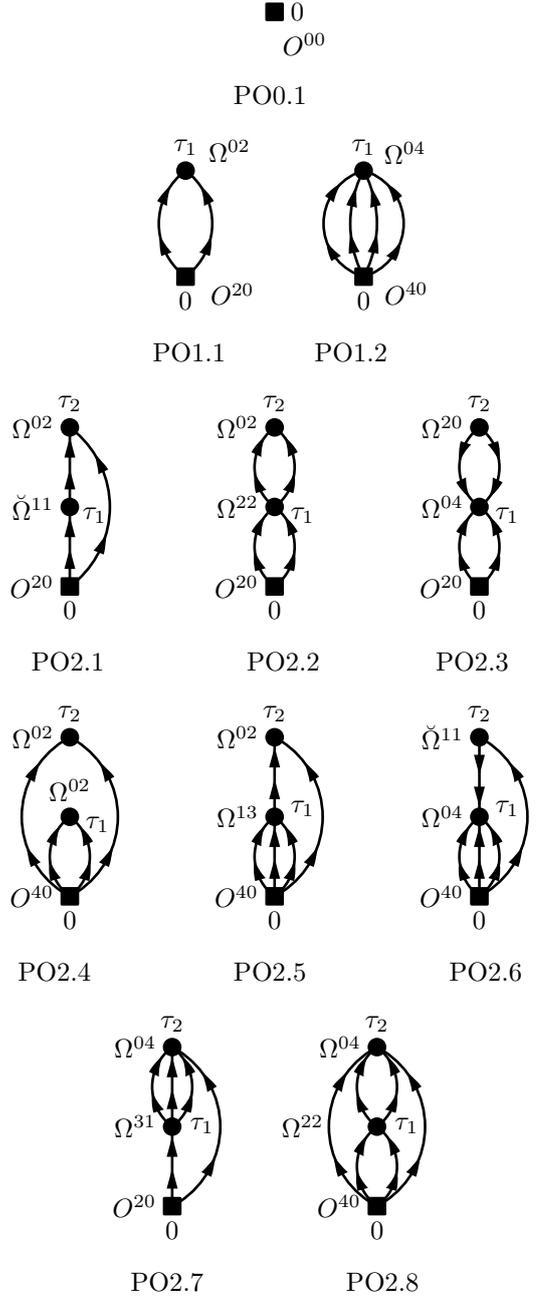

\subsection{Towards higher orders}

BMBPT diagrams of order $p=0,1$, and $2$ have been generated and evaluated manually within the NO2B approximation, i.e., excluding operators $O^{[k]}$ with $k>4$~\cite{Duguet:2015yle,Art18b}. The eleven corresponding diagrams are displayed in Fig.~\ref{diagBMBPT012} for illustration\footnote{These eleven BMBPT diagrams reduce to the corresponding 14 standard MBPT diagrams generated from two-body forces~\cite{Shavitt_Bartlett_2009} in the zero-pairing limit~\citep{Art18b}.}. They have been recently implemented numerically to perform \emph{ab initio} nuclear structure calculations of mid- and heavy-mass open-shell nuclei~\cite{Tichai:2018mll}. 

\begin{figure}[t!]
\begin{center}
\parbox{120pt}{
\begin{fmffile}{diag_arbitrary}
\begin{fmfgraph*}(80,120)
\fmfstraight
\fmftop{t0,t1,t2,t3}\fmfbottom{v0}
\fmf{phantom}{v0,v1}
\fmfcmd{style_def prop_pm expr p =
	draw_plain p;
	shrink(.7);
	cfill (marrow (p, .25));
	cfill (marrow (p, .75))
	endshrink;
	enddef;}
\fmfcmd{style_def half_prop expr p =
    draw_plain p;
    shrink(.7);
        cfill (marrow (p, .5))
    endshrink;
	enddef;}
\fmfv{d.shape=square,d.filled=full,d.size=3thick}{v0}
\fmf{phantom}{v1,v2}
\fmfv{d.shape=circle,d.filled=full,d.size=3thick}{v1}
\fmf{phantom}{v2,v3}
\fmfv{d.shape=circle,d.filled=full,d.size=3thick}{v2}
\fmfv{d.shape=circle,d.filled=full,d.size=3thick}{v3}
\fmfv{d.shape=circle,d.filled=full,d.size=3thick}{v4}
\fmfv{d.shape=circle,d.filled=full,d.size=3thick}{v5}
\fmfv{d.shape=circle,d.filled=full,d.size=3thick}{v6}
\fmf{phantom}{v4,v3}
\fmf{phantom}{v4,v5}
\fmf{phantom}{v5,v6}
\fmf{phantom}{v6,t0}
\fmf{phantom}{v6,t1}
\fmf{phantom}{v6,t2}
\fmf{phantom}{v6,t3}
\fmffreeze
\fmf{prop_pm,left=0.6}{v0,v5}
\fmf{prop_pm,right=0.6}{v0,v2}
\fmf{prop_pm,left=0.6}{v0,v3}
\fmf{prop_pm,right=0.6}{v0,v3}
\fmf{prop_pm,left=0.5}{v1,v2}
\fmf{prop_pm,right=0.5}{v1,v4}
\fmf{prop_pm,left=0.6}{v1,v3}
\fmf{prop_pm}{v3,v4}
\fmf{prop_pm,right=0.6}{v4,v6}
\fmf{prop_pm}{v5,v6}
\fmf{half_prop,right=0.2}{v5,t2}
\fmf{half_prop,right=0.2}{v4,t3}
\fmf{half_prop,left=0.2}{v5,t1}
\fmf{half_prop,left=0.2}{v1,t0}
\end{fmfgraph*}
\end{fmffile}
}
\end{center}
\caption{Lower-part of a possible arbitrary-order BMBPT diagram.}
\label{f:arbitrary_diag}
\end{figure}
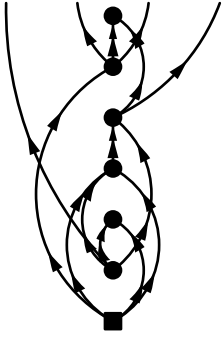

When going to higher orders and/or using vertices with more legs, the number of vertices and propagators grows out of proportion, as schematically illustrated in Fig.~\ref{f:arbitrary_diag}, with diagrams potentially displaying very involved topologies. The associated combinatorial makes generating all diagrams more difficult, cumbersome and prone to omissions. Algebraically, while the high order directly translates into the rise of the dimensionality of the time integral, the potentially complex topology of the diagram translates into the intricated structure of the time integrand dictated by the many step functions at play.  The development of automated tools to produce and evaluate high-order diagrams generated from vertices containing more, e.g., six, legs becomes thus essential.

Eventually, the expression of a generic diagram of order $p$ obtained from the application of Feynman's algebraic rules typically reads as
\begin{align}
D &= \lim\limits_{\tau \to \infty} \frac{(-1)^{a}}{2^b 3^c 4^d 5^e 6^f}\sum_{k_i}
  \Omega^{i_1j_1}_{k \ldots k} \ldots \Omega^{i_pj_p}_{k \ldots k} O^{m0}_{k \ldots k} \nonumber \\
 &\phantom{=} \,\,\, \times \int_{0}^{\tau}\!\!d\tau_{1}\ldots d\tau_{p}  \, \theta(\tau_q - \tau_r) \ldots \theta(\tau_u - \tau_v) \nonumber \\
&\phantom{= \,\,\,  \times \int_{0}^{\infty}} \times e^{-a_1\tau_1} \ldots e^{-a_p\tau_p} \ , \
\end{align}
where $(a,b,c,\dots)$ are integer numbers characterizing the topology of the diagram, $(q,r,\dots u,v)$ are integers between $1$ and $p$ and $(a_1,\dots,a_p)$ denote the sum/difference of quasi-particle energies associated with the lines entering/leaving each of the $p$ $\Omega^{ij}(\tau_k)$ vertices. Quasiparticle indices $k_i$ have been stripped of their labels in the matrix elements for the sake of concision. 

In the following, we detail the strategy, the algorithm and the code to automatically \emph{generate} and \emph{evaluate} all BMBPT diagrams appearing at an arbitrary order $p$.

\section{Automatic generation of BMBPT diagrams}
\label{secautomaticgeneration}

The automated generation of BMBPT Feynman diagrams is based on the use of graph theory, which is the domain of mathematics that focuses on studying graphs. Definitions and properties of quantities of interest, i.e., graph, walk and path on a graph, cycle, tree graph, adjacency matrix etc are detailed in~\ref{s:graph_theory}. Here, we limit ourselves to a brief and qualitative description of these notions and refer to the appendix for a more rigourous account.

\subsection{Basic elements}

The main notions of interest are
\begin{itemize}
\item A graph $G$ denotes a set of nodes and a set of edges, each edge being attached to a node or a pair of nodes.  
\item A walk on a graph is an alternative sequence of nodes and edges connecting them. A walk is closed (open) if the first and last nodes are (not) the same. The length of a walk corresponds to its number of edges.
\item A path is a walk whose nodes are all distinct.
\item A cycle is a closed walk where the initial/final node and the internal nodes are distinct.
\item A graph is connex if, for any pair of nodes, there exists a path connecting them.
\item A tree graph is a connex graph without cycle.
\item The oriented adjacency matrix $\tilde{A}(G)$ associated to a graph $G$ with labelled nodes $v_1...v_n$ is the matrix whose elements $\tilde{a}_{ij}$ indicate the number of edges going from node $v_i$ to node $v_j$.
\end{itemize}

\subsection{Oriented adjacency matrix and BMBPT diagram}
\label{subs:rules_mat}

\begin{figure}[t!]
\begin{center}
\parbox{80pt}{\begin{fmffile}{diag10}
\begin{fmfgraph*}(80,80)
\fmftop{v3}\fmfbottom{v0}
\fmfcmd{style_def prop_pm expr p =
	draw_plain p;
	shrink(.7);
	cfill (marrow (p, .25));
	cfill (marrow (p, .75))
	endshrink;
	enddef;}
\fmf{phantom}{v0,v1}
\fmfv{d.shape=square,d.filled=full,d.size=3thick}{v0}
\fmf{phantom}{v1,v2}
\fmfv{d.shape=circle,d.filled=full,d.size=3thick}{v1}
\fmf{phantom}{v2,v3}
\fmfv{d.shape=circle,d.filled=full,d.size=3thick}{v2}
\fmfv{d.shape=circle,d.filled=full,d.size=3thick}{v3}
\fmffreeze
\fmf{prop_pm,left=0.6}{v0,v2}
\fmf{prop_pm,right=0.6}{v0,v2}
\fmf{prop_pm,left=0.6}{v0,v3}
\fmf{prop_pm,right=0.6}{v0,v3}
\fmf{prop_pm,left=0.5}{v1,v2}
\fmf{prop_pm,right=0.5}{v1,v2}
\fmf{prop_pm,left=0.6}{v1,v3}
\fmf{prop_pm,right=0.6}{v1,v3}
\end{fmfgraph*}
\end{fmffile}}
\hspace{-10pt}$\Leftrightarrow$
$\begin{pmatrix}
0 & 0 & 2 & 2 \\
0 & 0 & 2 & 2 \\
0 & 0 & 0 & 0 \\
0 & 0 & 0 & 0
\end{pmatrix}$
\end{center}
\caption{A BMBPT diagram and its associated adjacency matrix. Vertices are indexed from bottom to top, such that the first line of the matrix corresponds to the propagators going out of the bottom vertex.}
\label{f:adj_mat_bmbpt}
\end{figure}
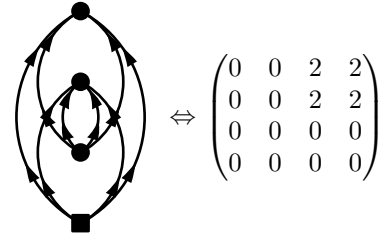

A BMBPT diagram being a connected graph with oriented edges, one can extract its oriented adjacency matrix, as exemplified in Fig.~\ref{f:adj_mat_bmbpt}. Feynman's topological rules characterizing valid BMBPT diagrams constrain the form of their oriented adjacency matrices.
\begin{enumerate}
\item A BMBPT diagram of order $p$, i.e., containing $p+1$ vertices, is associated to a $(p+1) \times (p+1)$ adjacency matrix.
\item As a BMBPT diagram is connected, its associated adjacency matrix cannot be recast into block-diagonal form through permutation of its rows and columns.
\item As each vertex $\Omega^{ij}$ involved has either an effective one-, two- or three-body character, thus exhibiting $i+j = 2,4 \text{ or } 6$, each matrix index $k$ fulfils the identity 
\begin{equation}
\sum_l (\tilde{a}_{kl}+\tilde{a}_{lk}) =  2, 4 \text{ or } 6 \, . \label{constraint}
\end{equation}
\item As only $O^{m0}(0)$ vertices with propagators going out contribute to BMBPT diagrams, the corresponding, e.g. first, column of a valid adjacency matrix is necessarily zero.
\item As no contraction of a vertex onto itself is possible, all diagonal elements $\tilde{a}_{ii}$ of a valid adjacency matrix are zero.
\item As no loop between two vertices is possible, matrix elements $\tilde{a}_{ij}$ and $\tilde{a}_{ji}$ of a valid adjacency matrix cannot be non-zero simultaneously.
\end{enumerate}

Producing the complete set of $(p+1) \times (p+1)$ matrices satisfying the above rules, one is sure to generate all possible BMBPT diagrams of order $p$. One must, however, further discard topologically equivalent diagrams. This can be done by performing simultaneous permutations of rows and columns and by comparing the result with other matrices in the set. However, as the generic operator $O$ is at fixed time $0$, it must not be considered in the process, i.e., its column and row must not be permuted with any other.

\subsection{Pedestrian generation of adjacency matrices}
\label{subs:brute_force_gen}

A simple way to generate all $(p+1) \times (p+1)$ adjacency matrices for a maximal vertex degree equal to \texttt{deg\_max}\footnote{Though the code allows to distinguish between \texttt{deg\_max} for the observable and \texttt{deg\_max} for the Hamiltonian, for the sake of clarity the same \texttt{deg\_max} is used for both in the following.} (i.e., 4 or 6 depending on the two- or three-body character of the operators) is to start with a set of matrices containing only one matrix fully initialized to zero and proceed as follows
\begin{enumerate}
\item Consider a matrix element,
\begin{enumerate}
\item Consider a matrix in the set,
\begin{enumerate}
\item Save the matrix,
\item Save copies of the matrix with the matrix element changed to every possible value from 1 to \texttt{deg\_max},
\end{enumerate}
\item Go back to (a) until all matrices in the set are exhausted,
\end{enumerate}
\item Go back to 1. until all matrix elements are exhausted.
\end{enumerate}
With this method, all possible BMBPT adjacency matrices are necessarily produced\footnote{In particular, the rule regarding $n_a=0$ detailed in Sec.~\ref{subs:diag_gene} is respected by construction, as the matrices are generated on a propagator-by-propagator basis, with no external leg. The burden is thus transferred on making sure only vertices with the appropriate one-, two- or three-body character are produced. A generalization of the process authorizing $n_a \geq 0$ thus has to be considered for off-diagonal BMBPT.}. One must, however, apply the set of tests necessary to only retain adjacency matrices that conform with the rules listed in Sec.~\ref{subs:rules_mat}.

\subsection{Optimized generation of adjacency matrices}

The pedestrian method detailed in Sec.~\ref{subs:brute_force_gen} is time and memory consuming from a numerical viewpoint. It is thus beneficial to integrate as many of the selection rules as possible into the very production process of the matrices. Doing so, time and memory are saved as less matrices are actually produced while the tests enforcing the selection rules become superfluous. This is particularly beneficial regarding the restriction to topologically distinct diagrams as the corresponding test scales factorially with the number of matrices in the set.

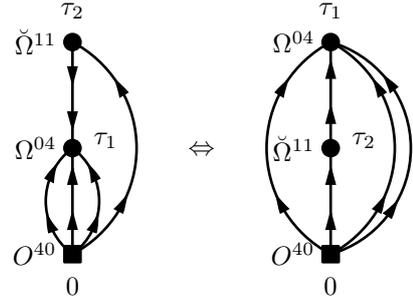
\begin{figure}[t!]
\begin{center}
\parbox{80pt}{\begin{fmffile}{PO2_6_big}
    \begin{fmfgraph*}(80,80)
    \fmftop{t1} \fmfbottom{b1}
    \fmfcmd{style_def prop_pm expr p =
	draw_plain p;
	shrink(.7);
	cfill (marrow (p, .25));
	cfill (marrow (p, .75))
	endshrink;
	enddef;}
    \fmf{prop_pm,tag=1}{b1,i1}
    \fmf{prop_pm,tag=2}{t1,i1}
    \fmffreeze
    \fmf{prop_pm,left=0.5,tag=3}{b1,i1}
    \fmf{prop_pm,right=0.5,tag=4}{b1,i1}
    \fmf{prop_pm,right=0.6,tag=5}{b1,t1}
    \fmfv{d.shape=square,d.filled=full,d.size=3thick,l=$O^{40}$,l.angle=180}{b1}
    \fmfv{d.shape=circle,d.filled=full,d.size=3thick,l=$\Omega^{04}$,l.angle=180}{i1}
    \fmfv{d.shape=circle,d.filled=full,d.size=3thick,l=$\breve{\Omega}^{11}$,l.angle=180}{t1}
    \fmfposition
    \fmfipath{p[]}
    \fmfiset{p1}{vpath1(__b1,__i1)}
    \fmfiset{p2}{vpath2(__t1,__i1)}
    \fmfiset{p3}{vpath3(__b1,__i1)}
    \fmfiset{p4}{vpath4(__b1,__i1)}
    \fmfiset{p5}{vpath5(__b1,__t1)}
    \fmfiv{label=$0$,l.angle=-90,l.dist=0.15w}{point 0 of p1}
    \fmfiv{label=$\tau_1$,l.angle=0,l.dist=0.1w}{point length(p2) of p2}
    \fmfiv{label=$\tau_2$,l.angle=90,l.dist=0.15w}{point 0of p2}
    \end{fmfgraph*}
    \end{fmffile}}
$\Leftrightarrow$
\parbox{80pt}{\begin{fmffile}{PO2_6_big_perm}
\begin{fmfgraph*}(80,80)
\fmftop{v2}\fmfbottom{v0}
\fmfcmd{style_def prop_pm expr p =
	draw_plain p;
	shrink(.7);
	cfill (marrow (p, .25));
	cfill (marrow (p, .75))
	endshrink;
	enddef;}
\fmf{phantom, tag=1}{v0,v1}
\fmfv{d.shape=square,d.filled=full,d.size=3thick,l=$O^{40}$,l.angle=180}{v0}
\fmf{phantom, tag=2}{v1,v2}
\fmfv{d.shape=circle,d.filled=full,d.size=3thick,l=$\breve{\Omega}^{11}$,l.angle=180}{v1}
\fmfv{d.shape=circle,d.filled=full,d.size=3thick,l=$\Omega^{04}$,l.angle=180}{v2}
\fmffreeze
\fmf{prop_pm}{v0,v1}
\fmf{prop_pm,right=0.75}{v0,v2}
\fmf{prop_pm,left=0.6}{v0,v2}
\fmf{prop_pm,right=0.6}{v0,v2}
\fmf{prop_pm}{v1,v2}
\fmfposition
    \fmfipath{p[]}
    \fmfiset{p1}{vpath1(__v0,__v1)}
    \fmfiset{p2}{vpath2(__v1,__v2)}
    \fmfiv{label=$0$,l.angle=-90,l.dist=0.15w}{point 0 of p1}
    \fmfiv{label=$\tau_2$,l.angle=0,l.dist=0.1w}{point 0 of p2}
    \fmfiv{label=$\tau_1$,l.angle=90,l.dist=0.15w}{point length(p2) of p2}
\end{fmfgraph*}
\end{fmffile}}
\end{center}

\caption{A BMBPT diagram drawn with some propagators going downwards (left) can be turned into an equivalent diagram with all its propagators going upwards (right) by moving the second vertex above the third one.}
\label{f:all_lines_upwards}
\end{figure}

The first obvious improvement is to avoid producing any matrix with a non-zero diagonal element. Second, the fact that BMBPT diagrams do not display oriented loops between any given number of vertices makes always possible, by moving vertices within the plane of the graph, to recast any BMBPT diagram as an equivalent diagram with all propagators moving upwards, as is exemplified in Fig.~\ref{f:all_lines_upwards}. Accordingly, one limits the generation to upper-triangular matrices, thus reducing their number drastically and discarding at the same time a whole set of topologically equivalent diagrams.

The number of considered matrices can be further reduced by checking the $k$-body character of the vertices on-the-fly. As the matrix elements are updated row by row, and thus vertex by vertex given that only upper-triangular matrices are considered, one can check directly after filling a row that the corresponding vertex has indeed a one-, two- or three-body character, i.e.,
 that it satisfies Eq.~\eqref{constraint}. If it is not the case, the matrix is rejected on-the-fly along with all matrices that would have spawn from it. For example, the adjacency matrix
\begin{equation}
\begin{pmatrix}
0 & 2 & 2 & 1 \\
0 & 0 & 0 & 0 \\
0 & 0 & 0 & 0 \\
0 & 0 & 0 & 0
\end{pmatrix}
\end{equation}
can be discarded right after filling the first row given that the sum of its matrix elements differ from 2, 4 or 6.

A similar reasoning applies to the disconnected character of the diagrams. For matrices associated to diagrams of order $p>0$, it is possible to test after the second (or any further) row is filled if the matrix is bound to span a disconnected diagram in the end. For example, the adjacency matrix
\begin{equation}
\begin{pmatrix}
0 & 2 & 0 & 0 \\
0 & 0 & 0 & 0 \\
0 & 0 & 0 & 0 \\
0 & 0 & 0 & 0
\end{pmatrix} \ ,
\end{equation}
if already filled on its two first rows, would result in vertices 1 and 2 being disconnected from vertices 3 and 4.
One can thus eliminate the matrix on the fly, along with all matrices that would have spawn from it.

\subsection{Drawing associated BMBPT diagrams}

Once all allowed $(p+1) \times (p+1)$ adjacency matrices have been produced, the corresponding BMBPT diagrams can be drawn by simply reading the matrices, as each matrix element encodes the number of propagators going from one vertex to another. As the number of generated diagrams quickly increases with $p$, it is of interest to design a program to do it automatically. The program reads the content of the matrix and writes in a text file the appropriate drawing instructions for \emph{FeynMP} \cite{Ohl:1995kr}, a \LaTeX\ package used to draw Feynman diagrams.

Although adjacency matrices are sufficient to draw the diagrams, one may want to perform operations on the diagrams. We thus chose to make use of a graph theory package for Python called \emph{NetworkX}~\cite{SciPyProceedings_11}. The package takes adjacency matrices as input and produces graph objects on which different operations can be performed. For example, the check for topologically equivalent diagrams is performed using the built-in \emph{NetworkX} function \texttt{is\_isomorphic} and its related interfaces. Combining the \texttt{MultiDiGraph} object from \emph{NetworkX} and object-oriented programming, it was possible to implement this test in a time-savy way, first checking that two graphs share basic structures properties (degree of the different nodes, two-body only or three-body operators character, etc.) before performing the costly permutations eventually needed to check if they are indeed topologically equivalent.

With topologically distinct \emph{NetworkX} BMBPT diagrams at hand, we could adapt the program running through adjacency matrices to make it iterate through the nodes of the graph and obtain the FeynMP instructions accordingly. As an example, the output displaying the drawing instructions of the BMBPT diagram displayed in Fig.~\ref{f:ex_bmbpt_diag} is given in Fig.~\ref{f:feynmf_inst}.
\begin{figure}[t!]
{\small
\begin{verbatim}
\begin{fmffile}{diag10}
\begin{fmfgraph*}(80,80)
\fmfcmd{style_def prop_pm expr p =
	draw_plain p;
	shrink(.7);
	cfill (marrow (p, .25));
	cfill (marrow (p, .75))
	endshrink;
	enddef;}
\fmftop{v3}\fmfbottom{v0}
\fmf{phantom}{v0,v1}
\fmfv{d.shape=square,d.filled=full,d.size=3thick}{v0}
\fmf{phantom}{v1,v2}
\fmfv{d.shape=circle,d.filled=full,d.size=3thick}{v1}
\fmf{phantom}{v2,v3}
\fmfv{d.shape=circle,d.filled=full,d.size=3thick}{v2}
\fmfv{d.shape=circle,d.filled=full,d.size=3thick}{v3}
\fmffreeze
\fmf{prop_pm,left=0.6}{v0,v2}
\fmf{prop_pm,right=0.6}{v0,v2}
\fmf{prop_pm,left=0.6}{v0,v3}
\fmf{prop_pm,right=0.6}{v0,v3}
\fmf{prop_pm,left=0.5}{v1,v2}
\fmf{prop_pm,right=0.5}{v1,v2}
\fmf{prop_pm,left=0.6}{v1,v3}
\fmf{prop_pm,right=0.6}{v1,v3}
\end{fmfgraph*}
\end{fmffile}
\end{verbatim} }
\vspace{-\baselineskip}
\caption{\emph{FeynMP} instructions to draw the BMBPT diagram displayed in Fig.~\ref{f:ex_bmbpt_diag}.}
\label{f:feynmf_inst}
\end{figure}

\section{Automatic evaluation of BMBPT diagrams}
\label{secautomaticevaluation}

Having the capacity to generate all BMBPT Feynman diagrams of order $p$, the next challenge is to systematically derive their expression. Doing so on the basis of Feynman's algebraic rules is rather straightforward. However, it leaves the $p$-tuple time integral to perform in order to obtain the time-integrated expression of interest. Finding an algorithm to do so without prior knowledge of the perturbative order or of the topology of the diagram constitutes an unsolved challenge to our knowledge. In the present section, we introduce a method to achieve this goal, eventually leading to the identification of a novel diagrammatic rule. 

\subsection{Time-structure diagram}
\label{subsubs:timediags}

In a BMBPT Feynman diagram, a time label is attached to each vertex. Given any two vertices, their time labels are ordered with respect to each other as soon as a propagator connects the vertices directly by virtue of the step function it carries (see Eq.~\eqref{e:propagators}). The time labels may also be ordered in a less obvious way if the two vertices are connected through a set of intermediate vertices and propagators.

Eventually, a BMBPT Feynman diagram exhibits an underlying time structure that translates into the specific form of the integrand of the $p$-tuple integral to be performed. This specific form is characterized by a string of step functions ordering a subset of the time variables that must be integrated over. In order to characterize the typical structure of the integrand and compute the corresponding integral, we choose to represent it diagrammatically by introducing the so-called time-structure diagram (TSD) of a given BMBPT diagram. As we shall see below, the algorithm to perform the time-integral strongly depends on the \emph{topology} of the TSD that happens to play a fundamental role. Consequently, we now introduce and characterize TSDs. 

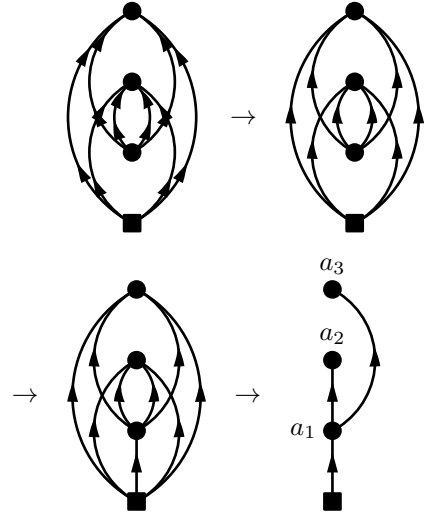
\begin{figure}[t!]
\begin{center}
$\phantom{\rightarrow}$
\hspace{-10pt}
\parbox{80pt}{\begin{fmffile}{diag_tsd_const_ex0}
\begin{fmfgraph*}(80,80)
\fmfcmd{style_def prop_pm expr p =
    draw_plain p;
    shrink(.7);
        cfill (marrow (p, .25));
        cfill (marrow (p, .75))
    endshrink;
	enddef;}
\fmftop{v3}\fmfbottom{v0}
\fmf{phantom}{v0,v1}
\fmfv{d.shape=square,d.filled=full,d.size=3thick}{v0}
\fmf{phantom}{v1,v2}
\fmfv{d.shape=circle,d.filled=full,d.size=3thick}{v1}
\fmf{phantom}{v2,v3}
\fmfv{d.shape=circle,d.filled=full,d.size=3thick}{v2}
\fmfv{d.shape=circle,d.filled=full,d.size=3thick}{v3}
\fmffreeze
\fmf{prop_pm,left=0.6}{v0,v2}
\fmf{prop_pm,right=0.6}{v0,v2}
\fmf{prop_pm,left=0.6}{v0,v3}
\fmf{prop_pm,right=0.6}{v0,v3}
\fmf{prop_pm,left=0.5}{v1,v2}
\fmf{prop_pm,right=0.5}{v1,v2}
\fmf{prop_pm,left=0.6}{v1,v3}
\fmf{prop_pm,right=0.6}{v1,v3}
\end{fmfgraph*}
\end{fmffile}}
\hspace{-10pt}
$\rightarrow$
\hspace{-10pt}
\parbox{80pt}{\begin{fmffile}{diag_tsd_const_ex1}
\begin{fmfgraph*}(80,80)
\fmfcmd{style_def half_prop expr p =
    draw_plain p;
    shrink(.7);
        cfill (marrow (p, .5))
    endshrink;
	enddef;}
\fmftop{v3}\fmfbottom{v0}
\fmf{phantom}{v0,v1}
\fmfv{d.shape=square,d.filled=full,d.size=3thick}{v0}
\fmf{phantom}{v1,v2}
\fmfv{d.shape=circle,d.filled=full,d.size=3thick}{v1}
\fmf{phantom}{v2,v3}
\fmfv{d.shape=circle,d.filled=full,d.size=3thick}{v2}
\fmfv{d.shape=circle,d.filled=full,d.size=3thick}{v3}
\fmffreeze
\fmf{half_prop,left=0.6}{v0,v2}
\fmf{half_prop,right=0.6}{v0,v2}
\fmf{half_prop,left=0.6}{v0,v3}
\fmf{half_prop,right=0.6}{v0,v3}
\fmf{half_prop,left=0.5}{v1,v2}
\fmf{half_prop,right=0.5}{v1,v2}
\fmf{half_prop,left=0.6}{v1,v3}
\fmf{half_prop,right=0.6}{v1,v3}
\end{fmfgraph*}
\end{fmffile}}
\hspace{-10pt}

\vspace{2\baselineskip}

$\rightarrow$
\hspace{-10pt}
\parbox{80pt}{\begin{fmffile}{diag_tsd_const_ex2}
\begin{fmfgraph*}(80,80)
\fmfcmd{style_def half_prop expr p =
    draw_plain p;
    shrink(.7);
        cfill (marrow (p, .5))
    endshrink;
	enddef;}
\fmftop{v3}\fmfbottom{v0}
\fmf{phantom}{v0,v1}
\fmfv{d.shape=square,d.filled=full,d.size=3thick}{v0}
\fmf{phantom}{v1,v2}
\fmfv{d.shape=circle,d.filled=full,d.size=3thick}{v1}
\fmf{phantom}{v2,v3}
\fmfv{d.shape=circle,d.filled=full,d.size=3thick}{v2}
\fmfv{d.shape=circle,d.filled=full,d.size=3thick}{v3}
\fmffreeze
\fmf{half_prop,left=0.6}{v0,v2}
\fmf{half_prop,right=0.6}{v0,v2}
\fmf{half_prop,left=0.6}{v0,v3}
\fmf{half_prop,right=0.6}{v0,v3}
\fmf{half_prop,left=0.5}{v1,v2}
\fmf{half_prop,right=0.5}{v1,v2}
\fmf{half_prop,left=0.6}{v1,v3}
\fmf{half_prop,right=0.6}{v1,v3}
\fmf{half_prop}{v0,v1}
\end{fmfgraph*}
\end{fmffile}}
\hspace{-10pt}
$\rightarrow$
\hspace{-20pt}
\parbox{80pt}{\begin{fmffile}{tsd_const_ex}
\begin{fmfgraph*}(80,80)
\fmfcmd{style_def half_prop expr p =
    draw_plain p;
    shrink(.7);
        cfill (marrow (p, .5))
    endshrink;
	enddef;}
\fmftop{v3}\fmfbottom{v0}
\fmfv{d.shape=square,d.filled=full,d.size=3thick}{v0}
\fmfv{d.shape=circle,d.filled=full,d.size=3thick,l=$a_1$,l.a=180}{v1}
\fmf{phantom}{v2,v3}
\fmfv{d.shape=circle,d.filled=full,d.size=3thick,l=$a_2$,l.a=90}{v2}
\fmfv{d.shape=circle,d.filled=full,d.size=3thick,l=$a_3$,l.a=90}{v3}
\fmf{half_prop}{v0,v1}
\fmf{half_prop}{v1,v2}
\fmffreeze
\fmf{half_prop,right=0.6}{v1,v3}
\end{fmfgraph*}
\end{fmffile}}
\end{center}
\caption{Production of the TSD associated with the third-order BMBPT diagram displayed in Fig.~\ref{f:ex_bmbpt_diag}.}
\label{f:tsd_production}
\end{figure}

\begin{enumerate}
\item The TSD associated to a BMBPT diagram of order $p$ is made out of the following building blocks
\begin{enumerate}
\item $p+1$ vertices representing operators in the interaction representation. While their positive time labels $(0, \tau_1, \ldots, \tau_p)$ are left implicit, vertices but the bottom one carry explicit energy factors $(a_1, \ldots, a_p)$.
\item oriented links representing ordering relations, i.e., step functions, between pairs of vertices. A link is oriented from  the vertex carrying the smaller time to the vertex carrying the larger time. Only the minimal set of links necessary to describe the time structure of the diagram is to be drawn, i.e.~only the longest path linking two different vertices is to be represented\footnote{This corresponds to omitting a step function $\theta(\tau_i-\tau_j)$ whenever a string of step functions $\theta(\tau_i-\tau_\alpha)\dots\theta(\tau_\omega-\tau_j)$ carrying the same information already appears.}.
\item BMBPT diagrams being connected, TSDs are necessarily connex.
\end{enumerate}
\item The expression of a TSD of order $p$ is extracted in the following way
\begin{enumerate}
\item each vertex $a_q$, $q=1,\ldots,p$, contributes a factor $e^{-a_q\tau_q}$,
\item each link\footnote{Links originating from the bottom vertex do not contribute an explicit step function given that the positivity of the running time labels is encoded into the boundary of the integral; see rule 2.(c).} oriented from vertex $a_u$ to vertex $a_v$ contributes the step function $\theta(\tau_v - \tau_u)$,
\item the $p$ time labels $\tau_1, \ldots, \tau_p$ are integrated over from $0$ to $\tau \rightarrow +\infty$,
\end{enumerate}
and thus typically reads as
\begin{align}
T &= \lim\limits_{\tau \to \infty} \int_{0}^{\tau}\!\!d\tau_{1}\ldots d\tau_{p} \, \theta(\tau_q - \tau_r) \ldots \theta(\tau_u - \tau_v) \nonumber \\
&\phantom{= \,\,\,\,\,\,\,  \times \int_{0}^{\infty}} \times e^{-a_1\tau_1} \ldots e^{-a_p\tau_p} \ , \
\end{align}
where $(q,r,\dots u,v)$ are integers between $1$ and $p$.
\item The TSD associated to a BMBPT diagram can be obtained from the latter through the following steps
\begin{enumerate}
\item copy the BMBPT diagram,
\item replace propagators by links,
\item add a link between the bottom vertex at time 0 and every other vertex if such a link does not exist\footnote{The operator vertex $O$ at time 0 entertains an ordering relation with every other vertex.},
\item for each pair of vertices, consider all possible paths linking them and only retain the longest one,
\item match $a_q$ to the sum/difference of quasi-particle energies associated with the lines entering/leaving the corresponding vertex in the BMBPT diagram.
\end{enumerate}

The procedure is illustrated in Fig.~\ref{f:tsd_production} for the BMBPT diagram originally displayed in Fig.~\ref{f:ex_bmbpt_diag}. Cleared of other informations, the TSD tranparently characterizes the time-ordering structure underlying the BMBPT diagram, i.e., the three $\Omega^{ij}$ vertices are at higher times than $O^{40}$ such that the two $\Omega^{04}$ vertices are at higher times than $\Omega^{40}$ without being ordered with respect to one another. From the graph theory viewpoint, the corresponding TSD is a tree, i.e., it contains no cycle, with two branches such that the vertices on the two branches are not ordered with respect to one another. 
\end{enumerate}

\subsection{Discussion}
\label{subsubs:discussion}

\begin{figure}[t!]
\begin{center}
\parbox{30pt}{\begin{fmffile}{time_0_0}
\begin{fmfgraph*}(30,30)
\fmftop{t1} \fmfbottom{b1}
    \fmf{phantom}{b1,i1}
    \fmf{phantom}{i1,t1}
    \fmfv{d.shape=square,d.filled=full,d.size=3thick}{i1}
\end{fmfgraph*}
\end{fmffile}}

T0.1
\vspace{2\baselineskip}

\parbox{30pt}{\begin{fmffile}{time_1_0}
\begin{fmfgraph*}(30,30)
\fmfcmd{style_def half_prop expr p =
draw_plain p;
shrink(.7);
    cfill (marrow (p, .5))
endshrink;
enddef;}
\fmftop{v1}\fmfbottom{v0}
\fmf{phantom}{v0,v1}
\fmfv{d.shape=square,d.filled=full,d.size=3thick}{v0}
\fmfv{d.shape=circle,d.filled=full,d.size=3thick,l=$a_1$}{v1}
\fmffreeze
\fmf{half_prop}{v0,v1}
\end{fmfgraph*}
\end{fmffile}}

\vspace{\baselineskip}
T1.1
\vspace{2\baselineskip}

\parbox{60pt}{\begin{fmffile}{time_2_0}
\begin{fmfgraph*}(60,60)
\fmfcmd{style_def half_prop expr p =
draw_plain p;
shrink(.7);
    cfill (marrow (p, .5))
endshrink;
enddef;}
\fmftop{v2}\fmfbottom{v0}
\fmf{phantom}{v0,v1}
\fmfv{d.shape=square,d.filled=full,d.size=3thick}{v0}
\fmf{phantom}{v1,v2}
\fmfv{d.shape=circle,d.filled=full,d.size=3thick,l=$a_1$}{v1}
\fmfv{d.shape=circle,d.filled=full,d.size=3thick,l=$a_2$}{v2}
\fmffreeze
\fmf{half_prop}{v0,v1}
\fmf{half_prop}{v1,v2}
\end{fmfgraph*}
\end{fmffile}}
 \hspace{-20pt}
\parbox{60pt}{\begin{fmffile}{time_2_1}
\begin{fmfgraph*}(60,60)
\fmfcmd{style_def half_prop expr p =
draw_plain p;
shrink(.7);
    cfill (marrow (p, .5))
endshrink;
enddef;}
\fmftop{v2}\fmfbottom{v0}
\fmf{phantom}{v0,v1}
\fmfv{d.shape=square,d.filled=full,d.size=3thick}{v0}
\fmf{phantom}{v1,v2}
\fmfv{d.shape=circle,d.filled=full,d.size=3thick,l=$a_1$,l.a=90}{v1}
\fmfv{d.shape=circle,d.filled=full,d.size=3thick,l=$a_2$}{v2}
\fmffreeze
\fmf{half_prop}{v0,v1}
\fmf{half_prop,right=0.6}{v0,v2}
\end{fmfgraph*}
\end{fmffile}}

\vspace{\baselineskip}
T2.1 \quad \quad \quad T2.2
\vspace{2\baselineskip}

\hspace{-40pt}
\parbox{80pt}{\begin{fmffile}{time_3_0}
\begin{fmfgraph*}(80,80)
\fmfcmd{style_def half_prop expr p =
draw_plain p;
shrink(.7);
    cfill (marrow (p, .5))
endshrink;
enddef;}
\fmftop{v3}\fmfbottom{v0}
\fmf{phantom}{v0,v1}
\fmfv{d.shape=square,d.filled=full,d.size=3thick}{v0}
\fmf{phantom}{v1,v2}
\fmfv{d.shape=circle,d.filled=full,d.size=3thick,l=$a_1$,l.a=0}{v1}
\fmf{phantom}{v2,v3}
\fmfv{d.shape=circle,d.filled=full,d.size=3thick,l=$a_2$,l.a=0}{v2}
\fmfv{d.shape=circle,d.filled=full,d.size=3thick,l=$a_3$}{v3}
\fmffreeze
\fmf{half_prop}{v0,v1}
\fmf{half_prop}{v1,v2}
\fmf{half_prop}{v2,v3}
\end{fmfgraph*}
\end{fmffile}}
\hspace{-35pt}
\parbox{80pt}{\begin{fmffile}{time_3_1}
\begin{fmfgraph*}(80,80)
\fmfcmd{style_def half_prop expr p =
draw_plain p;
shrink(.7);
    cfill (marrow (p, .5))
endshrink;
enddef;}
\fmftop{v3}\fmfbottom{v0}
\fmf{phantom}{v0,v1}
\fmfv{d.shape=square,d.filled=full,d.size=3thick}{v0}
\fmf{phantom}{v1,v2}
\fmfv{d.shape=circle,d.filled=full,d.size=3thick,l=$a_1$,l.a=-60}{v1}
\fmf{phantom}{v2,v3}
\fmfv{d.shape=circle,d.filled=full,d.size=3thick,l=$a_2$,l.a=90}{v2}
\fmfv{d.shape=circle,d.filled=full,d.size=3thick,l=$a_3$}{v3}
\fmffreeze
\fmf{half_prop}{v0,v1}
\fmf{half_prop}{v1,v2}
\fmf{half_prop,right=0.6}{v1,v3}
\end{fmfgraph*}
\end{fmffile}}
\hspace{-35pt}
\parbox{80pt}{\begin{fmffile}{time_3_2}
\begin{fmfgraph*}(80,80)
\fmfcmd{style_def half_prop expr p =
draw_plain p;
shrink(.7);
    cfill (marrow (p, .5))
endshrink;
enddef;}
\fmftop{v3}\fmfbottom{v0}
\fmf{phantom}{v0,v1}
\fmfv{d.shape=square,d.filled=full,d.size=3thick}{v0}
\fmf{phantom}{v1,v2}
\fmfv{d.shape=circle,d.filled=full,d.size=3thick,l=$a_1$,l.a=90}{v1}
\fmf{phantom}{v2,v3}
\fmfv{d.shape=circle,d.filled=full,d.size=3thick,l=$a_2$,l.a=60}{v2}
\fmfv{d.shape=circle,d.filled=full,d.size=3thick,l=$a_3$}{v3}
\fmffreeze
\fmf{half_prop}{v0,v1}
\fmf{half_prop,right=0.6}{v0,v2}
\fmf{half_prop}{v2,v3}
\end{fmfgraph*}
\end{fmffile}}
\hspace{-35pt}
\parbox{80pt}{\begin{fmffile}{time_3_3}
\begin{fmfgraph*}(80,80)
\fmfcmd{style_def half_prop expr p =
draw_plain p;
shrink(.7);
    cfill (marrow (p, .5))
endshrink;
enddef;}
\fmftop{v3}\fmfbottom{v0}
\fmf{phantom}{v0,v1}
\fmfv{d.shape=square,d.filled=full,d.size=3thick}{v0}
\fmf{phantom}{v1,v2}
\fmfv{d.shape=circle,d.filled=full,d.size=3thick,l=$a_1$,l.a=90}{v1}
\fmf{phantom}{v2,v3}
\fmfv{d.shape=circle,d.filled=full,d.size=3thick,l=$a_2$}{v2}
\fmfv{d.shape=circle,d.filled=full,d.size=3thick,l=$a_3$}{v3}
\fmffreeze
\fmf{half_prop}{v0,v1}
\fmf{half_prop,right=0.6}{v0,v2}
\fmf{half_prop,right=0.6}{v0,v3}
\end{fmfgraph*}
\end{fmffile}}
\hspace{-30pt}
\parbox{80pt}{\begin{fmffile}{time_3_4}
\begin{fmfgraph*}(80,80)
\fmfcmd{style_def half_prop expr p =
draw_plain p;
shrink(.7);
    cfill (marrow (p, .5))
endshrink;
enddef;}
\fmftop{v3}\fmfbottom{v0}
\fmf{phantom}{v0,v1}
\fmfv{d.shape=square,d.filled=full,d.size=3thick}{v0}
\fmf{phantom}{v1,v2}
\fmfv{d.shape=circle,d.filled=full,d.size=3thick,l=$a_1$,l.a=0}{v1}
\fmf{phantom}{v2,v3}
\fmfv{d.shape=circle,d.filled=full,d.size=3thick,l=$a_2$,l.a=-90}{v2}
\fmfv{d.shape=circle,d.filled=full,d.size=3thick,l=$a_3$}{v3}
\fmffreeze
\fmf{half_prop}{v0,v1}
\fmf{half_prop,left=0.6}{v0,v2}
\fmf{half_prop,right=0.6}{v1,v3}
\fmf{half_prop}{v2,v3}
\end{fmfgraph*}
\end{fmffile}}
\hspace{-30pt}

\vspace{\baselineskip}
T3.1 \quad \quad\quad T3.2 \quad \quad\quad T3.3 \quad \quad\quad T3.4 \quad \quad\quad T3.5
\vspace{\baselineskip}

\end{center}
\caption{Zero-, first-, second- and third-order TSDs corresponding to BMBPT diagrams generated from operators containing six legs at most, i.e., with \texttt{deg\_max}~$=6$.}
\label{diagTSD0123}
\end{figure}
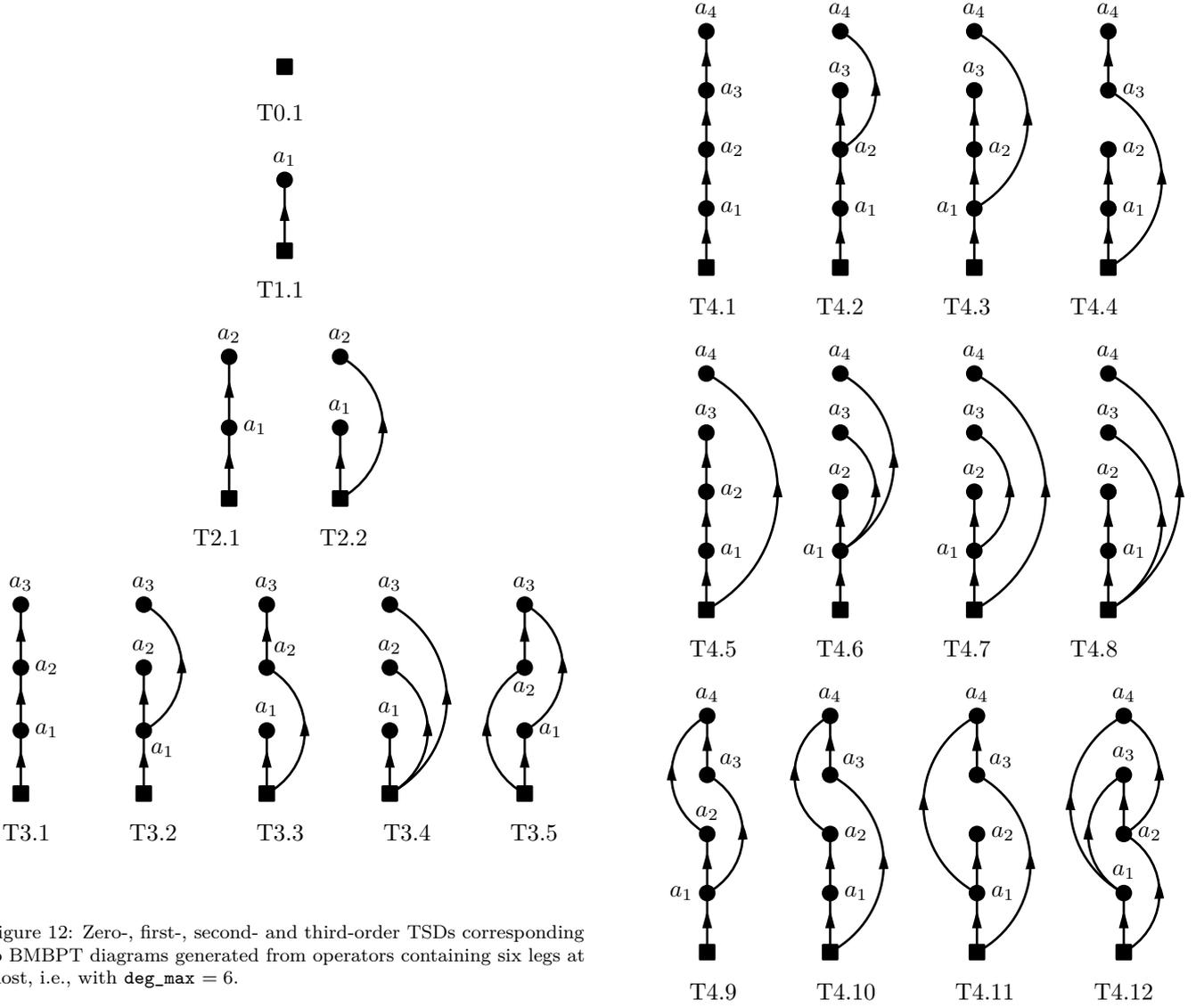

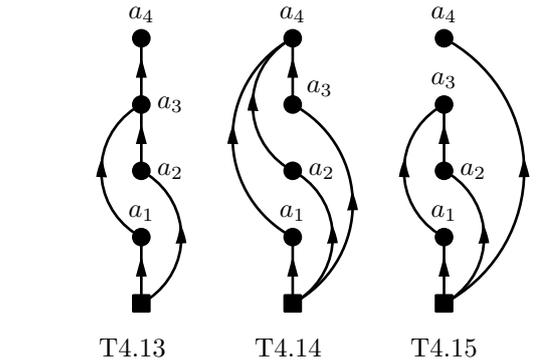
\begin{figure}[t!]
\begin{center}
\hspace{-20pt}
\parbox{100pt}{\begin{fmffile}{time_4_0}
\begin{fmfgraph*}(100,100)
\fmfcmd{style_def half_prop expr p =
draw_plain p;
shrink(.7);
     cfill (marrow (p, .5))
endshrink;
enddef;}
\fmftop{v4}\fmfbottom{v0}
\fmf{phantom}{v0,v1}
\fmfv{d.shape=square,d.filled=full,d.size=3thick}{v0}
\fmf{phantom}{v1,v2}
\fmfv{d.shape=circle,d.filled=full,d.size=3thick,l=$a_1$,l.a=0}{v1}
\fmf{phantom}{v2,v3}
\fmfv{d.shape=circle,d.filled=full,d.size=3thick,l=$a_2$}{v2}
\fmf{phantom}{v3,v4}
\fmfv{d.shape=circle,d.filled=full,d.size=3thick,l=$a_3$,l.a=0}{v3}
\fmfv{d.shape=circle,d.filled=full,d.size=3thick,l=$a_4$}{v4}
\fmffreeze
\fmf{half_prop}{v0,v1}
\fmf{half_prop}{v1,v2}
\fmf{half_prop}{v2,v3}
\fmf{half_prop}{v3,v4}
\end{fmfgraph*}
\end{fmffile}}
\hspace{-50pt}
\parbox{100pt}{\begin{fmffile}{time_4_1}
\begin{fmfgraph*}(100,100)
\fmfcmd{style_def half_prop expr p =
draw_plain p;
shrink(.7);
     cfill (marrow (p, .5))
endshrink;
enddef;}
\fmftop{v4}\fmfbottom{v0}
\fmf{phantom}{v0,v1}
\fmfv{d.shape=square,d.filled=full,d.size=3thick}{v0}
\fmf{phantom}{v1,v2}
\fmfv{d.shape=circle,d.filled=full,d.size=3thick,l=$a_1$,l.a=0}{v1}
\fmf{phantom}{v2,v3}
\fmfv{d.shape=circle,d.filled=full,d.size=3thick,l=$a_2$}{v2}
\fmf{phantom}{v3,v4}
\fmfv{d.shape=circle,d.filled=full,d.size=3thick,l=$a_3$}{v3}
\fmfv{d.shape=circle,d.filled=full,d.size=3thick,l=$a_4$}{v4}
\fmffreeze
\fmf{half_prop}{v0,v1}
\fmf{half_prop}{v1,v2}
\fmf{half_prop}{v2,v3}
\fmf{half_prop,right=0.6}{v2,v4}
\end{fmfgraph*}
\end{fmffile}}
\hspace{-50pt}
\parbox{100pt}{\begin{fmffile}{time_4_2}
\begin{fmfgraph*}(100,100)
\fmfcmd{style_def half_prop expr p =
draw_plain p;
shrink(.7);
     cfill (marrow (p, .5))
endshrink;
enddef;}
\fmftop{v4}\fmfbottom{v0}
\fmf{phantom}{v0,v1}
\fmfv{d.shape=square,d.filled=full,d.size=3thick}{v0}
\fmf{phantom}{v1,v2}
\fmfv{d.shape=circle,d.filled=full,d.size=3thick,l=$a_1$,l.a=180}{v1}
\fmf{phantom}{v2,v3}
\fmfv{d.shape=circle,d.filled=full,d.size=3thick,l=$a_2$}{v2}
\fmf{phantom}{v3,v4}
\fmfv{d.shape=circle,d.filled=full,d.size=3thick,l=$a_3$}{v3}
\fmfv{d.shape=circle,d.filled=full,d.size=3thick,l=$a_4$}{v4}
\fmffreeze
\fmf{half_prop}{v0,v1}
\fmf{half_prop}{v1,v2}
\fmf{half_prop,right=0.6}{v1,v4}
\fmf{half_prop}{v2,v3}
\end{fmfgraph*}
\end{fmffile}}
\hspace{-50pt}
\parbox{100pt}{\begin{fmffile}{time_4_3}
\begin{fmfgraph*}(100,100)
\fmfcmd{style_def half_prop expr p =
draw_plain p;
shrink(.7);
     cfill (marrow (p, .5))
endshrink;
enddef;}
\fmftop{v4}\fmfbottom{v0}
\fmf{phantom}{v0,v1}
\fmfv{d.shape=square,d.filled=full,d.size=3thick}{v0}
\fmf{phantom}{v1,v2}
\fmfv{d.shape=circle,d.filled=full,d.size=3thick,l=$a_1$,l.a=0}{v1}
\fmf{phantom}{v2,v3}
\fmfv{d.shape=circle,d.filled=full,d.size=3thick,l=$a_2$}{v2}
\fmf{phantom}{v3,v4}
\fmfv{d.shape=circle,d.filled=full,d.size=3thick,l=$a_3$,l.a=0}{v3}
\fmfv{d.shape=circle,d.filled=full,d.size=3thick,l=$a_4$}{v4}
\fmffreeze
\fmf{half_prop}{v0,v1}
\fmf{half_prop,right=0.6}{v0,v3}
\fmf{half_prop}{v1,v2}
\fmf{half_prop}{v3,v4}
\end{fmfgraph*}
\end{fmffile}}

\vspace{\baselineskip}
\hspace{-20pt} T4.1 \quad \quad\quad T4.2 \quad \quad\quad T4.3 \quad 
\quad\quad T4.4
\vspace{2\baselineskip}

\hspace{-20pt}
\parbox{100pt}{\begin{fmffile}{time_4_4}
\begin{fmfgraph*}(100,100)
\fmfcmd{style_def half_prop expr p =
draw_plain p;
shrink(.7);
     cfill (marrow (p, .5))
endshrink;
enddef;}
\fmftop{v4}\fmfbottom{v0}
\fmf{phantom}{v0,v1}
\fmfv{d.shape=square,d.filled=full,d.size=3thick}{v0}
\fmf{phantom}{v1,v2}
\fmfv{d.shape=circle,d.filled=full,d.size=3thick,l=$a_1$,l.a=0}{v1}
\fmf{phantom}{v2,v3}
\fmfv{d.shape=circle,d.filled=full,d.size=3thick,l=$a_2$,l.a=0}{v2}
\fmf{phantom}{v3,v4}
\fmfv{d.shape=circle,d.filled=full,d.size=3thick,l=$a_3$}{v3}
\fmfv{d.shape=circle,d.filled=full,d.size=3thick,l=$a_4$}{v4}
\fmffreeze
\fmf{half_prop}{v0,v1}
\fmf{half_prop,right=0.6}{v0,v4}
\fmf{half_prop}{v1,v2}
\fmf{half_prop}{v2,v3}
\end{fmfgraph*}
\end{fmffile}}
\hspace{-50pt}
\parbox{100pt}{\begin{fmffile}{time_4_5}
\begin{fmfgraph*}(100,100)
\fmfcmd{style_def half_prop expr p =
draw_plain p;
shrink(.7);
     cfill (marrow (p, .5))
endshrink;
enddef;}
\fmftop{v4}\fmfbottom{v0}
\fmf{phantom}{v0,v1}
\fmfv{d.shape=square,d.filled=full,d.size=3thick}{v0}
\fmf{phantom}{v1,v2}
\fmfv{d.shape=circle,d.filled=full,d.size=3thick,l=$a_1$,l.a=180}{v1}
\fmf{phantom}{v2,v3}
\fmfv{d.shape=circle,d.filled=full,d.size=3thick,l=$a_2$,l.a=90}{v2}
\fmf{phantom}{v3,v4}
\fmfv{d.shape=circle,d.filled=full,d.size=3thick,l=$a_3$}{v3}
\fmfv{d.shape=circle,d.filled=full,d.size=3thick,l=$a_4$}{v4}
\fmffreeze
\fmf{half_prop}{v0,v1}
\fmf{half_prop}{v1,v2}
\fmf{half_prop,right=0.6}{v1,v3}
\fmf{half_prop,right=0.6}{v1,v4}
\end{fmfgraph*}
\end{fmffile}}
\hspace{-50pt}
\parbox{100pt}{\begin{fmffile}{time_4_6}
\begin{fmfgraph*}(100,100)
\fmfcmd{style_def half_prop expr p =
draw_plain p;
shrink(.7);
     cfill (marrow (p, .5))
endshrink;
enddef;}
\fmftop{v4}\fmfbottom{v0}
\fmf{phantom}{v0,v1}
\fmfv{d.shape=square,d.filled=full,d.size=3thick}{v0}
\fmf{phantom}{v1,v2}
\fmfv{d.shape=circle,d.filled=full,d.size=3thick,l=$a_1$,l.a=180}{v1}
\fmf{phantom}{v2,v3}
\fmfv{d.shape=circle,d.filled=full,d.size=3thick,l=$a_2$,l.a=90}{v2}
\fmf{phantom}{v3,v4}
\fmfv{d.shape=circle,d.filled=full,d.size=3thick,l=$a_3$}{v3}
\fmfv{d.shape=circle,d.filled=full,d.size=3thick,l=$a_4$}{v4}
\fmffreeze
\fmf{half_prop}{v0,v1}
\fmf{half_prop,right=0.6}{v0,v4}
\fmf{half_prop}{v1,v2}
\fmf{half_prop,right=0.6}{v1,v3}
\end{fmfgraph*}
\end{fmffile}}
\hspace{-50pt}
\parbox{100pt}{\begin{fmffile}{time_4_7_forgotten}
\begin{fmfgraph*}(100,100)
\fmfcmd{style_def half_prop expr p =
draw_plain p;
shrink(.7);
     cfill (marrow (p, .5))
endshrink;
enddef;}
\fmftop{v4}\fmfbottom{v0}
\fmf{phantom}{v0,v1}
\fmfv{d.shape=square,d.filled=full,d.size=3thick}{v0}
\fmf{phantom}{v1,v2}
\fmfv{d.shape=circle,d.filled=full,d.size=3thick,l=$a_1$,l.a=0}{v1}
\fmf{phantom}{v2,v3}
\fmfv{d.shape=circle,d.filled=full,d.size=3thick,l=$a_2$,l.a=90}{v2}
\fmf{phantom}{v3,v4}
\fmfv{d.shape=circle,d.filled=full,d.size=3thick,l=$a_3$,l.a=90}{v3}
\fmfv{d.shape=circle,d.filled=full,d.size=3thick,l=$a_4$,l.a=90}{v4}
\fmffreeze
\fmf{half_prop}{v0,v1}
\fmf{half_prop,right=0.6}{v0,v3}
\fmf{half_prop,right=0.6}{v0,v4}
\fmf{half_prop}{v1,v2}
\end{fmfgraph*}
\end{fmffile}}

\vspace{\baselineskip}
\hspace{-20pt} T4.5 \quad \quad\quad T4.6 \quad \quad\quad T4.7 \quad 
\quad\quad T4.8
\vspace{2\baselineskip}

\hspace{-40pt}
\parbox{100pt}{\begin{fmffile}{time_4_7}
\begin{fmfgraph*}(100,100)
\fmfcmd{style_def half_prop expr p =
draw_plain p;
shrink(.7);
     cfill (marrow (p, .5))
endshrink;
enddef;}
\fmftop{v4}\fmfbottom{v0}
\fmf{phantom}{v0,v1}
\fmfv{d.shape=square,d.filled=full,d.size=3thick}{v0}
\fmf{phantom}{v1,v2}
\fmfv{d.shape=circle,d.filled=full,d.size=3thick,l=$a_1$,l.a=180}{v1}
\fmf{phantom}{v2,v3}
\fmfv{d.shape=circle,d.filled=full,d.size=3thick,l=$a_2$,l.a=90}{v2}
\fmf{phantom}{v3,v4}
\fmfv{d.shape=circle,d.filled=full,d.size=3thick,l=$a_3$,l.a=30}{v3}
\fmfv{d.shape=circle,d.filled=full,d.size=3thick,l=$a_4$}{v4}
\fmffreeze
\fmf{half_prop}{v0,v1}
\fmf{half_prop}{v1,v2}
\fmf{half_prop,right=0.6}{v1,v3}
\fmf{half_prop,left=0.6}{v2,v4}
\fmf{half_prop}{v3,v4}
\end{fmfgraph*}
\end{fmffile}}
\hspace{-55pt}
\parbox{100pt}{\begin{fmffile}{time_4_8}
\begin{fmfgraph*}(100,100)
\fmfcmd{style_def half_prop expr p =
draw_plain p;
shrink(.7);
     cfill (marrow (p, .5))
endshrink;
enddef;}
\fmftop{v4}\fmfbottom{v0}
\fmf{phantom}{v0,v1}
\fmfv{d.shape=square,d.filled=full,d.size=3thick}{v0}
\fmf{phantom}{v1,v2}
\fmfv{d.shape=circle,d.filled=full,d.size=3thick,l=$a_1$,l.a=0}{v1}
\fmf{phantom}{v2,v3}
\fmfv{d.shape=circle,d.filled=full,d.size=3thick,l=$a_2$}{v2}
\fmf{phantom}{v3,v4}
\fmfv{d.shape=circle,d.filled=full,d.size=3thick,l=$a_3$,l.a=30}{v3}
\fmfv{d.shape=circle,d.filled=full,d.size=3thick,l=$a_4$}{v4}
\fmffreeze
\fmf{half_prop}{v0,v1}
\fmf{half_prop,right=0.6}{v0,v3}
\fmf{half_prop}{v1,v2}
\fmf{half_prop,left=0.6}{v2,v4}
\fmf{half_prop}{v3,v4}
\end{fmfgraph*}
\end{fmffile}}
\hspace{-45pt}
\parbox{100pt}{\begin{fmffile}{time_4_9}
\begin{fmfgraph*}(100,100)
\fmfcmd{style_def half_prop expr p =
draw_plain p;
shrink(.7);
     cfill (marrow (p, .5))
endshrink;
enddef;}
\fmftop{v4}\fmfbottom{v0}
\fmf{phantom}{v0,v1}
\fmfv{d.shape=square,d.filled=full,d.size=3thick}{v0}
\fmf{phantom}{v1,v2}
\fmfv{d.shape=circle,d.filled=full,d.size=3thick,l=$a_1$,l.a=0}{v1}
\fmf{phantom}{v2,v3}
\fmfv{d.shape=circle,d.filled=full,d.size=3thick,l=$a_2$}{v2}
\fmf{phantom}{v3,v4}
\fmfv{d.shape=circle,d.filled=full,d.size=3thick,l=$a_3$,l.a=30}{v3}
\fmfv{d.shape=circle,d.filled=full,d.size=3thick,l=$a_4$}{v4}
\fmffreeze
\fmf{half_prop}{v0,v1}
\fmf{half_prop,right=0.6}{v0,v3}
\fmf{half_prop}{v1,v2}
\fmf{half_prop,left=0.6}{v1,v4}
\fmf{half_prop}{v3,v4}
\end{fmfgraph*}
\end{fmffile}}
\hspace{-45pt}
\parbox{100pt}{\begin{fmffile}{time_4_10}
\begin{fmfgraph*}(100,100)
\fmfcmd{style_def half_prop expr p =
draw_plain p;
shrink(.7);
     cfill (marrow (p, .5))
endshrink;
enddef;}
\fmftop{v4}\fmfbottom{v0}
\fmf{phantom}{v0,v1}
\fmfv{d.shape=square,d.filled=full,d.size=3thick}{v0}
\fmf{phantom}{v1,v2}
\fmfv{d.shape=circle,d.filled=full,d.size=3thick,l=$a_1$,l.a=90}{v1}
\fmf{phantom}{v2,v3}
\fmfv{d.shape=circle,d.filled=full,d.size=3thick,l=$a_2$}{v2}
\fmf{phantom}{v3,v4}
\fmfv{d.shape=circle,d.filled=full,d.size=3thick,l=$a_3$}{v3}
\fmfv{d.shape=circle,d.filled=full,d.size=3thick,l=$a_4$}{v4}
\fmffreeze
\fmf{half_prop}{v0,v1}
\fmf{half_prop,right=0.6}{v0,v2}
\fmf{half_prop,left=0.6}{v1,v3}
\fmf{half_prop,left=0.6}{v1,v4}
\fmf{half_prop}{v2,v3}
\fmf{half_prop,right=0.6}{v2,v4}
\end{fmfgraph*}
\end{fmffile}}
\hspace{-30pt}

\vspace{\baselineskip}
\hspace{-5pt} T4.9 \quad \quad\quad T4.10 \quad \quad\quad T4.11 \quad 
\quad\quad T4.12
\vspace{2\baselineskip}

\hspace{-10pt}
\parbox{100pt}{\begin{fmffile}{time_4_11}
\begin{fmfgraph*}(100,100)
\fmfcmd{style_def half_prop expr p =
draw_plain p;
shrink(.7);
     cfill (marrow (p, .5))
endshrink;
enddef;}
\fmftop{v4}\fmfbottom{v0}
\fmf{phantom}{v0,v1}
\fmfv{d.shape=square,d.filled=full,d.size=3thick}{v0}
\fmf{phantom}{v1,v2}
\fmfv{d.shape=circle,d.filled=full,d.size=3thick,l=$a_1$,l.a=90}{v1}
\fmf{phantom}{v2,v3}
\fmfv{d.shape=circle,d.filled=full,d.size=3thick,l=$a_2$}{v2}
\fmf{phantom}{v3,v4}
\fmfv{d.shape=circle,d.filled=full,d.size=3thick,l=$a_3$,l.a=0}{v3}
\fmfv{d.shape=circle,d.filled=full,d.size=3thick,l=$a_4$}{v4}
\fmffreeze
\fmf{half_prop}{v0,v1}
\fmf{half_prop,right=0.6}{v0,v2}
\fmf{half_prop,left=0.6}{v1,v3}
\fmf{half_prop}{v2,v3}
\fmf{half_prop}{v3,v4}
\end{fmfgraph*}
\end{fmffile}}
\hspace{-50pt}
\parbox{100pt}{\begin{fmffile}{time_4_12}
\begin{fmfgraph*}(100,100)
\fmfcmd{style_def half_prop expr p =
draw_plain p;
shrink(.7);
     cfill (marrow (p, .5))
endshrink;
enddef;}
\fmftop{v4}\fmfbottom{v0}
\fmf{phantom}{v0,v1}
\fmfv{d.shape=square,d.filled=full,d.size=3thick}{v0}
\fmf{phantom}{v1,v2}
\fmfv{d.shape=circle,d.filled=full,d.size=3thick,l=$a_1$,l.a=90}{v1}
\fmf{phantom}{v2,v3}
\fmfv{d.shape=circle,d.filled=full,d.size=3thick,l=$a_2$}{v2}
\fmf{phantom}{v3,v4}
\fmfv{d.shape=circle,d.filled=full,d.size=3thick,l=$a_3$,l.a=30}{v3}
\fmfv{d.shape=circle,d.filled=full,d.size=3thick,l=$a_4$}{v4}
\fmffreeze
\fmf{half_prop}{v0,v1}
\fmf{half_prop,right=0.6}{v0,v2}
\fmf{half_prop,right=0.6}{v0,v3}
\fmf{half_prop,left=0.6}{v1,v4}
\fmf{half_prop,left=0.6}{v2,v4}
\fmf{half_prop}{v3,v4}
\end{fmfgraph*}
\end{fmffile}}
\hspace{-50pt}
\parbox{100pt}{\begin{fmffile}{time_4_13}
\begin{fmfgraph*}(100,100)
\fmfcmd{style_def half_prop expr p =
draw_plain p;
shrink(.7);
     cfill (marrow (p, .5))
endshrink;
enddef;}
\fmftop{v4}\fmfbottom{v0}
\fmf{phantom}{v0,v1}
\fmfv{d.shape=square,d.filled=full,d.size=3thick}{v0}
\fmf{phantom}{v1,v2}
\fmfv{d.shape=circle,d.filled=full,d.size=3thick,l=$a_1$,l.a=90}{v1}
\fmf{phantom}{v2,v3}
\fmfv{d.shape=circle,d.filled=full,d.size=3thick,l=$a_2$}{v2}
\fmf{phantom}{v3,v4}
\fmfv{d.shape=circle,d.filled=full,d.size=3thick,l=$a_3$}{v3}
\fmfv{d.shape=circle,d.filled=full,d.size=3thick,l=$a_4$}{v4}
\fmffreeze
\fmf{half_prop}{v0,v1}
\fmf{half_prop,right=0.6}{v0,v2}
\fmf{half_prop,right=0.6}{v0,v4}
\fmf{half_prop,left=0.6}{v1,v3}
\fmf{half_prop}{v2,v3}
\end{fmfgraph*}
\end{fmffile}}

\vspace{\baselineskip}
\hspace{-10pt} T4.13 \quad \quad\quad T4.14 \quad \quad\quad T4.15
\vspace{\baselineskip}

\end{center}

\caption{Fourth-order TSDs corresponding to BMBPT diagrams generated from operators containing six legs at most, i.e., with \texttt{deg\_max}~$=6$.}
\label{diagTSD4}
\end{figure}

It is mandatory to generate the TSDs \emph{from} the underlying BMBPT diagrams. Indeed, only in the latter can the maximum degree \texttt{deg\_max} of the operators at play be employed to constrain the topology of the diagrams, eventually dictating the topology of allowed TSDs. With this in mind and following the above rules, the $1/1/2/5/15$ TSDs of order $0/1/2/3/4$ corresponding to BMBPT diagrams generated from operators with \texttt{deg\_max}~$=6$, i.e., containing effective three-body terms, have been produced and systematically displayed in Figs.~\ref{diagTSD0123} and~\ref{diagTSD4}. One notices that the first TSD containing a cycle is the third-order TSD labelled as T3.5 in Fig.~\ref{diagTSD0123}, i.e., all TSDs up to order $2$ ($3$) are trees (but one). At order $4$, seven out of fourteen TSDs contain cycles. Obviously, the higher the order, the more complex the topology can be.    

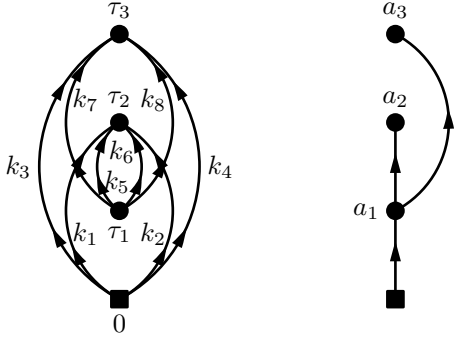
\begin{figure}[t!]
\begin{center}
\parbox{100pt}{\begin{fmffile}{diag_ex_labelled}
\begin{fmfgraph*}(100,100)
\fmfcmd{style_def prop_pm expr p =
    draw_plain p;
    shrink(.7);
        cfill (marrow (p, .25));
        cfill (marrow (p, .75))
    endshrink;
	enddef;}
\fmftop{v3}\fmfbottom{v0}
\fmf{phantom}{v0,v1}
\fmfv{d.shape=square,d.filled=full,d.size=3thick,l=$0$}{v0}
\fmf{phantom}{v1,v2}
\fmfv{d.shape=circle,d.filled=full,d.size=3thick,l=$\tau_1$}{v1}
\fmf{phantom}{v2,v3}
\fmfv{d.shape=circle,d.filled=full,d.size=3thick,l=$\tau_2$}{v2}
\fmfv{d.shape=circle,d.filled=full,d.size=3thick,l=$\tau_3$}{v3}
\fmffreeze
\fmf{prop_pm,left=0.6,tag=1}{v0,v2}
\fmf{prop_pm,right=0.6,tag=2}{v0,v2}
\fmf{prop_pm,left=0.6,tag=3}{v0,v3}
\fmf{prop_pm,right=0.6,tag=4}{v0,v3}
\fmf{prop_pm,left=0.5,tag=5}{v1,v2}
\fmf{prop_pm,right=0.5,tag=6}{v1,v2}
\fmf{prop_pm,left=0.6,tag=7}{v1,v3}
\fmf{prop_pm,right=0.6,tag=8}{v1,v3}
\fmfposition
	\fmfipath{p[]}
	\fmfiset{p1}{vpath1(__v0,__v2)}
	\fmfiset{p2}{vpath2(__v0,__v2)}
	\fmfiset{p3}{vpath3(__v0,__v3)}
	\fmfiset{p4}{vpath4(__v0,__v3)}
	\fmfiset{p5}{vpath5(__v1,__v2)}
	\fmfiset{p6}{vpath6(__v1,__v2)}
	\fmfiset{p7}{vpath7(__v1,__v3)}
	\fmfiset{p8}{vpath8(__v1,__v3)}
	\fmfiv{label=$k_1$,l.dist=.05w,l.a=-60}{point length(p1)/2 of p1}
	\fmfiv{label=$k_2$,l.dist=.05w,l.a=-120}{point length(p2)/2 of p2}
	\fmfiv{label=$k_3$,l.dist=.03w}{point length(p3)/2 of p3}
	\fmfiv{label=$k_4$,l.dist=.03w}{point length(p4)/2 of p4}
	\fmfiv{label=$k_5$,l.dist=.03w,l.a=-30}{point length(p5)/2 of p5}
	\fmfiv{label=$k_6$,l.dist=.03w,l.a=150}{point length(p6)/2 of p6}
	\fmfiv{label=$k_7$,l.dist=.05w,l.a=60}{point length(p7)/2 of p7}
	\fmfiv{label=$k_8$,l.dist=.05w,l.a=120}{point length(p8)/2 of p8}
\end{fmfgraph*}
\end{fmffile}}
\parbox{100pt}{\begin{fmffile}{tsd_ex_labelled}
\begin{fmfgraph*}(100,100)
\fmfcmd{style_def half_prop expr p =
    draw_plain p;
    shrink(.7);
        cfill (marrow (p, .5))
    endshrink;
	enddef;}
\fmftop{v3}\fmfbottom{v0}
\fmfv{d.shape=square,d.filled=full,d.size=3thick}{v0}
\fmfv{d.shape=circle,d.filled=full,d.size=3thick,l=$a_1$,l.a=180}{v1}
\fmf{phantom}{v2,v3}
\fmfv{d.shape=circle,d.filled=full,d.size=3thick,l=$a_2$}{v2}
\fmfv{d.shape=circle,d.filled=full,d.size=3thick,l=$a_3$}{v3}
\fmf{half_prop}{v0,v1}
\fmf{half_prop}{v1,v2}
\fmffreeze
\fmf{half_prop,right=0.6}{v1,v3}
\end{fmfgraph*}
\end{fmffile}}
\end{center}
\caption{Fully-labelled third-order BMBPT diagram displayed in Fig.~\ref{f:ex_bmbpt_diag} and its associated TSD.}
\label{f:tsd_labelling}
\end{figure}

In the end, \emph{different} BMBPT diagrams of order $p$ can have the \emph{same} TSD, i.e., the same underlying time structure. For instance, the eleven BMBPT diagrams of orders $0$, $1$ and $2$ displayed in Fig.~\ref{diagBMBPT012} translate into only four TSDs in Fig.~\ref{diagTSD0123}. However, the time integral eventually turns into a different result for each associated BMBPT diagram given that the energy labels in terms of which the result is expressed have a different meaning in each case. Back to our example of Fig.~\ref{f:tsd_production}, the full labelling of the BMBPT diagram provided in Fig.~\ref{f:tsd_labelling} allows us to identify the actual expression of the vertex labels
\begin{align*}
a_1 &= \epsilon^{k_{5}k_{6}k_{7}k_{8}}_{}  \ , \\
a_2 &= \epsilon_{k_{1}k_{2}k_{5}k_{6}}^{} \ , \\
a_3 &= \epsilon_{k_{3}k_{4}k_{7}k_{8}}^{}  \ ,
\end{align*}
to be used in the final outcome of the TSD. Another fourth-order diagram with the same TSD would associate other combinations of quasi-particle energies to the energy labels $a_1$, $a_2$ and $a_3$.

\subsection{Calculation of tree TSDs}
\label{subs:trees}

Tree TSDs happen to play an instrumental role in the present context. Indeed, they constitute the category for which a direct algorithm can be found to evaluate the corresponding $p$-tuple time integral. Building on it, non-tree TSDs (i.e., starting with order $p=3$) will be treated by re-expressing them as a sum of tree TSDs. 

The identification of the rule to compute a tree TSD relies on a recursive procedure, i.e., starting from a tree TSD of order $p$, whose expression is considered to be known, a vertex $\Omega$ carrying label $a_{p+1}$ is added by connecting it to one of the vertices of the original TSD. Having generated a new TSD of order $p+1$, its expression is obtained.

\subsubsection{Minimal tree TSD}
\label{subsubs:minimaltree}

One starts with the minimal tree TSD, i.e., the single TSD of order $0$ denoted as T0.1 in Fig.~\ref{diagTSD0123}. It is built from the sole vertex representing the operator $O$ and does not carry any running time label. It looks like
\begin{center}
\parbox{30pt}{\begin{fmffile}{minimaltreetsd}
\begin{fmfgraph*}(30,30)
\fmftop{v1}\fmfbottom{v0}
\fmfv{d.shape=square,d.filled=full,d.size=3thick}{i1}
\fmf{phantom}{v0,i1}
\fmf{phantom}{i1,v1}
\end{fmfgraph*}
\end{fmffile}}
\end{center}
and its expression is nothing but
\begin{align*}
\text{T0.1} &= 1 \ .
\end{align*}

\subsubsection{First-order TSD}

The single first-order TSD, denoted as T1.1 in Fig.~\ref{diagTSD0123}, is generated from the minimal tree graph by connecting one $\Omega$ vertex carrying label $a_1$ to the vertex $O$
\begin{center}
\vspace{\baselineskip}
\parbox{30pt}{\begin{fmffile}{minimaltimediag}
\begin{fmfgraph*}(30,30)
\fmfcmd{style_def half_prop expr p =
    draw_plain p;
    shrink(.7);
        cfill (marrow (p, .5))
    endshrink;
	enddef;}
\fmftop{v1}\fmfbottom{v0}
\fmfv{d.shape=square,d.filled=full,d.size=3thick}{v0}
\fmfv{d.shape=circle,d.filled=full,d.size=3thick,l=$a_1$}{v1}
\fmf{half_prop}{v0,v1}
\end{fmfgraph*}
\end{fmffile}}
\vspace{\baselineskip}
\end{center}
The expression of this TSD is given by the single integral
\begin{align*}
\text{T1.1} &= \lim\limits_{\tau \to \infty} \int_0^{\tau} d\tau_1 \, e^{-a_1\tau_1} \\
&= \frac{1}{a_1} \ ,
\end{align*}
such that the end denominator is simply equal to the energy factor $a_1$\footnote{The finiteness of the result relies on the fact that the energy factor $a_1$ is taken to be positive. In the following, all prefactors at play in a given integral will be assumed to be positive, which will eventually be justified for BMBPT diagrams in Sec.~\ref{subs:tsd_exp_to_goldstone}.}. One trivially observes that the end result could have been obtained directly by adding the factor $a_1$ associated to the new vertex to the denominator of the minimal tree TSD.

\subsubsection{Second-order TSDs}

As is visible in Fig.~\ref{diagTSD0123}, two second-order TSDs denoted as T2.1 and T2.2 can be built from the first-order TSD. The first one is obtained by connecting the new vertex to the one labelled by $a_1$. It provides the \emph{linear} tree TSD
\begin{center}
\vspace{\baselineskip}
\parbox{60pt}{\begin{fmffile}{order2a}
\begin{fmfgraph*}(60,60)
\fmfcmd{style_def half_prop expr p =
    draw_plain p;
    shrink(.7);
        cfill (marrow (p, .5))
    endshrink;
	enddef;}
\fmftop{v2}\fmfbottom{v0}
\fmfv{d.shape=square,d.filled=full,d.size=3thick}{v0}
\fmfv{d.shape=circle,d.filled=full,d.size=3thick,l=$a_1$}{v1}
\fmfv{d.shape=circle,d.filled=full,d.size=3thick,l=$a_2$}{v2}
\fmf{half_prop}{v0,v1}
\fmf{half_prop}{v1,v2}
\end{fmfgraph*}
\end{fmffile}}
\vspace{\baselineskip}
\end{center}
in which all vertices belong to the same branch and are thus sequentially ordered in time. As a result, the double time integral displays two nested integrals, the one over the earlier time depending on the result of the one over the later time that is thus performed first, i.e.
\begin{align*}
\text{T2.1} &= \lim\limits_{\tau \to \infty} \int_0^{\tau} d\tau_1 d\tau_2 \, \theta(\tau_2 - \tau_1) e^{-a_1\tau_1} e^{-a_2\tau_2} \\
&= \lim\limits_{\tau \to \infty} \int_0^{\tau} d\tau_1 e^{-a_1\tau_1} \int_{\tau_1}^{\tau} d\tau_2 e^{-a_2\tau_2} \\
&= \lim\limits_{\tau \to \infty} -\frac{1}{a_2}\int_0^{\tau} d\tau_1 e^{-a_1\tau_1} \left( e^{-a_2\tau} - e^{-a_2\tau_1} \right) \\
&= \lim\limits_{\tau \to \infty} \frac{1}{a_2} \left( \int_0^{\tau} d\tau_1 e^{-(a_1+a_2)\tau_1}  -e^{-a_2\tau} \int_0^{\tau} d\tau_1 e^{-a_1\tau_1} \right) \\
&= \frac{1}{a_2(a_1+a_2)} \, .
\end{align*}
One observes that the end result could have been obtained directly by adding the factor $a_2$ associated to the new vertex to the denominator of T1.1 and by further replacing $a_1$ by $a_1+a_2$. This is because the first integration over $\tau_2$ trivially brings the factor $a_2$ to the denominator and makes at the same time the variable $a_1+a_2$ become the factor in front of $\tau_1$ in the part of the subsequent integration that eventually remains in the limit $\tau\rightarrow +\infty$.

The alternative way to generate a second-order TSD from the first-order one is to connect the new vertex to the bottom vertex. This gives a tree TSD with two branches
\begin{center}
\vspace{\baselineskip}
\parbox{60pt}{\begin{fmffile}{order2b}
\begin{fmfgraph*}(60,60)
\fmfcmd{style_def half_prop expr p =
    draw_plain p;
    shrink(.7);
        cfill (marrow (p, .5))
    endshrink;
	enddef;}
\fmftop{v2}\fmfbottom{v0}
\fmfv{d.shape=square,d.filled=full,d.size=3thick}{v0}
\fmfv{d.shape=circle,d.filled=full,d.size=3thick,l=$a_1$,l.a=90}{v1}
\fmfv{d.shape=circle,d.filled=full,d.size=3thick,l=$a_2$}{v2}
\fmf{half_prop}{v0,v1}
\fmf{phantom}{v1,v2}
\fmffreeze
\fmf{half_prop,right=0.6}{v0,v2}
\end{fmfgraph*}
\end{fmffile}}
\vspace{\baselineskip}
\end{center}
such that the vertices in the two branches are not time ordered with respect to each other. Consequently, the double time integral reduces to the product of two independent single integrals, i.e.
\begin{align*}
\text{T2.2} &= \lim\limits_{\tau \to \infty} \int_0^{\tau} d\tau_1 d\tau_2  \, e^{-a_1\tau_1} e^{-a_2\tau_2} \\
&= \lim\limits_{\tau \to \infty} \left(\int_0^{\tau} d\tau_1 e^{-a_1\tau_1}\right) \left(\int_0^{\tau} d\tau_2 e^{-a_2\tau_2} \right)\\
&= \lim\limits_{\tau \to \infty} \frac{1}{a_1a_2}\left(e^{-a_1\tau} - 1\right) \left( e^{-a_2\tau} - 1 \right) \\
&= \frac{1}{a_1a_2} \, .
\end{align*}
One observes that the end result could have been obtained directly by adding the factor $a_2$ associated to the new vertex to the denominator of T1.1, leaving $a_1$ unaffected. Indeed, while the first integration over $\tau_2$ trivially brings the factor $a_2$ to the denominator, it leaves the second integration unaffected as the two are independent.

\subsubsection{Order-$p$ TSDs}

The procedure described above can be extended to compute any tree TSD of order $p>0$ in terms of a reference TSD of order $p-1$. Indeed, any tree TSD of order $p>0$ can be obtained via the addition of a vertex $a_p$ to a reference TSD of order $p-1$. The only three options to do so are to add vertex $a_p$ (i) through a link originating from a pre-existing vertex $a_{q}$, $q =1,\ldots, p-1$, such that $a_q$ (i1) continues an existing branch or (i2) initiates a new branch containing a single vertex (i.e., itself), or (ii) through a link originating from the bottom vertex at fixed time $0$. In all three cases, the integral over $\tau_p$ trivially brings the factor $a_p$ to the final denominator. In cases (i1) and (i2), the prefactor $a_{q}$ in the integration over $\tau_q$ in the reference TSD is replaced by the factor $a_{q}+a_{p}$ in the part of the integral that eventually remains in the limit $\tau\rightarrow +\infty$. Contrarily, all factors $a_q$, $q=1,\ldots,p-1$, involved in the subsequent integrations are left unaffected in case (ii). As a result, the denominator of the TSD of interest can be calculated through the following steps
\begin{enumerate}
\item start from the denominator expression of the reference TSD of order $p-1$,
\item add the factor $a_{p}$,
\item replace every occurrence of $a_{q}$ by $a_{q}+a_{p}$ except if the new vertex is linked to the bottom vertex.
\end{enumerate}

\subsubsection{Algorithm}
\label{subsubs:gen_tree_rules}

Given a general tree TSD of order $p$, the above procedure can be used iteratively to calculate its expression, i.e., the end denominator. Starting from the vertices located at the very end of each branch of the tree, one can indeed iterate the above algorithm to remove them one by one back to the minimal tree TSD. In doing so, the merging of branches is naturally handled. Each step of the way, the most external vertex of a branch is treated as if it had been added to a tree TSD of one order less. Applying the above algorithm, one elementary step results into (i) choping off the treated vertex, (ii) storing a contribution to the end denominator equal to the effective label carried by the removed vertex and (iii) adding the effective label of the removed vertex to the label of the vertex it was linked to, except if the latter is the bottom vertex, in which case the procedure associated to that branch stops. The end expression of the denominator contains $p$ factors resulting from the $p$ steps necessary to iterate through all the vertices. Eventually, the iterative procedure induces the rule to be employed to generate the denominator of any tree TSD of order $p$, i.e.
\begin{enumerate}
\item Consider a vertex $a_q$, $q=1,\ldots,p$, in the TSD,
\begin{enumerate}
\item find all its \emph{descendants}, i.e., all the vertices that are reachable from $a_q$ by following links upward,
\item sum label $a_q$ to the labels of all its descendants,
\item add the corresponding factor to the denominator expression,
\end{enumerate}
\item Go back to 1. until all vertices have been exhausted.
\end{enumerate}
Let us illustrate the diagrammatic rule for tree TSDs by computing the denominator associated with the third-order TSD displayed in Fig.~\ref{f:tsd_labelling}
\begin{enumerate}
\item Starting with vertex $a_1$, vertices $a_2$ and $a_3$ are reached by following two different sets of links upward\footnote{In the present example, each path followed contains only one link.} corresponding to the two branches of the tree TSD. Consequently, the factor $a_1+a_2+a_3$ is associated to vertex $a_1$. 
\item Moving to vertex $a_2$, no other vertex is reachable from it. Thus, the plain factor $a_2$ is associated to it. 
\item Similarly, the plain factor $a_3$ must be associated to vertex $a_3$. 
\item Eventually, the denominator is formed by the product of the factors associated with vertices $a_1$, $a_2$ and $a_3$; i.e., it is equal to $(a_1+a_2+a_3)a_2a_3$. One correctly recovers the result derived in Sec.~\ref{subs:feynman_to_goldstone} via the explicit integration of the corresponding triple time integral.
\end{enumerate}
We, thus, have a rule at hand to compute the time-integrated expression of \emph{any} tree TSD, independently of its perturbative order and of its topology. Although TSDs including at least one cycle are excluded at this point, a tree TSD of arbitrary order $p$ may already correspond to a complex BMBPT diagram displaying any number of branches and sub-branches of arbitrary lengths. 

\subsubsection{Output of the \texttt{ADG} program}

A typical output for a tree TSD looks like

\paragraph{Time-structure diagram T1:}

\begin{center}
\parbox{60pt}{\begin{fmffile}{time_0}
\begin{fmfgraph*}(60,60)
\fmfcmd{style_def half_prop expr p =
    draw_plain p;
    shrink(.7);
        cfill (marrow (p, .5))
    endshrink;
	enddef;}
\fmftop{v2}\fmfbottom{v0}
\fmf{phantom}{v0,v1}
\fmfv{d.shape=square,d.filled=full,d.size=3thick}{v0}
\fmf{phantom}{v1,v2}
\fmfv{d.shape=circle,d.filled=full,d.size=3thick}{v1}
\fmfv{d.shape=circle,d.filled=full,d.size=3thick}{v2}
\fmffreeze
\fmf{half_prop}{v0,v1}
\fmf{half_prop}{v1,v2}
\end{fmfgraph*}
\end{fmffile}}

\end{center}

\begin{equation*}
\text{T1} = \frac{1}{(a_1+ a_2)a_2}
\end{equation*}
Related Feynman diagrams: 8, 6, 5, 4, 3, 2, 1. \\

The TSD is displayed along with its associated expression and the labels of all the BMBPT diagrams it corresponds to.

\subsubsection{From a tree TSD back to BMBPT diagrams}
\label{subs:tsd_exp_to_goldstone}

Once the expression of a tree TSD of order $p$ has been obtained, the goal is to generate the actual time-integrated expression of the BMBPT diagrams associated to it. One obvious way consists of replacing the factors $a_q$, $q=1,\ldots,p$, by their expressions for each BMBPT diagram. However, while these factors constitute the natural variables to write the integrand associated with the Feynman diagram, the time-integrated expression rather depends on specific combinations of them that eventually lead to remarkable cancellations between the terms. It is, thus, more satisfactory to identify what these combinations actually correspond to and formulate the final rule directly in terms of them.

To do so, we introduce the notion of subdiagram, or subgraph, of a diagram as a diagram composed by a subset of vertices plus the propagators that are exchanged between them. As each vertex label $a_q$ in a TSD eventually stands for the sum/difference of quasi-particle energies associated with the lines entering/leaving the \emph{vertex} in the associated BMBPT diagram, a combination of these labels denotes the sum/difference of quasi-particle energies associated with the lines entering/leaving the \emph{subdiagram} grouping the corresponding vertices.

Let us illustrate this notion by coming back to the BMBPT diagram displayed in Fig.~\ref{f:tsd_labelling}. The expression of the associated TSD denominator includes a factor $a_1+a_2+a_3= \epsilon_{k_{1}k_{2}k_{3}k_{4}}$. Considering the subdiagram grouping vertices $a_1$, $a_2$ and $a_3$, one observes that this factor indeed corresponds to the sum/difference of quasi-particle energies associated with the lines entering/leaving it, which in the present case reduces to the sum of $E_{k_1}$, $E_{k_2}$, $E_{k_3}$ and $E_{k_4}$ corresponding to the four entering lines, i.e., there is no line leaving the subdiagram.

The above example underlines a fundamental point. Because each effective factor entering the end denominator sums the label of a given vertex with the labels of all its descendants, the corresponding BMBPT subdiagram only displays entering lines. This results into the effective factors being positive sums of quasi-particle energies. This key feature is responsible for the finiteness of all the encountered time integrals in the limit $\tau\rightarrow +\infty$, as alluded to in footnote 16. 

Eventually, the energy denominator of a BMBPT diagram associated with a tree TSD is obtained in the following way
\begin{enumerate}
\item Consider a vertex but the lowest one in the BMBPT diagram,
\begin{enumerate}
\item determine all its descendants using the TSD diagram,
\item form a subdiagram using the vertex and its descendants,
\item sum the quasi-particle energies corresponding to the lines entering the subdiagram,
\item add the corresponding factor to the denominator expression,
\end{enumerate}
\item Go back to 1. until all vertices have been exhausted.
\end{enumerate}

Let us illustrate the final diagrammatic rule by coming back to the BMBPT diagram displayed in Fig.~\ref{f:tsd_labelling}.
\begin{enumerate}
\item The vertex at time $\tau_1$ in the BMBPT diagram corresponds to vertex $a_1$ in the TSD. Its descendants are vertices $a_2$ and $a_3$ corresponding to BMBPT vertices at times $\tau_2$ and $\tau_3$, respectively. The sum of quasi-particle energies associated to the lines entering the subgraph grouping the three vertices is $\epsilon_{k_{1}k_{2}k_{3}k_{4}}$, thus, providing the first factor entering the denominator.
\item The vertex at time $\tau_2$ in the BMBPT diagram correspond to vertex $a_2$ in the TSD. It has no descendant such that the corresponding subgraph reduces to itself. The sum of quasi-particle energies associated to the lines entering the subgraph is $\epsilon_{k_{1}k_{2}k_{5}k_{6}}$, thus. providing the second factor entering the denominator.
\item The vertex at time $\tau_3$ in the BMBPT diagram correspond to vertex $a_3$ in the TSD. It has no descendant such that the corresponding subgraph reduces to itself. The sum of quasi-particle energies associated to the lines entering the subgraph is $\epsilon_{k_{3}k_{4}k_{7}k_{8}}$, thus, providing the last factor entering the denominator. 
\item Eventually, the complete denominator reads as
\begin{equation*}
\epsilon_{k_1 k_2 k_3 k_4} \, \epsilon_{k_1 k_2 k_5 k_6} \, \epsilon_{k_3 k_4 k_7 k_8} \ ,
\end{equation*}
where each factor corresponds to a positive sum of quasi-particle energies.
\end{enumerate}
The result does indeed match the one obtained in Sec.~\ref{subs:feynman_to_goldstone}.

\subsection{Calculation of non-tree TSDs}
\label{nontreeSec}

Having a direct method at hand to compute the time-integrated form of any BMBPT diagram associated with a tree TSD, one is left with the important task to find an algorithm to tackle diagrams corresponding to non-tree TSDs, i.e., to TSDs containing at least one cycle. As no direct rule applies to them, the strategy consists of commuting any non-tree TSD into a sum of tree TSDs to which the above diagrammatic rule applies.

\subsubsection{Minimal non-tree TSD}

To familiarize ourselves with non-tree TSDs, let us focus on the simplest of them, the third-order TSD denoted as T3.5 in Fig.~\ref{diagTSD0123}
\begin{center}
\vspace{\baselineskip}
\parbox{80pt}{\begin{fmffile}{cycle}
\begin{fmfgraph*}(80,80)
\fmfcmd{style_def half_prop expr p =
    draw_plain p;
    shrink(.7);
        cfill (marrow (p, .5))
    endshrink;
	enddef;}
\fmftop{v3}\fmfbottom{v0}
\fmfv{d.shape=square,d.filled=full,d.size=3thick}{v0}
\fmfv{d.shape=circle,d.filled=full,d.size=3thick,l=$a_1$,l.angle=90}{v1}
\fmfv{d.shape=circle,d.filled=full,d.size=3thick,l=$a_2$,l.angle=30}{v2}
\fmfv{d.shape=circle,d.filled=full,d.size=3thick,l=$a_3$}{v3}
\fmf{half_prop}{v0,v1}
\fmf{half_prop}{v2,v3}
\fmf{phantom}{v1,v2}
\fmffreeze
\fmf{half_prop,right=0.6}{v0,v2}
\fmf{half_prop,left=0.6}{v1,v3}
\end{fmfgraph*}
\end{fmffile}}
\vspace{\baselineskip}
\end{center}
One first notices that vertices $a_1$ and $a_2$ are not time ordered with respect to each other. While this could be dealt with if the two vertices were situated on different branches of a tree TSD, the fact that they are time ordered with respect to $a_3$, i.e., that their time labels run from $0$ to $\tau_3$ and not from $0$ to $\tau\rightarrow +\infty$, prevents a direct treatment. In the following, we use the term \emph{cycle}\footnote{It actually corresponds to an undirected cycle as defined in~\ref{s:graph_theory}.} to describe structures where some vertices are not time ordered with respect to one another while being time ordered with respect to a vertex or a set of vertices at higher times. Calculating the triple time integral associated to the TSD leads to
\begin{align}
\text{T3.5} &=  \lim\limits_{\tau \to \infty} \int_0^{\tau} d\tau_1 d\tau_2 d\tau_3 \, \theta(\tau_3 - \tau_1) \, \theta(\tau_3 - \tau_2) \nonumber \\
& \hspace{2.5cm} \times e^{-a_1\tau_1} e^{-a_2\tau_2} e^{-a_3\tau_3} \nonumber \\
&= \frac{1}{(a_1+a_3)(a_2+a_3)} \left[ \frac{1}{a_3} + \frac{1}{a_1+a_2+a_3} \right] \nonumber \\
&= \frac{1}{a_3(a_1+a_2+a_3)} \left[ \frac{1}{a_1+a_3} + \frac{1}{a_2+a_3} \right] \ . \label{nontreedenominator}
\end{align}
Because the lines inside a cycle do not constitute separate branches, the corresponding time integrals are not independent from one another. It implies that the vertices inside a cycle need to be ordered explicitly in all possible ways. It is what is actually behind the two terms appearing in Eq.~\eqref{nontreedenominator} that were generated via the time partitioning $1 = \theta(\tau_1 - \tau_2) + \theta(\tau_2 - \tau_1)$ in the integrals over $\tau_1$ and $\tau_2$ thus producing the two ways of ordering $a_1$ and $a_2$.

Diagrammatically, employing such a partitioning for two vertices inside a cycle corresponds to generating the sum of two TSDs through the following steps
\begin{enumerate}
\item Select the two internal vertices that are not connected by a link.
\item Connect them via an oriented link and only keep the maximal length paths between each pair of vertices in the graph. This generates a first new graph.
\item Proceeding similarly but with the added link pointing in the opposite direction generates a second new graph. 
\end{enumerate}
In the minimal non-tree graph, the two vertices to be ordered are $a_1$ and $a_2$. Applying the ordering procedure leads to
\begin{center}
\vspace{\baselineskip}
\begin{fmffile}{cycle_as_sam}
\parbox{80pt}{
\begin{fmfgraph*}(80,80)
\fmfcmd{style_def half_prop expr p =
    draw_plain p;
    shrink(.7);
        cfill (marrow (p, .5))
    endshrink;
	enddef;}
\fmftop{v3}\fmfbottom{v0}
\fmfv{d.shape=square,d.filled=full,d.size=3thick}{v0}
\fmfv{d.shape=circle,d.filled=full,d.size=3thick,l=$a_1$,l.angle=90}{v1}
\fmfv{d.shape=circle,d.filled=full,d.size=3thick,l=$a_2$,l.angle=30}{v2}
\fmfv{d.shape=circle,d.filled=full,d.size=3thick,l=$a_3$}{v3}
\fmf{half_prop}{v0,v1}
\fmf{half_prop}{v2,v3}
\fmf{phantom}{v1,v2}
\fmffreeze
\fmf{half_prop,right=0.6}{v0,v2}
\fmf{half_prop,left=0.6}{v1,v3}
\end{fmfgraph*}}
\hspace{-10pt}$=$\hspace{-20pt}
\parbox{80pt}{
\begin{fmfgraph*}(80,80)
\fmfcmd{style_def half_prop expr p =
    draw_plain p;
    shrink(.7);
        cfill (marrow (p, .5))
    endshrink;
	enddef;}
\fmftop{v3}\fmfbottom{v0}
\fmfv{d.shape=square,d.filled=full,d.size=3thick}{v0}
\fmfv{d.shape=circle,d.filled=full,d.size=3thick,l=$a_1$,l.angle=0}{v1}
\fmfv{d.shape=circle,d.filled=full,d.size=3thick,l=$a_2$,l.angle=0}{v2}
\fmfv{d.shape=circle,d.filled=full,d.size=3thick,l=$a_3$}{v3}
\fmf{half_prop}{v0,v1}
\fmf{half_prop}{v2,v3}
\fmf{half_prop}{v1,v2}
\end{fmfgraph*}}
\hspace{-20pt}$+$\hspace{-10pt}
\parbox{80pt}{
\begin{fmfgraph*}(80,80)
\fmfcmd{style_def half_prop expr p =
    draw_plain p;
    shrink(.7);
        cfill (marrow (p, .5))
    endshrink;
	enddef;}
\fmftop{v3}\fmfbottom{v0}
\fmfv{d.shape=square,d.filled=full,d.size=3thick}{v0}
\fmfv{d.shape=circle,d.filled=full,d.size=3thick,l=$a_1$}{v1}
\fmfv{d.shape=circle,d.filled=full,d.size=3thick,l=$a_2$}{v2}
\fmfv{d.shape=circle,d.filled=full,d.size=3thick,l=$a_3$}{v3}
\fmf{phantom}{v0,v1}
\fmf{phantom}{v2,v3}
\fmf{half_prop}{v2,v1}
\fmffreeze
\fmf{half_prop,right=0.6}{v0,v2}
\fmf{half_prop,left=0.6}{v1,v3}
\end{fmfgraph*}}
\end{fmffile}
\vspace{\baselineskip}
\end{center}
such that the non-tree graph T3.5 is nothing but the sum of twice the linear tree graph T3.1 with the order of $a_1$ and $a_2$ exchanged\footnote{It is obviously necessary to move $a_1$ above $a_2$ into the second tree graph to realize that it also displays the topology of T3.1.}.

Applying the algorithm detailed in Sec.~\ref{subs:trees} to each of the two resulting tree TSDs provides
\begin{align}
\text{T3.5} &= \frac{1}{a_3(a_2+a_3)(a_1+a_2+a_3)}  \nonumber \\
&\,\,+ \frac{1}{a_3(a_1+a_3)(a_1+a_2+a_3)} \nonumber \\
&= \frac{1}{a_3(a_1+a_2+a_3)} \left[ \frac{1}{a_1+a_3} + \frac{1}{a_2+a_3} \right] \ , \label{cycleT35}
\end{align}
which is indeed the result of Eq.~\eqref{nontreedenominator}.

\subsubsection{Cycles identification}
\label{subsubs:cycle_finding_alg}

In order to treat non-tree TSDs, one must first identify all the cycles, i.e., the end nodes of the cycles, possibly contained in a given TSD. Exploiting the \emph{NetworkX} graph, it is done by applying the following algorithm
\begin{enumerate}
\item Consider \texttt{node\_a} with \texttt{out\_degree} $\geq$ 2,
\begin{enumerate}
\item Consider \texttt{node\_b} different from \texttt{node\_a},
\begin{itemize}
\item Check for paths going from \texttt{node\_a} to \texttt{node\_b},
\item If \texttt{in\_degree(node\_b)} $\geq$ 2 and \texttt{nb\_paths} $\geq$ 2, \texttt{node\_a} and \texttt{node\_b} are end nodes of a cycle,
\item Check that the two paths share only their end nodes,
\end{itemize}
\item Go back to (a) until all nodes are exhausted,
\end{enumerate}
\item Go back to 1. until all nodes are exhausted,
\end{enumerate}
where the \texttt{in\_degree} (\texttt{out\_degree}) of a node, or vertex, denotes the number of incoming (outgoing) lines and \texttt{nb\_paths} denotes the number of different paths going from \texttt{node\_a} to \texttt{node\_b}.

\begin{figure}[t!]
\begin{center}
\parbox{120pt}{\begin{fmffile}{time_6_0}
\begin{fmfgraph*}(120,120)
\fmfcmd{style_def half_prop expr p =
    draw_plain p;
    shrink(.7);
        cfill (marrow (p, .5))
    endshrink;
	enddef;}
\fmftop{v5}\fmfbottom{v0}
\fmf{phantom}{v0,v1}
\fmfv{d.shape=square,d.filled=full,d.size=3thick,l=$0$}{v0}
\fmf{phantom}{v1,v2}
\fmfv{d.shape=circle,d.filled=full,d.size=3thick,l=$1$,l.angle=180,l.d=.04w}{v1}
\fmf{phantom}{v2,v3}
\fmfv{d.shape=circle,d.filled=full,d.size=3thick,l=$2$,l.angle=0,l.d=.04w}{v2}
\fmf{phantom}{v3,v4}
\fmfv{d.shape=circle,d.filled=full,d.size=3thick,l=$3$,l.angle=180,l.d=.04w}{v3}
\fmf{phantom}{v4,v5}
\fmfv{d.shape=circle,d.filled=full,d.size=3thick,l=$4$}{v4}
\fmfv{d.shape=circle,d.filled=full,d.size=3thick,l=$5$}{v5}
\fmffreeze
\fmf{half_prop}{v0,v1}
\fmf{half_prop,left=0.6}{v0,v2}
\fmf{half_prop,right=0.6}{v0,v3}
\fmf{half_prop,right=0.6}{v1,v4}
\fmf{half_prop,left=0.6}{v2,v4}
\fmf{half_prop,left=0.6}{v2,v5}
\fmf{half_prop}{v3,v4}
\fmf{half_prop,right=0.6}{v3,v5}
\end{fmfgraph*}
\end{fmffile}}
\end{center}
\caption{A fifth-order non-tree TSD.}
\label{f:tsd_cycle_finding}
\end{figure}
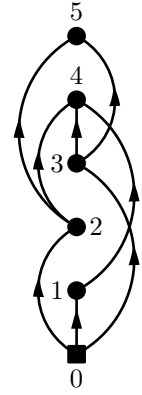

Let us illustrate the algorithm by applying it to the fifth-order non-tree TSD displayed in Fig.~\ref{f:tsd_cycle_finding}.
\begin{enumerate}
\item \texttt{node\_0} has \texttt{out\_degree} $\geq2$.
\begin{itemize}
\item \texttt{node\_1}, \texttt{node\_2} and \texttt{node\_3} have \texttt{in\_degree} $=1$.
\item \texttt{node\_4} has \texttt{in\_degree} $=3$ and \texttt{nb\_paths} $=3$, so \texttt{node\_0} and \texttt{node\_4} are end nodes of a cycle, since the paths do not share other nodes.
\item \texttt{node\_5} has \texttt{in\_degree} $=2$ and \texttt{nb\_paths} $=2$, so \texttt{node\_0} and \texttt{node\_5} are end nodes of a cycle, since the paths do not share other nodes.
\end{itemize}
\item \texttt{node\_1} has \texttt{out\_degree} $=1$.
\item \texttt{node\_2} has \texttt{out\_degree} $\geq2$.
\begin{itemize}
\item \texttt{node\_4} has only one path coming from \texttt{node\_2}.
\item \texttt{node\_5} has only one path coming from \texttt{node\_2}.
\end{itemize}
\item \texttt{node\_3} has \texttt{out\_degree} $\geq2$.
\begin{itemize}
\item \texttt{node\_4} has only one path coming from \texttt{node\_3}.
\item \texttt{node\_5} has only one path coming from \texttt{node\_3}.
\end{itemize}
\end{enumerate}
Eventually, the TSD comprises two cycles, one with end nodes \texttt{node\_0} and \texttt{node\_4}, and one with end nodes \texttt{node\_0} and \texttt{node\_5}.

\subsubsection{Cycles treatment}

Once the cycles of a TSD have been identified, they must be traded for a sum of tree TSDs via the systematic ordering of their internal vertices. Starting from the \emph{NetworkX} diagram, the two end nodes of the cycle, and the two paths connecting the end nodes, one applies the following algorithm
\begin{enumerate}
\item Set \texttt{node\_to\_insert} as the first node of \texttt{path\_1} after start node.

\item For each \texttt{daughter\_node} in \texttt{path\_2} but the starting node
\begin{enumerate}
\item Make a copy of the graph,
\item Add a link from \texttt{node\_to\_insert} to \texttt{daughter\_node},
\item Set \texttt{mother\_node} as the node preceeding \texttt{daughter\_node} in \texttt{path\_2},
\item Add a link from \texttt{mother\_node} to \texttt{daughter\_node},
\item Remove the links carrying unnecessary information.
\end{enumerate}
\end{enumerate}

\begin{figure}[t!]
\begin{center}
\parbox{100pt}{\begin{fmffile}{diag_time_8}
\begin{fmfgraph*}(100,100)
\fmfcmd{style_def half_prop expr p =
    draw_plain p;
    shrink(.7);
        cfill (marrow (p, .5))
    endshrink;
	enddef;}
\fmftop{v4}\fmfbottom{v0}
\fmf{phantom}{v0,v1}
\fmfv{d.shape=square,d.filled=full,d.size=3thick,l=$0$}{v0}
\fmf{phantom}{v1,v2}
\fmfv{d.shape=circle,d.filled=full,d.size=3thick,l=$1$,l.a=180}{v1}
\fmf{phantom}{v2,v3}
\fmfv{d.shape=circle,d.filled=full,d.size=3thick,l=$2$,l.a=180}{v2}
\fmf{phantom}{v3,v4}
\fmfv{d.shape=circle,d.filled=full,d.size=3thick,l=$3$,l.a=180}{v3}
\fmfv{d.shape=circle,d.filled=full,d.size=3thick,l=$4$}{v4}
\fmffreeze
\fmf{half_prop}{v0,v1}
\fmf{half_prop,right=0.6}{v0,v3}
\fmf{half_prop}{v1,v2}
\fmf{half_prop,right=0.6}{v2,v4}
\fmf{half_prop}{v3,v4}
\end{fmfgraph*}
\end{fmffile}}
\end{center}
\caption{A fourth order non-tree TSD.}
\label{f:ex_cycle_treatment}
\end{figure}
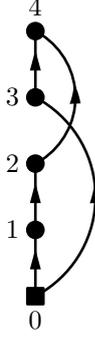

\begin{figure}[t!]
\begin{center}
\parbox{100pt}{\begin{fmffile}{time_equivalent8_1}
\begin{fmfgraph*}(100,100)
\fmfcmd{style_def half_prop expr p =
    draw_plain p;
    shrink(.7);
        cfill (marrow (p, .5))
    endshrink;
	enddef;}
\fmftop{v4}\fmfbottom{v0}
\fmf{phantom}{v0,v1}
\fmfv{d.shape=square,d.filled=full,d.size=3thick}{v0}
\fmf{phantom}{v1,v2}
\fmfv{d.shape=circle,d.filled=full,d.size=3thick,l=$a_1$}{v1}
\fmf{phantom}{v2,v3}
\fmfv{d.shape=circle,d.filled=full,d.size=3thick,l=$a_2$,l.a=90}{v2}
\fmf{phantom}{v3,v4}
\fmfv{d.shape=circle,d.filled=full,d.size=3thick,l=$a_3$}{v3}
\fmfv{d.shape=circle,d.filled=full,d.size=3thick,l=$a_4$}{v4}
\fmffreeze
\fmf{half_prop,right=0.6}{v0,v3}
\fmf{half_prop}{v1,v2}
\fmf{half_prop,right=0.6}{v2,v4}
\fmf{half_prop,right=0.6}{v3,v1}
\end{fmfgraph*}
\end{fmffile}}
\hspace{-30pt}
\parbox{100pt}{\begin{fmffile}{time_equivalent8_2}
\begin{fmfgraph*}(100,100)
\fmfcmd{style_def half_prop expr p =
    draw_plain p;
    shrink(.7);
        cfill (marrow (p, .5))
    endshrink;
	enddef;}
\fmftop{v4}\fmfbottom{v0}
\fmf{phantom}{v0,v1}
\fmfv{d.shape=square,d.filled=full,d.size=3thick}{v0}
\fmf{phantom}{v1,v2}
\fmfv{d.shape=circle,d.filled=full,d.size=3thick,l=$a_1$,l.a=180}{v1}
\fmf{phantom}{v2,v3}
\fmfv{d.shape=circle,d.filled=full,d.size=3thick,l=$a_2$,l.a=-90}{v2}
\fmf{phantom}{v3,v4}
\fmfv{d.shape=circle,d.filled=full,d.size=3thick,l=$a_3$}{v3}
\fmfv{d.shape=circle,d.filled=full,d.size=3thick,l=$a_4$}{v4}
\fmffreeze
\fmf{half_prop}{v0,v1}
\fmf{half_prop,right=0.6}{v1,v3}
\fmf{half_prop,right=0.6}{v2,v4}
\fmf{half_prop}{v3,v2}
\end{fmfgraph*}
\end{fmffile}}
\hspace{-30pt}
\parbox{100pt}{\begin{fmffile}{time_equivalent8_0}
\begin{fmfgraph*}(100,100)
\fmfcmd{style_def half_prop expr p =
    draw_plain p;
    shrink(.7);
        cfill (marrow (p, .5))
    endshrink;
	enddef;}
\fmftop{v4}\fmfbottom{v0}
\fmf{half_prop}{v0,v1}
\fmfv{d.shape=square,d.filled=full,d.size=3thick}{v0}
\fmf{half_prop}{v1,v2}
\fmfv{d.shape=circle,d.filled=full,d.size=3thick,l=$a_1$,l.a=0}{v1}
\fmf{half_prop}{v2,v3}
\fmfv{d.shape=circle,d.filled=full,d.size=3thick,l=$a_2$,l.a=0}{v2}
\fmf{half_prop}{v3,v4}
\fmfv{d.shape=circle,d.filled=full,d.size=3thick,l=$a_3$,l.a=0}{v3}
\fmfv{d.shape=circle,d.filled=full,d.size=3thick,l=$a_4$,l.a=0}{v4}
\end{fmfgraph*}
\end{fmffile}}

\vspace{\baselineskip}
$\Downarrow$
\vspace{\baselineskip}

\parbox{100pt}{\begin{fmffile}{time_equivalent8_1_straight}
\begin{fmfgraph*}(100,100)
\fmfcmd{style_def half_prop expr p =
    draw_plain p;
    shrink(.7);
        cfill (marrow (p, .5))
    endshrink;
	enddef;}
\fmftop{v4}\fmfbottom{v0}
\fmf{half_prop}{v0,v1}
\fmfv{d.shape=square,d.filled=full,d.size=3thick}{v0}
\fmf{half_prop}{v1,v2}
\fmfv{d.shape=circle,d.filled=full,d.size=3thick,l=$a_3$,l.a=0}{v1}
\fmf{half_prop}{v2,v3}
\fmfv{d.shape=circle,d.filled=full,d.size=3thick,l=$a_1$,l.a=0}{v2}
\fmf{half_prop}{v3,v4}
\fmfv{d.shape=circle,d.filled=full,d.size=3thick,l=$a_2$,l.a=0}{v3}
\fmfv{d.shape=circle,d.filled=full,d.size=3thick,l=$a_4$,l.a=0}{v4}
\end{fmfgraph*}
\end{fmffile}}
\hspace{-30pt}
\parbox{100pt}{\begin{fmffile}{time_equivalent8_2_straight}
\begin{fmfgraph*}(100,100)
\fmfcmd{style_def half_prop expr p =
    draw_plain p;
    shrink(.7);
        cfill (marrow (p, .5))
    endshrink;
	enddef;}
\fmftop{v4}\fmfbottom{v0}
\fmf{half_prop}{v0,v1}
\fmfv{d.shape=square,d.filled=full,d.size=3thick}{v0}
\fmf{half_prop}{v1,v2}
\fmfv{d.shape=circle,d.filled=full,d.size=3thick,l=$a_1$,l.a=0}{v1}
\fmf{half_prop}{v2,v3}
\fmfv{d.shape=circle,d.filled=full,d.size=3thick,l=$a_3$,l.a=0}{v2}
\fmf{half_prop}{v3,v4}
\fmfv{d.shape=circle,d.filled=full,d.size=3thick,l=$a_2$,l.a=0}{v3}
\fmfv{d.shape=circle,d.filled=full,d.size=3thick,l=$a_1$,l.a=0}{v4}
\end{fmfgraph*}
\end{fmffile}}
\hspace{-30pt}
\parbox{100pt}{\begin{fmffile}{time_equivalent8_0_straight}
\begin{fmfgraph*}(100,100)
\fmfcmd{style_def half_prop expr p =
    draw_plain p;
    shrink(.7);
        cfill (marrow (p, .5))
    endshrink;
	enddef;}
\fmftop{v4}\fmfbottom{v0}
\fmf{half_prop}{v0,v1}
\fmfv{d.shape=square,d.filled=full,d.size=3thick}{v0}
\fmf{half_prop}{v1,v2}
\fmfv{d.shape=circle,d.filled=full,d.size=3thick,l=$a_1$,l.a=0}{v1}
\fmf{half_prop}{v2,v3}
\fmfv{d.shape=circle,d.filled=full,d.size=3thick,l=$a_2$,l.a=0}{v2}
\fmf{half_prop}{v3,v4}
\fmfv{d.shape=circle,d.filled=full,d.size=3thick,l=$a_3$,l.a=0}{v3}
\fmfv{d.shape=circle,d.filled=full,d.size=3thick,l=$a_4$,l.a=0}{v4}
\end{fmfgraph*}
\end{fmffile}}
\end{center}
\caption{Tree TSDs generated by applying the cycle treatment algorithm to the non-tree TSD displayed in Fig.~\ref{f:ex_cycle_treatment} with the choice of setting \texttt{daughter\_node} to 1, 2 and 4, drawn with the original vertex oredering (top) or after reordering the vertices ascendingly (bottom).}
\label{f:res_cycle_treatment}
\end{figure}
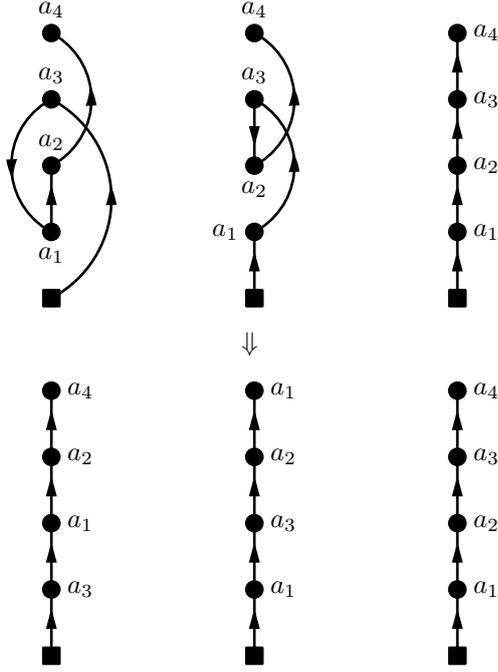

Let us illustrate the previous algorithm by applying it to the fourth-order TSD displayed in Fig.~\ref{f:ex_cycle_treatment}. One chooses here to set \texttt{path\_1} as $\{0,3,4\}$ and \texttt{path\_2} as $\{0,1,2,4\}$.
\begin{enumerate}
\item Set \texttt{node\_to\_insert} to \texttt{node\_3},
\item Set \texttt{daughter\_node} to \texttt{node\_1}
\begin{enumerate}
\item Copy the graph,
\item Add a link from \texttt{node\_3} to \texttt{node\_1},
\item \texttt{mother\_node} is set to \texttt{node\_0},
\item Add a link from \texttt{node\_0} to \texttt{node\_3},
\item Remove the links from \texttt{node\_0} to \texttt{node\_1} and from \texttt{node\_3} to \texttt{node\_4},
\end{enumerate}
\item Set \texttt{daughter\_node} to \texttt{node\_2}
\begin{enumerate}
\item Copy the graph,
\item Add a link from \texttt{node\_3} to \texttt{node\_2},
\item Set \texttt{mother\_node} to \texttt{node\_1},
\item Add a link from \texttt{node\_1} to \texttt{node\_3},
\item Remove the links from \texttt{node\_1} to \texttt{node\_2} and from \texttt{node\_3} to \texttt{node\_4},
\end{enumerate}
\item Set \texttt{daughter\_node} to \texttt{node\_4}
\begin{enumerate}
\item Copy the graph,
\item Add a link from \texttt{node\_3} to \texttt{node\_4},
\item Set \texttt{mother\_node} to \texttt{node\_2},
\item Add a link from \texttt{node\_2} to \texttt{node\_3},
\item Remove the links from \texttt{node\_0} to \texttt{node\_3} and one of the links from \texttt{node\_3} to \texttt{node\_4}.
\end{enumerate}
\end{enumerate}
The tree TSDs thus generated are displayed in Fig.~\ref{f:res_cycle_treatment}. In the present case, they all correspond to the linear tree TSD of order 4, denoted as T4.1 in Fig.~\ref{diagTSD4}, with different orderings of vertices $a_1$, $a_2$ and $a_3$.

Though applying once the algorithm exemplified above does not guarantee to obtain only tree TSDs, the three diagrams obtained presently are indeed tree TSDs. Whenever it is not the case, one must repeatedly apply the cycle (identification and treatment) algorithms to the TSDs generated at each step until only tree TSDs are obtained. In the above example, initially inverting \texttt{path\_1} and \texttt{path\_2} would have required more than one step.

All tree TSDs corresponding to an initial non-tree TSD being generated, the algorithm detailed in Sec.~\ref{subs:trees} can be applied to each of them. The expression of the non-tree TSD is nothing but the sum of the individual contributions thus obtained.

\subsubsection{Output of the \texttt{ADG} program}

A typical output of \texttt{ADG} for a non-tree TSD looks like

\paragraph{Time-structure diagram T4:}

\begin{center}
\parbox{80pt}{\begin{fmffile}{time_3}
\begin{fmfgraph*}(80,80)
\fmfcmd{style_def half_prop expr p =
    draw_plain p;
    shrink(.7);
        cfill (marrow (p, .5))
    endshrink;
	enddef;}
\fmftop{v3}\fmfbottom{v0}
\fmf{phantom}{v0,v1}
\fmfv{d.shape=square,d.filled=full,d.size=3thick}{v0}
\fmf{phantom}{v1,v2}
\fmfv{d.shape=circle,d.filled=full,d.size=3thick}{v1}
\fmf{phantom}{v2,v3}
\fmfv{d.shape=circle,d.filled=full,d.size=3thick}{v2}
\fmfv{d.shape=circle,d.filled=full,d.size=3thick}{v3}
\fmffreeze
\fmf{half_prop}{v0,v1}
\fmf{half_prop,right=0.6}{v0,v2}
\fmf{half_prop,right=0.6}{v1,v3}
\fmf{half_prop}{v2,v3}
\end{fmfgraph*}
\end{fmffile}}

\end{center}

\begin{equation*}
\text{T4} = \frac{1}{(a_1+ a_3)(a_2+ a_1+ a_3)a_3} + \frac{1}{(a_1+ a_2+ a_3)(a_2+ a_3)a_3}
\end{equation*}
Equivalent tree diagrams:  T1, T1.

\begin{center}
\parbox{80pt}{\begin{fmffile}{equivalent3_0}
\begin{fmfgraph*}(80,80)
\fmfcmd{style_def half_prop expr p =
    draw_plain p;
    shrink(.7);
        cfill (marrow (p, .5))
    endshrink;
	enddef;}
\fmftop{v3}\fmfbottom{v0}
\fmf{phantom}{v0,v1}
\fmfv{d.shape=square,d.filled=full,d.size=3thick}{v0}
\fmf{phantom}{v1,v2}
\fmfv{d.shape=circle,d.filled=full,d.size=3thick}{v1}
\fmf{phantom}{v2,v3}
\fmfv{d.shape=circle,d.filled=full,d.size=3thick}{v2}
\fmfv{d.shape=circle,d.filled=full,d.size=3thick}{v3}
\fmffreeze
\fmf{half_prop,right=0.6}{v0,v2}
\fmf{half_prop,right=0.6}{v1,v3}
\fmf{half_prop}{v2,v1}
\end{fmfgraph*}
\end{fmffile}}
\parbox{80pt}{\begin{fmffile}{equivalent3_1}
\begin{fmfgraph*}(80,80)
\fmfcmd{style_def half_prop expr p =
    draw_plain p;
    shrink(.7);
        cfill (marrow (p, .5))
    endshrink;
	enddef;}
\fmftop{v3}\fmfbottom{v0}
\fmf{phantom}{v0,v1}
\fmfv{d.shape=square,d.filled=full,d.size=3thick}{v0}
\fmf{phantom}{v1,v2}
\fmfv{d.shape=circle,d.filled=full,d.size=3thick}{v1}
\fmf{phantom}{v2,v3}
\fmfv{d.shape=circle,d.filled=full,d.size=3thick}{v2}
\fmfv{d.shape=circle,d.filled=full,d.size=3thick}{v3}
\fmffreeze
\fmf{half_prop}{v0,v1}
\fmf{half_prop}{v1,v2}
\fmf{half_prop}{v2,v3}
\end{fmfgraph*}
\end{fmffile}}

\end{center}

Related Feynman diagrams: 6, 9, 20, 27, 31, 39, 40, 43, 48, 50. \\

The TSD is displayed along with its expression and a graphical representation of the tree TSDs obtained by disentangling its cycles. The list of associated BMBPT Feynman diagrams is also provided.

\subsection{Final ouput of the \texttt{ADG} program}

Exploiting the findings detailed in the previous sections, we are now in position to provide the typical output of the \texttt{ADG} program. For a BMBPT diagram associated with a tree TSD, the output looks like

\paragraph{Diagram 1:}
\begin{align*}
\text{PO}3.1
&= \lim\limits_{\tau \to \infty}\frac{(-1)^2 }{(2!)^3}\sum_{k_i}O^{40}_{k_{1}k_{2}k_{3}k_{4}} \Omega^{22}_{k_{5}k_{6}k_{1}k_{2}} \Omega^{04}_{k_{5}k_{6}k_{3}k_{4}} \\
&\phantom{=} \times \int_{0}^{\tau}\mathrm{d}\tau_1\mathrm{d}\tau_2\theta(\tau_2-\tau_1) e^{-\tau_1 \epsilon_{k_{1}k_{2}}^{k_{5}k_{6}}}e^{-\tau_2 \epsilon_{k_{3}k_{4}k_{5}k_{6}}^{}}  \\
&= \frac{(-1)^2 }{(2!)^3}\sum_{k_i}\frac{O^{40}_{k_{1}k_{2}k_{3}k_{4}} \Omega^{22}_{k_{5}k_{6}k_{1}k_{2}} \Omega^{04}_{k_{5}k_{6}k_{3}k_{4}} }{\epsilon_{k_{1}k_{2}k_{3}k_{4}}^{}\ \epsilon_{k_{3}k_{4}k_{5}k_{6}}^{}\ } 
\end{align*}

\begin{center}
\parbox{60pt}{\begin{fmffile}{diag_0}
\begin{fmfgraph*}(60,60)
\fmfcmd{style_def prop_pm expr p =
    draw_plain p;
    shrink(.7);
        cfill (marrow (p, .25));
        cfill (marrow (p, .75))
    endshrink;
	enddef;}
\fmftop{v2}\fmfbottom{v0}
\fmf{phantom}{v0,v1}
\fmfv{d.shape=square,d.filled=full,d.size=3thick}{v0}
\fmf{phantom}{v1,v2}
\fmfv{d.shape=circle,d.filled=full,d.size=3thick}{v1}
\fmfv{d.shape=circle,d.filled=full,d.size=3thick}{v2}
\fmffreeze
\fmf{prop_pm,left=0.5}{v0,v1}
\fmf{prop_pm,right=0.5}{v0,v1}
\fmf{prop_pm,left=0.6}{v0,v2}
\fmf{prop_pm,right=0.6}{v0,v2}
\fmf{prop_pm,left=0.5}{v1,v2}
\fmf{prop_pm,right=0.5}{v1,v2}
\end{fmfgraph*}
\end{fmffile}}
\hspace{10pt} $\rightarrow$ \hspace{10pt} T1:\parbox{60pt}{\begin{fmffile}{time_0}
\begin{fmfgraph*}(60,60)
\fmfcmd{style_def half_prop expr p =
    draw_plain p;
    shrink(.7);
        cfill (marrow (p, .5))
    endshrink;
	enddef;}
\fmftop{v2}\fmfbottom{v0}
\fmf{phantom}{v0,v1}
\fmfv{d.shape=square,d.filled=full,d.size=3thick}{v0}
\fmf{phantom}{v1,v2}
\fmfv{d.shape=circle,d.filled=full,d.size=3thick}{v1}
\fmfv{d.shape=circle,d.filled=full,d.size=3thick}{v2}
\fmffreeze
\fmf{half_prop}{v0,v1}
\fmf{half_prop}{v1,v2}
\end{fmfgraph*}
\end{fmffile}}

\end{center}

\begin{equation*}
\text{T}1 = \frac{1}{(a_1+ a_2)a_2}
\end{equation*}
\begin{align*}
a_1 &= \epsilon_{k_{1}k_{2}}^{k_{5}k_{6}}\\
a_2 &= \epsilon_{k_{3}k_{4}k_{5}k_{6}}^{}
\end{align*}
whereas for a BMBPT diagram associated with a non-tree TSD, it looks like

\paragraph{Diagram 6:}
\begin{align*}
\text{PO}4.6
&= \lim\limits_{\tau \to \infty}\frac{-(-1)^3 }{(3!)^2}\sum_{k_i}O^{40}_{k_{1}k_{2}k_{3}k_{4}} \Omega^{31}_{k_{5}k_{6}k_{7}k_{1}} \Omega^{13}_{k_{8}k_{2}k_{3}k_{4}} \Omega^{04}_{k_{8}k_{5}k_{6}k_{7}} \\
&\phantom{=} \times \int_{0}^{\tau}\mathrm{d}\tau_1\mathrm{d}\tau_2\mathrm{d}\tau_3\theta(\tau_3-\tau_1) \theta(\tau_3-\tau_2) \\
&\phantom{= \times \int_{0}^{\tau}} \times e^{-\tau_1 \epsilon_{k_{1}}^{k_{5}k_{6}k_{7}}}e^{-\tau_2 \epsilon_{k_{2}k_{3}k_{4}}^{k_{8}}}e^{-\tau_3 \epsilon_{k_{5}k_{6}k_{7}k_{8}}^{}}  \\
&= \frac{-(-1)^3 }{(3!)^2}\sum_{k_i}O^{40}_{k_{1}k_{2}k_{3}k_{4}} \Omega^{31}_{k_{5}k_{6}k_{7}k_{1}} \Omega^{13}_{k_{8}k_{2}k_{3}k_{4}} \Omega^{04}_{k_{8}k_{5}k_{6}k_{7}} \\
&\phantom{=} \times \left[\frac{1}{\epsilon_{k_{1}k_{8}}^{}\ \epsilon_{k_{1}k_{2}k_{3}k_{4}}^{}\ \epsilon_{k_{5}k_{6}k_{7}k_{8}}^{}\ } \right. \\
&\phantom{= \times} \left. + \frac{1}{\epsilon_{k_{1}k_{2}k_{3}k_{4}}^{}\ \epsilon_{k_{2}k_{3}k_{4}k_{5}k_{6}k_{7}}^{}\ \epsilon_{k_{5}k_{6}k_{7}k_{8}}^{}\ } \right]
\end{align*}

\begin{center}
\parbox{80pt}{\begin{fmffile}{diag_5}
\begin{fmfgraph*}(80,80)
\fmfcmd{style_def prop_pm expr p =
    draw_plain p;
    shrink(.7);
        cfill (marrow (p, .25));
        cfill (marrow (p, .75))
    endshrink;
	enddef;}
\fmftop{v3}\fmfbottom{v0}
\fmf{phantom}{v0,v1}
\fmfv{d.shape=square,d.filled=full,d.size=3thick}{v0}
\fmf{phantom}{v1,v2}
\fmfv{d.shape=circle,d.filled=full,d.size=3thick}{v1}
\fmf{phantom}{v2,v3}
\fmfv{d.shape=circle,d.filled=full,d.size=3thick}{v2}
\fmfv{d.shape=circle,d.filled=full,d.size=3thick}{v3}
\fmffreeze
\fmf{prop_pm}{v0,v1}
\fmf{prop_pm,right=0.75}{v0,v2}
\fmf{prop_pm,left=0.6}{v0,v2}
\fmf{prop_pm,right=0.6}{v0,v2}
\fmf{prop_pm,right=0.75}{v1,v3}
\fmf{prop_pm,left=0.6}{v1,v3}
\fmf{prop_pm,right=0.6}{v1,v3}
\fmf{prop_pm}{v2,v3}
\end{fmfgraph*}
\end{fmffile}}
\hspace{10pt} $\rightarrow$ \hspace{10pt} T4:\parbox{80pt}{\begin{fmffile}{time_3}
\begin{fmfgraph*}(80,80)
\fmfcmd{style_def half_prop expr p =
    draw_plain p;
    shrink(.7);
        cfill (marrow (p, .5))
    endshrink;
	enddef;}
\fmftop{v3}\fmfbottom{v0}
\fmf{phantom}{v0,v1}
\fmfv{d.shape=square,d.filled=full,d.size=3thick}{v0}
\fmf{phantom}{v1,v2}
\fmfv{d.shape=circle,d.filled=full,d.size=3thick}{v1}
\fmf{phantom}{v2,v3}
\fmfv{d.shape=circle,d.filled=full,d.size=3thick}{v2}
\fmfv{d.shape=circle,d.filled=full,d.size=3thick}{v3}
\fmffreeze
\fmf{half_prop}{v0,v1}
\fmf{half_prop,right=0.6}{v0,v2}
\fmf{half_prop,right=0.6}{v1,v3}
\fmf{half_prop}{v2,v3}
\end{fmfgraph*}
\end{fmffile}}

\end{center}

\begin{equation*}
\text{T}4 = \frac{1}{(a_1+ a_3)(a_2+ a_1+ a_3)a_3} + \frac{1}{(a_1+ a_2+ a_3)(a_2 + a_3)a_3}
\end{equation*}
\begin{align*}
a_1 &= \epsilon_{k_{1}}^{k_{5}k_{6}k_{7}}\\
a_2 &= \epsilon_{k_{2}k_{3}k_{4}}^{k_{8}}\\
a_3 &= \epsilon_{k_{5}k_{6}k_{7}k_{8}}^{}
\end{align*}

The BMBPT diagram and its associated TSD are displayed. The original Feynman expression, its time-integrated expression and the expression of the TSD it derives from are added before listing the correspondence between the vertex labels in the TSD and the sum of quasiparticle energies in the BMBPT diagram.

\section{Connection to time-ordered diagrammatics}
\label{resolventrule}

The formal and numerical developments presented in this paper rely on the time-dependent formulation of (B)MBPT. It is, however, more customary to design MBPT on the basis of a time-independent formalism~\cite{Shavitt_Bartlett_2009}. While the end result is necessarily the same, the partitioning\footnote{A valid partitioning relates to splitting the complete order $p$ in a sum of terms that are individually proportional to a fraction of the form $1/(\epsilon_{k_i\ldots k_j}\ldots\epsilon_{k_u\ldots k_v})$ with $p$ energy factors in the denominator. Any other form does not constitute a valid partitioning in the present context.} of the complete order-$p$ contribution to the observable $\text{O}^{\text{A}}_0$ differs in both approaches. 

\subsection{Combinatorics}
\label{subs:combin}

The main characteristic of the time-dependent formalism is to authorize each diagram to capture as many different time orderings of the vertices as possible. While the contractions linking the vertices explicitly order a subset of the vertices, some vertices are left unordered in the integrand such that the diagram eventually seizes, i.e., sums, all remaining orderings at once. The combinatorics of these remaining orderings depends on the diagram and relates to the topology of the associated TSD.
\begin{enumerate}
\item Vertices belonging to the \emph{linear} tree TSD of order $p$ are fully ordered in time such that no further reordering is possible.
\item A \emph{non-linear} tree TSD contains several branches. The vertices on a given branch are fully ordered with respect to each other and with respect to the vertices located on the trunk the branch emerges from. However, the vertices on a branch are not ordered with respect to those belonging to another branch. Correspondingly, one can define the combinatorics "$n_{\text{branches}}$" as the total number of ways to order the vertices on the various branches. This corresponds to the number of fully time-ordered diagrams (i.e.\@ linear tree TSDs) the Feynman diagram captures.
\item A \emph{non-tree} TSD further contains cycles. The vertices on a branch inside a cycle are fully ordered  with respect to each other and with respect to the vertices located below (above) the starting (end) node of the cycle. However, the vertices on the various branches of the cycle are not ordered with respect to each other. The combinatorics "$n_{\text{cycle}}$" relates to ordering the vertices on the various branches of the cycle in all possible ways. The corresponding algorithm was discussed at length in Sec.~\ref{nontreeSec}. Performing this ordering for all cycles in a given non-tree TSD generates a set of tree TSDs.
\end{enumerate}

Contrarily, the  main characteristic of the time-independent formalism is to operate with fully time-ordered diagrams from the outset, i.e., to associate one diagram per possible time ordering of all the vertices. Correspondingly, there is no point invoking TSDs in this diagrammatic given that each time-ordered diagram of order $p$ trivially relates to the linear tree TSD of order $p$. 

\begin{table}
\begin{center}
\begin{tabular}{|l|l|c|c|c|c|c|}
\hline  Order & & 0 & 1 & 2 & 3 & 4  \\
  \hline\hline \texttt{deg\_max}~$=4$ & TSD & 1 & 1 & 2 & 4 & 14 \\
  \hline                        	  & BMBPT & 1 & 2 & 8 & 59 & 568 \\
  \hline\hline \texttt{deg\_max}~$=6$ & TSD & 1 & 1 & 2 & 5 & 15 \\
  \hline 							  & BMBPT & 1 & 3 & 23 & 396 & 10 716 \\
  \hline
\end{tabular}
\end{center}
\caption{Number of \emph{time-unordered} diagrams generated from operators containing at most four (\texttt{deg\_max}~$=4$) or six (\texttt{deg\_max}~$=6$) legs.}
\label{Tab1}
\end{table}

\begin{table}
\begin{center}
\begin{tabular}{|l|l|c|c|c|c|c|}
\hline  Order & & 0 & 1 & 2 & 3 & 4  \\
  \hline\hline \texttt{deg\_max}~$=4$ & TSD & 1 & 1 & 1 & 1 & 1 \\
  \hline                        	  & BMBPT & 1 & 2 & 9 & 87 & 1 377 \\
  \hline\hline \texttt{deg\_max}~$=6$ & TSD & 1 & 1 & 1 & 1 & 1 \\
  \hline 							  & BMBPT & 1 & 3 & 25 & 551 & 21 814 \\
  \hline
\end{tabular}
\end{center}
\caption{Number of \emph{time-ordered} diagrams generated from operators containing at most four (\texttt{deg\_max}~$=4$) or six (\texttt{deg\_max}~$=6$) legs.}
\label{Tab2}
\end{table}

Obviously, the main difference between both diagrammatics relates to the number of diagrams partitioning the complete order $p$. The number of time-unordered\footnote{As explained above, {\it time-unordered} diagrams do contain a certain degree of time ordering among a subset of vertices but this degree is \emph{minimal}. One could thus better refer to the \emph{minimally-ordered} diagrammatic.} (time-ordered) BMBPT diagrams and associated TSDs generated from operators containing at most four or six legs are provided in Tab.~\ref{Tab1} (Tab.~\ref{Tab2}) for perturbative orders $p=0,1,2,3$ and $4$. While the difference is not significant at low-order and/or for low \texttt{deg\_max}, it obviously increases with $p$ and \texttt{deg\_max}.

A key interest of the present work is to demonstrate that (i) a direct and systematic calculation of any Feynman BMBPT diagram associated with a tree TSD is possible and that (ii) the treatment of diagrams associated with non-tree TSD does require an explicit reordering of the vertices inside a given cycle. On the one hand, point (ii) underlines that the smaller number of time-unordered diagrams is partially illusory given that some explicit ordering (with combinatorial factor "$n_{\text{cycle}}$") of the vertices is actually mandatory to compute the diagrams. On the other, point (i) stresses that the large combinatorics of the fully-time-ordered diagrammatic is an overkill that can be avoided given that explicitly ordering the vertices on the various branches of tree diagrams (with combinatorial factor "$n_{\text{branches}}$") is superfluous.

\begin{table}
\begin{center}
\begin{tabular}{|l|l|c|c|c|c|c|}
\hline  Order & & 0 & 1 & 2 & 3 & 4  \\
  \hline\hline \texttt{deg\_max}~$=4$ & TSD & 1 & 1 & 2 & 3 & 7 \\
  \hline                        	  & BMBPT & 1 & 2 & 8 & 69 & 866 \\
  \hline\hline \texttt{deg\_max}~$=6$ & TSD & 1 & 1 & 2 & 4 & 8 \\
  \hline 							  & BMBPT & 1 & 3 & 23 & 449 & 15 250 \\
  \hline
\end{tabular}
\end{center}
\caption{Number of \emph{partially-time-ordered} diagrams generated from operators containing at most four (\texttt{deg\_max}~$=4$) or six (\texttt{deg\_max}~$=6$) legs.}
\label{Tab3}
\end{table}

\begin{table*}
\begin{center}
\begin{tabular}{|l|c|c|c|c|c|c|c|c|c|c|}
\hline Order & \multicolumn{2}{|c|}{0} & \multicolumn{2}{|c|}{1} & \multicolumn{2}{|c|}{2} & \multicolumn{2}{|c|}{3} & \multicolumn{2}{|c|}{4} \\ 
\hline
\hline & TSD & BMBPT & TSD & BMBPT & TSD & BMBPT & TSD & BMBPT & TSD & BMBPT \\ 
\hline
\hline  Linear tree             & 1 & 1 & 1 & 2 & 1 & 7 & 1 & 35 & 1 & 205 \\
  \hline Non-linear tree     & 0 & 0 & 0 & 0 & 1 & 1 & 2 & 14 & 6 & 147 \\
  \hline Non-tree                 & 0 & 0 & 0 & 0 & 0 & 0 & 1 & 10 & 7 & 216  \\
  \hline
\end{tabular}
\end{center}
\caption{Number of TSDs and BMBPT diagrams per topological category generated from operators containing at most four legs (\texttt{deg\_max}~$=4$).} 
\label{Tab4}
\end{table*}

\begin{table*}
\begin{center}
\begin{tabular}{|l|c|c|c|c|c|c|c|c|c|c|}
\hline Order & \multicolumn{2}{|c|}{0} & \multicolumn{2}{|c|}{1} & \multicolumn{2}{|c|}{2} & \multicolumn{2}{|c|}{3} & \multicolumn{2}{|c|}{4} \\ 
\hline
\hline & TSD & BMBPT & TSD & BMBPT & TSD & BMBPT & TSD & BMBPT & TSD & BMBPT \\ 
\hline
\hline  Linear tree             & 1 & 1 & 1 & 3 & 1 & 21 & 1 & 267 & 1 & 4970 \\
  \hline Non-linear tree     & 0 & 0 & 0 & 0 & 1 & 2  & 3 & 76 & 7 & 2311 \\
  \hline Non-tree                 & 0 & 0 & 0 & 0 & 0 & 0  & 1 & 53 & 7 & 3435  \\
  \hline
\end{tabular}
\end{center}
\caption{Number of TSDs and BMBPT diagrams per topological category generated from operators containing at most six legs (\texttt{deg\_max}~$=6$).} 
\label{Tab5}
\end{table*}

\begin{figure}
\begin{center}
\includegraphics[scale=1.0]{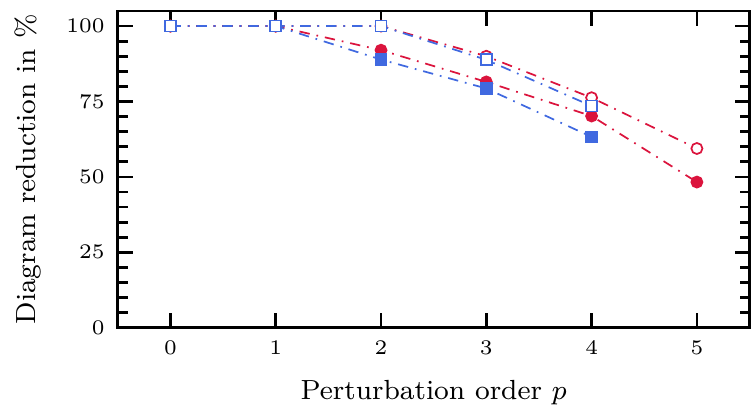}
\vspace{-\baselineskip}
\end{center}
\caption{Resummation efficiency expressed as the number of time-unordered BMBPT diagrams with respect to fully-ordered BMBPT diagrams. Red dots (blue squares) correspond to using vertices with four (six) legs at most. Empty symbols correspond to canonical diagrams only.}
\label{f:efficiency}
\end{figure}

As a result of the above, the optimal, i.e., minimal, number of BMBPT diagrams and associated TSDs one must eventually handle after ordering the vertices inside cycles is given in Tab.~\ref{Tab3}. This corresponds to what can be denoted as the \emph{partially-time-ordered} diagrammatic whose combinatorics is obviously in between those appearing in Tabs.~\ref{Tab1} and~\ref{Tab2}. The number of diagrams typically is comprised between 40\% and 90\% of those at play in the fully-time-ordered diagrammatic as illustrated in Fig.~\ref{f:efficiency}, and reduces with growing perturbative order. This minimal number of diagrams that must effectively be dealt with is of course dictated by how many of the original time-unordered diagrams relates to (i) a linear tree, (ii) a non-linear tree or (iii) a non-tree TSD. Indeed, how many of the original diagrams are in fact already fully time-ordered limits how much one can take advantage of not fully ordering the other ones.  For orientation, this partitioning of the diagrams is given in Tabs.~\ref{Tab4} and~\ref{Tab5} for perturbative orders $p=0,1,2,3$ and $4$. Beyond the lowest orders, the number of BMBPT Feynman diagrams that are not fully time-ordered to begin with grows radiply with both $p$ and \texttt{deg\_max}. 

\subsection{Resolvent rule}

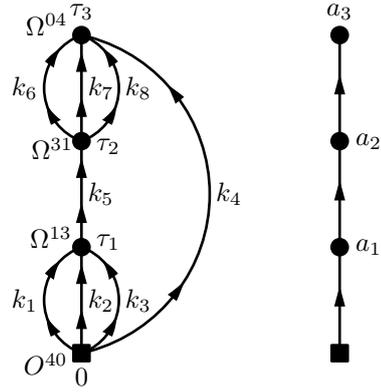
\begin{figure}[t!]
\begin{center}
\parbox{120pt}{\begin{fmffile}{diag_1_1}
\begin{fmfgraph*}(120,120)
\fmfcmd{style_def prop_pm expr p =
    draw_plain p;
    shrink(.7);
        cfill (marrow (p, .25));
        cfill (marrow (p, .75))
    endshrink;
	enddef;}
\fmftop{v3}\fmfbottom{v0}
\fmf{phantom, tag=1}{v0,v1}
\fmfv{d.shape=square,d.filled=full,d.size=3thick,l=$0$}{v0}
\fmf{phantom, tag=2}{v1,v2}
\fmfv{d.shape=circle,d.filled=full,d.size=3thick,l=$\tau_1$,l.a=20}{v1}
\fmf{phantom,tag=3}{v2,v3}
\fmfv{d.shape=circle,d.filled=full,d.size=3thick,l=$\tau_2$,l.a=-20}{v2}
\fmfv{d.shape=circle,d.filled=full,d.size=3thick,l=$\tau_3$}{v3}
\fmffreeze
\fmf{prop_pm,label=$k_2$,label.dist=.02w}{v0,v1}
\fmf{prop_pm,left=0.7,label=$k_1$,label.dist=.02w}{v0,v1}
\fmf{prop_pm,right=0.7,label=$k_3$,label.dist=.02w}{v0,v1}
\fmf{prop_pm,right=0.8,label=$k_4$,label.dist=.02w}{v0,v3}
\fmf{prop_pm,label=$k_5$,label.dist=.02w}{v1,v2}
\fmf{prop_pm,label=$k_7$,label.dist=.02w}{v2,v3}
\fmf{prop_pm,left=0.7,label=$k_6$,label.dist=.02w}{v2,v3}
\fmf{prop_pm,right=0.7,label=$k_8$,label.dist=.02w}{v2,v3}
\fmfposition
     \fmfipath{p[]}
       \fmfiset{p1}{vpath1(__v0,__v1)}
       \fmfiset{p2}{vpath2(__v1,__v2)}
       \fmfiset{p3}{vpath3(__v2,__v3)}
       \fmfiv{label=$O^{40}$,l.dist=.05w,l.a=-150}{point 0 of p1}
       \fmfiv{label=$\Omega^{13}$,l.dist=.03w,l.a=170}{point 0 of p2}
       \fmfiv{label=$\Omega^{31}$,l.dist=.03w,l.a=-150}{point 0 of p3}
       \fmfiv{label=$\Omega^{04}$,l.dist=.05w,l.a=150}{point length(p3) of p3}
\end{fmfgraph*}
\end{fmffile}}
\hspace{-30pt}
\parbox{120pt}{\begin{fmffile}{time_1_1}
\begin{fmfgraph*}(120,120)
\fmfcmd{style_def half_prop expr p =
    draw_plain p;
    shrink(.7);
        cfill (marrow (p, .5))
    endshrink;
	enddef;}
\fmftop{v3}\fmfbottom{v0}
\fmf{phantom}{v0,v1}
\fmfv{d.shape=square,d.filled=full,d.size=3thick}{v0}
\fmf{phantom}{v1,v2}
\fmfv{d.shape=circle,d.filled=full,d.size=3thick,l=$a_1$,l.a=0}{v1}
\fmf{phantom}{v2,v3}
\fmfv{d.shape=circle,d.filled=full,d.size=3thick,l=$a_2$,l.a=0}{v2}
\fmfv{d.shape=circle,d.filled=full,d.size=3thick,l=$a_3$}{v3}
\fmffreeze
\fmf{half_prop}{v0,v1}
\fmf{half_prop}{v1,v2}
\fmf{half_prop}{v2,v3}
\end{fmfgraph*}
\end{fmffile}}
\end{center}
\caption{A third-order BMBPT diagram and its associated linear tree TSD.}
\label{f:link_goldstone_rule}
\end{figure}

In the time-independent formulation of MBPT, the expression of each time-ordered diagram is derived via the application of the so-called resolvent rule~\cite{Shavitt_Bartlett_2009}. It is of interest to realize that the diagrammatic rule presently identified to compute any generic tree TSD in the time-unordered diagrammatic reduces to the resolvent rule for linear trees, i.e., for TSDs corresponding to BMBPT diagrams that are in fact fully time ordered.

Let us illustrate this feature on the basis of the third-order BMBPT diagram and its associated linear tree TSD displayed in Fig.~\ref{f:link_goldstone_rule}. The expression of the diagram reads, via the application of our diagrammatic rule based on the identification of the subdiagram associated to each vertex and its descendants, as
\begin{equation}
D = \frac{(-1)^3 }{(3!)^2}\sum_{k_i}\frac{ O^{40}_{k_{1}k_{2}k_{3}k_{4}} \Omega^{13}_{k_{5}k_{1}k_{2}k_{3}} \Omega^{31}_{k_{6}k_{7}k_{8}k_{5}} \Omega^{04}_{k_{6}k_{7}k_{8}k_{4}} }{ \epsilon_{k_1 k_2 k_3 k_4} \, \epsilon_{k_4 k_5} \, \epsilon_{k_4 k_6 k_7 k_8} } \ .
\label{eq:linear_tree_denom}
\end{equation}
While keeping all other elements unchanged, let us work out the denominator via the resolvent rule
\begin{enumerate}
\item drawing a line between the two lowest vertices, four lines carrying quasi-particle indices $k_1$, $k_2$, $k_3$ and $k_4$ are crossed. As a result, the resolvent rule contributes a factor $\epsilon_{k_1 k_2 k_3 k_4}$ to the denominator.
\item Repeating the procedure for the resolvent located between second and third (third and fourth) vertices, a factor $\epsilon_{k_4 k_5}$ ($\epsilon_{k_4 k_6 k_7 k_8}$) is found to contribute to the denominator.
\end{enumerate}
The overall denominator is thus the same as in Eq.~\eqref{eq:linear_tree_denom}. This result is easily understandable given that the lines between any two successive vertices of a linear tree are nothing but those entering the subdiagram formed by the second vertex and all its descendants.

\subsection{Diagrammatic resummation}
\label{subs:resummation}

As mentioned in Sec.~\ref{subs:combin}, the minimally-ordered BMBPT diagrammatics allows for a certain resummation of \emph{linear tree} TSDs into a more \emph{general tree} TSD associated to a combinatorial factor $n_{\text{branches}}$. Indeed the diagrammatic rule identified in Sec.~\ref{subs:trees} permits to sum at once, i.e., from a single BMBPT Feynman diagram associated to a tree TSD, the whole class of fully-time-ordered diagrams that derive from it, leading to a significant compactification of the computation. The number of fully-time-ordered diagrams generated from a tree TSD of order $p$ denoted as Tp.k is
\begin{equation}
n_{\text{branches}}(\text{Tp.k}) = \frac{p!}{\prod_{i=1}^{p} p_i} \, ,
\end{equation}
where $p_i$ denotes the effective order of the subdiagram associated with vertex $i$, i.e. the number of vertices in the subgraph made out of vertex $i$ and all its descendants. The product in the denominator accounts for all the combinations that are {\it not} summed into the tree TSD due to the vertices being partially ordered to begin with. The degree of resummation is maximal for a tree TSD in which all perturbative vertices are unordered with respect to each other, i.e. for a TSD containing $p$ independent branches associated with $p$ vertices directly connected to the bottom vertex. Indeed, $p_i=1$ for $i=1,\ldots,p$ in this case such that $n_{\text{branches}}(\text{Tp.k}) =p!$. Contrarily, all vertices belonging to the same branch in a linear tree, the successive $p_i$ coefficients are equal to $1, 2, \dots, p$ as one runs through the branch from top to bottom such that $n_{\text{branches}}(\text{Tp.k}) =1$ as expected. 

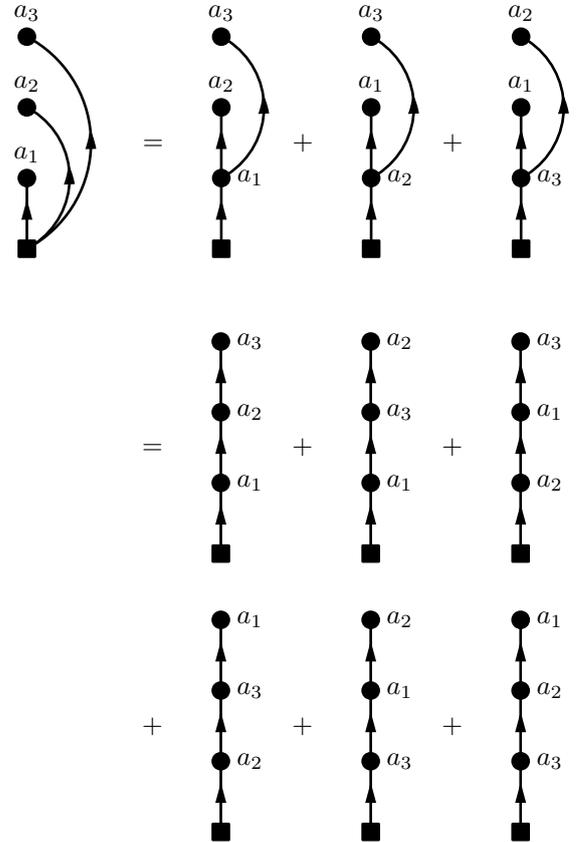
\begin{figure}[t!]
\begin{center}
\hspace{-30pt}
\parbox{80pt}{\begin{fmffile}{time_3_3bis}
\begin{fmfgraph*}(80,80)
\fmfcmd{style_def half_prop expr p =
draw_plain p;
shrink(.7);
	cfill (marrow (p, .5))
endshrink;
enddef;}
\fmftop{v3}\fmfbottom{v0}
\fmf{phantom}{v0,v1}
\fmfv{d.shape=square,d.filled=full,d.size=3thick}{v0}
\fmf{phantom}{v1,v2}
\fmfv{d.shape=circle,d.filled=full,d.size=3thick,l=$a_1$,l.a=90}{v1}
\fmf{phantom}{v2,v3}
\fmfv{d.shape=circle,d.filled=full,d.size=3thick,l=$a_2$}{v2}
\fmfv{d.shape=circle,d.filled=full,d.size=3thick,l=$a_3$}{v3}
\fmffreeze
\fmf{half_prop}{v0,v1}
\fmf{half_prop,right=0.6}{v0,v2}
\fmf{half_prop,right=0.6}{v0,v3}
\end{fmfgraph*}
\end{fmffile}}
$=$ \hspace{-25pt}
\parbox{80pt}{\begin{fmffile}{time_3_3_rs1}
\begin{fmfgraph*}(80,80)
\fmfcmd{style_def half_prop expr p =
draw_plain p;
shrink(.7);
	cfill (marrow (p, .5))
endshrink;
enddef;}
\fmftop{v3}\fmfbottom{v0}
\fmf{half_prop}{v0,v1}
\fmfv{d.shape=square,d.filled=full,d.size=3thick}{v0}
\fmf{half_prop}{v1,v2}
\fmfv{d.shape=circle,d.filled=full,d.size=3thick,l=$a_1$,l.a=0}{v1}
\fmf{phantom}{v2,v3}
\fmfv{d.shape=circle,d.filled=full,d.size=3thick,l=$a_2$,l.a=90}{v2}
\fmfv{d.shape=circle,d.filled=full,d.size=3thick,l=$a_3$,l.a=90}{v3}
\fmffreeze
\fmf{half_prop,right=0.6}{v1,v3}
\end{fmfgraph*}
\end{fmffile}}
\hspace{-20pt} $+$ \hspace{-25pt}
\parbox{80pt}{\begin{fmffile}{time_3_3_rs2}
\begin{fmfgraph*}(80,80)
\fmfcmd{style_def half_prop expr p =
draw_plain p;
shrink(.7);
	cfill (marrow (p, .5))
endshrink;
enddef;}
\fmftop{v3}\fmfbottom{v0}
\fmf{half_prop}{v0,v1}
\fmfv{d.shape=square,d.filled=full,d.size=3thick}{v0}
\fmf{half_prop}{v1,v2}
\fmfv{d.shape=circle,d.filled=full,d.size=3thick,l=$a_2$,l.a=0}{v1}
\fmf{phantom}{v2,v3}
\fmfv{d.shape=circle,d.filled=full,d.size=3thick,l=$a_1$,l.a=90}{v2}
\fmfv{d.shape=circle,d.filled=full,d.size=3thick,l=$a_3$,l.a=90}{v3}
\fmffreeze
\fmf{half_prop,right=0.6}{v1,v3}
\end{fmfgraph*}
\end{fmffile}}
\hspace{-20pt} $+$ \hspace{-25pt}
\parbox{80pt}{\begin{fmffile}{time_3_3_rs3}
\begin{fmfgraph*}(80,80)
\fmfcmd{style_def half_prop expr p =
draw_plain p;
shrink(.7);
	cfill (marrow (p, .5))
endshrink;
enddef;}
\fmftop{v3}\fmfbottom{v0}
\fmf{half_prop}{v0,v1}
\fmfv{d.shape=square,d.filled=full,d.size=3thick}{v0}
\fmf{half_prop}{v1,v2}
\fmfv{d.shape=circle,d.filled=full,d.size=3thick,l=$a_3$,l.a=0}{v1}
\fmf{phantom}{v2,v3}
\fmfv{d.shape=circle,d.filled=full,d.size=3thick,l=$a_1$,l.a=90}{v2}
\fmfv{d.shape=circle,d.filled=full,d.size=3thick,l=$a_2$,l.a=90}{v3}
\fmffreeze
\fmf{half_prop,right=0.6}{v1,v3}
\end{fmfgraph*}
\end{fmffile}}
\end{center}

\vspace{2\baselineskip}

\hspace{60pt} $=$ \hspace{-25pt}
\parbox{80pt}{\begin{fmffile}{time_3_3_s1}
\begin{fmfgraph*}(80,80)
\fmfcmd{style_def half_prop expr p =
draw_plain p;
shrink(.7);
	cfill (marrow (p, .5))
endshrink;
enddef;}
\fmftop{v3}\fmfbottom{v0}
\fmf{half_prop}{v0,v1}
\fmfv{d.shape=square,d.filled=full,d.size=3thick}{v0}
\fmf{half_prop}{v1,v2}
\fmfv{d.shape=circle,d.filled=full,d.size=3thick,l=$a_1$,l.a=0}{v1}
\fmf{half_prop}{v2,v3}
\fmfv{d.shape=circle,d.filled=full,d.size=3thick,l=$a_2$,l.a=0}{v2}
\fmfv{d.shape=circle,d.filled=full,d.size=3thick,l=$a_3$,l.a=0}{v3}
\end{fmfgraph*}
\end{fmffile}}
\hspace{-20pt} $+$ \hspace{-25pt}
\parbox{80pt}{\begin{fmffile}{time_3_3_s2}
\begin{fmfgraph*}(80,80)
\fmfcmd{style_def half_prop expr p =
draw_plain p;
shrink(.7);
	cfill (marrow (p, .5))
endshrink;
enddef;}
\fmftop{v3}\fmfbottom{v0}
\fmf{half_prop}{v0,v1}
\fmfv{d.shape=square,d.filled=full,d.size=3thick}{v0}
\fmf{half_prop}{v1,v2}
\fmfv{d.shape=circle,d.filled=full,d.size=3thick,l=$a_1$,l.a=0}{v1}
\fmf{half_prop}{v2,v3}
\fmfv{d.shape=circle,d.filled=full,d.size=3thick,l=$a_3$,l.a=0}{v2}
\fmfv{d.shape=circle,d.filled=full,d.size=3thick,l=$a_2$,l.a=0}{v3}
\end{fmfgraph*}
\end{fmffile}}
\hspace{-20pt} $+$ \hspace{-25pt}
\parbox{80pt}{\begin{fmffile}{time_3_3_s3}
\begin{fmfgraph*}(80,80)
\fmfcmd{style_def half_prop expr p =
draw_plain p;
shrink(.7);
	cfill (marrow (p, .5))
endshrink;
enddef;}
\fmftop{v3}\fmfbottom{v0}
\fmf{half_prop}{v0,v1}
\fmfv{d.shape=square,d.filled=full,d.size=3thick}{v0}
\fmf{half_prop}{v1,v2}
\fmfv{d.shape=circle,d.filled=full,d.size=3thick,l=$a_2$,l.a=0}{v1}
\fmf{half_prop}{v2,v3}
\fmfv{d.shape=circle,d.filled=full,d.size=3thick,l=$a_1$,l.a=0}{v2}
\fmfv{d.shape=circle,d.filled=full,d.size=3thick,l=$a_3$,l.a=0}{v3}
\end{fmfgraph*}
\end{fmffile}}

\vspace{2\baselineskip}

\hspace{60pt} $+$ \hspace{-25pt}
\parbox{80pt}{\begin{fmffile}{time_3_3_s4}
\begin{fmfgraph*}(80,80)
\fmfcmd{style_def half_prop expr p =
draw_plain p;
shrink(.7);
	cfill (marrow (p, .5))
endshrink;
enddef;}
\fmftop{v3}\fmfbottom{v0}
\fmf{half_prop}{v0,v1}
\fmfv{d.shape=square,d.filled=full,d.size=3thick}{v0}
\fmf{half_prop}{v1,v2}
\fmfv{d.shape=circle,d.filled=full,d.size=3thick,l=$a_2$,l.a=0}{v1}
\fmf{half_prop}{v2,v3}
\fmfv{d.shape=circle,d.filled=full,d.size=3thick,l=$a_3$,l.a=0}{v2}
\fmfv{d.shape=circle,d.filled=full,d.size=3thick,l=$a_1$,l.a=0}{v3}
\end{fmfgraph*}
\end{fmffile}}
\hspace{-20pt} $+$ \hspace{-25pt}
\parbox{80pt}{\begin{fmffile}{time_3_3_s5}
\begin{fmfgraph*}(80,80)
\fmfcmd{style_def half_prop expr p =
draw_plain p;
shrink(.7);
	cfill (marrow (p, .5))
endshrink;
enddef;}
\fmftop{v3}\fmfbottom{v0}
\fmf{half_prop}{v0,v1}
\fmfv{d.shape=square,d.filled=full,d.size=3thick}{v0}
\fmf{half_prop}{v1,v2}
\fmfv{d.shape=circle,d.filled=full,d.size=3thick,l=$a_3$,l.a=0}{v1}
\fmf{half_prop}{v2,v3}
\fmfv{d.shape=circle,d.filled=full,d.size=3thick,l=$a_1$,l.a=0}{v2}
\fmfv{d.shape=circle,d.filled=full,d.size=3thick,l=$a_2$,l.a=0}{v3}
\end{fmfgraph*}
\end{fmffile}}
\hspace{-20pt} $+$ \hspace{-25pt}
\parbox{80pt}{\begin{fmffile}{time_3_3_s6}
\begin{fmfgraph*}(80,80)
\fmfcmd{style_def half_prop expr p =
draw_plain p;
shrink(.7);
	cfill (marrow (p, .5))
endshrink;
enddef;}
\fmftop{v3}\fmfbottom{v0}
\fmf{half_prop}{v0,v1}
\fmfv{d.shape=square,d.filled=full,d.size=3thick}{v0}
\fmf{half_prop}{v1,v2}
\fmfv{d.shape=circle,d.filled=full,d.size=3thick,l=$a_3$,l.a=0}{v1}
\fmf{half_prop}{v2,v3}
\fmfv{d.shape=circle,d.filled=full,d.size=3thick,l=$a_2$,l.a=0}{v2}
\fmfv{d.shape=circle,d.filled=full,d.size=3thick,l=$a_1$,l.a=0}{v3}
\end{fmfgraph*}
\end{fmffile}}

\caption{Decomposition of T3.4 into a sum of linear tree TSDs.}
\label{f:tsd34_resum}
\end{figure}

\begin{figure}[t]
\begin{center}
\hspace{-30pt}
\parbox{80pt}{\begin{fmffile}{time_3_2}
\begin{fmfgraph*}(80,80)
\fmfcmd{style_def half_prop expr p =
draw_plain p;
shrink(.7);
    cfill (marrow (p, .5))
endshrink;
enddef;}
\fmftop{v3}\fmfbottom{v0}
\fmf{phantom}{v0,v1}
\fmfv{d.shape=square,d.filled=full,d.size=3thick}{v0}
\fmf{phantom}{v1,v2}
\fmfv{d.shape=circle,d.filled=full,d.size=3thick,l=$a_1$,l.a=90}{v1}
\fmf{phantom}{v2,v3}
\fmfv{d.shape=circle,d.filled=full,d.size=3thick,l=$a_2$,l.a=60}{v2}
\fmfv{d.shape=circle,d.filled=full,d.size=3thick,l=$a_3$}{v3}
\fmffreeze
\fmf{half_prop}{v0,v1}
\fmf{half_prop,right=0.6}{v0,v2}
\fmf{half_prop}{v2,v3}
\end{fmfgraph*}
\end{fmffile}}
\hspace{-15pt} $=$ \hspace{-25pt}
\parbox{80pt}{\begin{fmffile}{time_3_2_rs1}
\begin{fmfgraph*}(80,80)
\fmfcmd{style_def half_prop expr p =
draw_plain p;
shrink(.7);
    cfill (marrow (p, .5))
endshrink;
enddef;}
\fmftop{v3}\fmfbottom{v0}
\fmf{half_prop}{v0,v1}
\fmfv{d.shape=square,d.filled=full,d.size=3thick}{v0}
\fmf{half_prop}{v1,v2}
\fmfv{d.shape=circle,d.filled=full,d.size=3thick,l=$a_1$,l.a=0}{v1}
\fmf{half_prop}{v2,v3}
\fmfv{d.shape=circle,d.filled=full,d.size=3thick,l=$a_2$,l.a=0}{v2}
\fmfv{d.shape=circle,d.filled=full,d.size=3thick,l=$a_3$}{v3}
\end{fmfgraph*}
\end{fmffile}}
\hspace{-20pt} $+$ \hspace{-25pt}
\parbox{80pt}{\begin{fmffile}{time_3_2_rs2}
\begin{fmfgraph*}(80,80)
\fmfcmd{style_def half_prop expr p =
draw_plain p;
shrink(.7);
    cfill (marrow (p, .5))
endshrink;
enddef;}
\fmftop{v3}\fmfbottom{v0}
\fmf{half_prop}{v0,v1}
\fmfv{d.shape=square,d.filled=full,d.size=3thick}{v0}
\fmf{half_prop}{v1,v2}
\fmfv{d.shape=circle,d.filled=full,d.size=3thick,l=$a_2$,l.a=0}{v1}
\fmf{phantom}{v2,v3}
\fmfv{d.shape=circle,d.filled=full,d.size=3thick,l=$a_1$,l.a=90}{v2}
\fmfv{d.shape=circle,d.filled=full,d.size=3thick,l=$a_3$}{v3}
\fmffreeze
\fmf{half_prop,right=0.6}{v1,v3}
\end{fmfgraph*}
\end{fmffile}}
\hspace{50pt}

\vspace{2\baselineskip}

\hspace{50pt} $=$ \hspace{-25pt}
\parbox{80pt}{\begin{fmffile}{time_3_2_s1}
\begin{fmfgraph*}(80,80)
\fmfcmd{style_def half_prop expr p =
draw_plain p;
shrink(.7);
    cfill (marrow (p, .5))
endshrink;
enddef;}
\fmftop{v3}\fmfbottom{v0}
\fmf{half_prop}{v0,v1}
\fmfv{d.shape=square,d.filled=full,d.size=3thick}{v0}
\fmf{half_prop}{v1,v2}
\fmfv{d.shape=circle,d.filled=full,d.size=3thick,l=$a_1$,l.a=0}{v1}
\fmf{half_prop}{v2,v3}
\fmfv{d.shape=circle,d.filled=full,d.size=3thick,l=$a_2$,l.a=0}{v2}
\fmfv{d.shape=circle,d.filled=full,d.size=3thick,l=$a_3$}{v3}
\end{fmfgraph*}
\end{fmffile}}
\hspace{-20pt} $+$ \hspace{-25pt}
\parbox{80pt}{\begin{fmffile}{time_3_2_s2}
\begin{fmfgraph*}(80,80)
\fmfcmd{style_def half_prop expr p =
draw_plain p;
shrink(.7);
    cfill (marrow (p, .5))
endshrink;
enddef;}
\fmftop{v3}\fmfbottom{v0}
\fmf{half_prop}{v0,v1}
\fmfv{d.shape=square,d.filled=full,d.size=3thick}{v0}
\fmf{half_prop}{v1,v2}
\fmfv{d.shape=circle,d.filled=full,d.size=3thick,l=$a_2$,l.a=0}{v1}
\fmf{half_prop}{v2,v3}
\fmfv{d.shape=circle,d.filled=full,d.size=3thick,l=$a_1$,l.a=0}{v2}
\fmfv{d.shape=circle,d.filled=full,d.size=3thick,l=$a_3$}{v3}
\end{fmfgraph*}
\end{fmffile}}
\hspace{-20pt} $+$ \hspace{-25pt}
\parbox{80pt}{\begin{fmffile}{time_3_2_s3}
\begin{fmfgraph*}(80,80)
\fmfcmd{style_def half_prop expr p =
draw_plain p;
shrink(.7);
    cfill (marrow (p, .5))
endshrink;
enddef;}
\fmftop{v3}\fmfbottom{v0}
\fmf{half_prop}{v0,v1}
\fmfv{d.shape=square,d.filled=full,d.size=3thick}{v0}
\fmf{half_prop}{v1,v2}
\fmfv{d.shape=circle,d.filled=full,d.size=3thick,l=$a_2$,l.a=0}{v1}
\fmf{half_prop}{v2,v3}
\fmfv{d.shape=circle,d.filled=full,d.size=3thick,l=$a_3$,l.a=0}{v2}
\fmfv{d.shape=circle,d.filled=full,d.size=3thick,l=$a_1$}{v3}
\end{fmfgraph*}
\end{fmffile}}

\end{center}

\caption{Decomposition of T3.3 into a sum of linear tree TSDs.}
\label{f:tsd33_resum}
\end{figure}
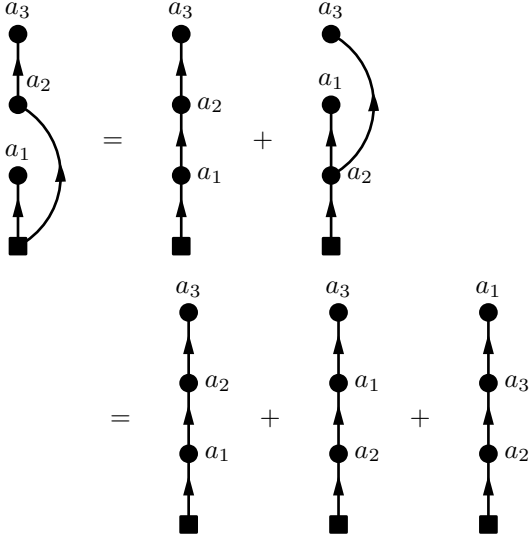

Let us illustrate the above for the two third-order tree TSDs denoted as T3.3 and T3.4 in Fig.~\ref{diagTSD0123}. Their decomposition into fully-time-ordered linear trees is displayed in Figs.~\ref{f:tsd34_resum} and~\ref{f:tsd33_resum}, respectively. The number of fully-time-ordered diagrams resummed into T3.4 is $n_{\text{branches}}(\text{T3.4}) =3!$. It corresponds to a maximal degree of resummation as T3.4 is made out of three independent vertices directly connected to the bottom vertex. Proceeding similarly with T3.3, the number of resummed fully-time-ordered diagrams is not maximal and equal to $n_{\text{branches}}(\text{T3.3}) =3!/2=3$ in this case. This relates to the fact that two vertices are ordered with respect to each other to begin with.

The maximal degree of resummation of tree TSDs of order $p$ generated from vertices containing at most four (\texttt{deg\_max}~$=4$) or six (\texttt{deg\_max}~$=6$) legs is given in Tab.~\ref{t:resum_power}. It is compared to the hypothetical maximal value of $p!$. The reason why the maximal degree becomes systematically smaller than the ideal value $p!$ as $p$ increases is that the number of independent branches authorized at a given order $p$ is drastically constrained by the value of \texttt{deg\_max}.

\begin{table}
\begin{center}
\begin{tabular}{|l|c|c|c|c|c|c|c|c|}
\hline  Order & 0 & 1 & 2 & 3 & 4  & 5 & 6 & 7 \\
  \hline
  \hline \texttt{deg\_max}~$=4$ & 1 & 1 & 2 & 3 & 8 & 30 & 90 & 420 \\
  \hline \texttt{deg\_max}~$=6$ & 1 & 1 & 2 & 6 & 12 & 40 & 180 & 1 008\\
  \hline
  \hline $p!$ & 1 & 1 & 2 & 6 & 24 & 120 & 720 & 5 040 \\
  \hline
\end{tabular}
\end{center}
\caption{Maximal degree of resummation of tree TSDs associated with BMBPT diagrams generated from operators containing at most four (\texttt{deg\_max}~$=4$) or six (\texttt{deg\_max}~$=6$) legs (estimations for orders 6 and 7 and for \texttt{deg\_max}~$=6$ at order 5). Factorial values are also provided for comparison.}
\label{t:resum_power}
\end{table}

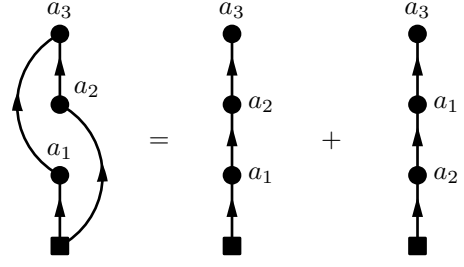
\begin{figure}[t]
\begin{center}
\vspace{\baselineskip}
\begin{fmffile}{cycle_as_sam2}
\parbox{80pt}{
\begin{fmfgraph*}(80,80)
\fmfcmd{style_def half_prop expr p =
    draw_plain p;
    shrink(.7);
        cfill (marrow (p, .5))
    endshrink;
	enddef;}
\fmftop{v3}\fmfbottom{v0}
\fmfv{d.shape=square,d.filled=full,d.size=3thick}{v0}
\fmfv{d.shape=circle,d.filled=full,d.size=3thick,l=$a_1$,l.angle=90}{v1}
\fmfv{d.shape=circle,d.filled=full,d.size=3thick,l=$a_2$,l.angle=30}{v2}
\fmfv{d.shape=circle,d.filled=full,d.size=3thick,l=$a_3$}{v3}
\fmf{half_prop}{v0,v1}
\fmf{half_prop}{v2,v3}
\fmf{phantom}{v1,v2}
\fmffreeze
\fmf{half_prop,right=0.6}{v0,v2}
\fmf{half_prop,left=0.6}{v1,v3}
\end{fmfgraph*}}
\hspace{-10pt}$=$\hspace{-20pt}
\parbox{80pt}{
\begin{fmfgraph*}(80,80)
\fmfcmd{style_def half_prop expr p =
    draw_plain p;
    shrink(.7);
        cfill (marrow (p, .5))
    endshrink;
	enddef;}
\fmftop{v3}\fmfbottom{v0}
\fmfv{d.shape=square,d.filled=full,d.size=3thick}{v0}
\fmfv{d.shape=circle,d.filled=full,d.size=3thick,l=$a_1$,l.angle=0}{v1}
\fmfv{d.shape=circle,d.filled=full,d.size=3thick,l=$a_2$,l.angle=0}{v2}
\fmfv{d.shape=circle,d.filled=full,d.size=3thick,l=$a_3$}{v3}
\fmf{half_prop}{v0,v1}
\fmf{half_prop}{v2,v3}
\fmf{half_prop}{v1,v2}
\end{fmfgraph*}}
\hspace{-10pt}$+$\hspace{-15pt}
\parbox{80pt}{
\begin{fmfgraph*}(80,80)
\fmfcmd{style_def half_prop expr p =
    draw_plain p;
    shrink(.7);
        cfill (marrow (p, .5))
    endshrink;
	enddef;}
\fmftop{v3}\fmfbottom{v0}
\fmfv{d.shape=square,d.filled=full,d.size=3thick}{v0}
\fmfv{d.shape=circle,d.filled=full,d.size=3thick,l=$a_2$,l.angle=0}{v1}
\fmfv{d.shape=circle,d.filled=full,d.size=3thick,l=$a_1$,l.angle=0}{v2}
\fmfv{d.shape=circle,d.filled=full,d.size=3thick,l=$a_3$}{v3}
\fmf{half_prop}{v0,v1}
\fmf{half_prop}{v2,v3}
\fmf{half_prop}{v1,v2}
\end{fmfgraph*}}
\end{fmffile}
\vspace{\baselineskip}
\end{center}
\caption{Decomposition of T3.5 into a sum of linear tree TSDs.}
\label{f:resum_fail}
\end{figure}

The capacity of tree TSDs to resum large classes of fully-time-ordered diagrams translates algebraically into the remarkable fact that the sum of associated fractions factorizes into a single fraction whose factors in the denominator are obtained by invoking a specific set of subdiagrams. Starting with T3.4, and following\footnote{The second equality provides an extra intermediate step to better visualize how the decomposition (factorization) operates between the first and last step.} the diagrammatic process displayed in Fig.~\ref{f:tsd34_resum}, its expression is decomposed by steps into a sum of fractions corresponding to partially- and eventually fully-time-ordered, i.e. linear, trees
\begin{align*}
\text{T}3.4 &= \frac{1}{a_1 a_2 a_3} \\
&= \frac{1}{a_1 + a_2 +a_3} \left[ \frac{1}{a_2 a_3} + \frac{1}{a_1 a_3} +  \frac{1}{a_1 a_2} \right] \\
&= \frac{1}{a_1 + a_2 +a_3} \left[ \frac{1}{a_2 + a_3} \left( \frac{1}{a_3} + \frac{1}{a_2} \right) \right. \\
&\phantom{= \frac{1}{a_1 + a_2 +a_3} \left[ \right.} + \frac{1}{a_1 + a_3} \left( \frac{1}{a_3} + \frac{1}{a_1} \right)  \\
&\phantom{= \frac{1}{a_1 + a_2 +a_3} \left[ \right.} \left. +  \frac{1}{a_1 + a_2} \left( \frac{1}{a_2} +  \frac{1}{a_1} \right) \right] \\
&= \frac{1}{(a_1 + a_2 +a_3) (a_2 + a_3) a_3}  \\
&\phantom{=}  +  \frac{1}{(a_1 + a_2 +a_3) (a_2 + a_3) a_2}  \\
&\phantom{=}  +  \frac{1}{(a_1 + a_2 +a_3) (a_1 + a_3) a_3}   \\
&\phantom{=}  + \frac{1}{(a_1 + a_2 +a_3) (a_1 + a_3) a_1} \\
&\phantom{=}  +  \frac{1}{(a_1 + a_2 +a_3) (a_1 + a_2) a_2}  \\
&\phantom{=}  +  \frac{1}{(a_1 + a_2 +a_3) (a_1 + a_2) a_1}  \ .
\end{align*}
Proceeding similarly with T3.3, the decomposition of the fraction operates as
\begin{align*}
\text{T}3.3 &= \frac{1}{a_1 (a_2 + a_3) a_3} \\
&= \frac{1}{(a_1 + a_2 +a_3) a_3} \left[ \frac{1}{a_2 + a_3} + \frac{1}{a_1} \right] \\
&= \frac{1}{a_1 + a_2 +a_3} \left[ \frac{1}{(a_2 + a_3) a_3} +  \frac{1}{a_1 + a_3} \left( \frac{1}{a_3} +  \frac{1}{a_1} \right) \right] \\
&= \frac{1}{(a_1 + a_2 +a_3) (a_2 + a_3) a_3}   \\
&\phantom{=}  + \frac{1}{(a_1 + a_2 +a_3) (a_1 + a_3) a_3}  \\
&\phantom{=}  +  \frac{1}{(a_1 + a_2 +a_3) (a_1 + a_3) a_1}  \ .
\end{align*}
In order to illustrate why a given set of fractions associated with linear trees may or may not be resummed into a single fraction, we now compare T3.3 with T3.5 whose decomposition into linear tree TSDs is displayed in Fig.~\ref{f:resum_fail} and whose expression was already given in Eq.~\eqref{cycleT35}. The non-tree TSD T3.5 sums one less linear tree than T3.3 whose associated fraction is necessary to factorize and cancel the longest factor $(a_1 + a_2 +a_3)$ appearing in all third-order linear trees to eventually obtain a single term. The linear tree in question corresponds to $a_1$ being at higher times than both $a_2$ and $a_3$. It is missing from T3.5 because $a_1$ and $a_2$ belong to a cycle and are thus unordered with respect to each other while being both ordered with respect to $a_3$ that is at a higher time. This situation corresponding to non-tree TSDs typically lead to missing terms that are necessary for the complete factorization to occur.  

Eventually, the resummation of $n_\text{branches}(\text{Tp.k})$ fully-time-ordered TSDs (fractions) into a single time-unordered TSD (fraction) can be generically written as
\begin{multline}
\sum^{n_\text{branches}(\text{Tp.k})}_{i=1} \hspace{-0.6cm} \frac{1}{(a_{i1} \dots a_{ip}) (a_{i2} \dots a_{ip}) \dots a_{ip}} \\
= \frac{1}{{\displaystyle\prod_{j=1}^{p}} (a_{j1} + \dots + a_{jp_{j}})} \ ,
\end{multline}
where $a_{i1}, \dots, a_{ip}$ label the $p$ vertices from bottom to top in each of the $i=1,\ldots,n_\text{branches}(\text{Tp.k})$ summed linear tree TSDs, whereas $a_{j1}, \dots, a_{jp_{j}}$ label the $p_j$ vertices in the subgraph of Tp.k made out of vertex $j$ and all its descendants.

\section{Use of the \texttt{ADG} program}
\label{programuse}

\texttt{ADG} has been designed to work on any computer with a Python2.7 distribution, and successfully tested on recent GNU/Linux distributions and on MacOS. Additionally to Python, \emph{setuptools} and \emph{distutils} packages must already be installed, which is the case on most standard recent distributions. Having \emph{pip} installed eases the process but is not technically required. The \emph{NumPy}, \emph{NetworkX} and \emph{SciPy} libraries are automatically downloaded during the install process. Additionally, one needs a \LaTeX\ distribution installed with the PDF\LaTeX\ compiler for \texttt{ADG} to produce the pdf file associated to the output if desired.

\subsection{Installation}

\subsubsection{From the Python Package Index}

The easiest way to install \texttt{ADG} is to obtain it from the Python Package Index\footnote{\url{https://pypi.org/project/adg/}} by entering the following command
\begin{verbatim}
pip2 install adg
\end{verbatim}
Provided \emph{setuptools} is already installed, \emph{pip} takes care of downloading and installing \texttt{ADG} as well as \emph{NumPy} and \emph{NetworkX}. Once a new version of \texttt{ADG} is released, one can install it by entering the command
\begin{verbatim}
pip2 install --upgrade adg
\end{verbatim}

\subsubsection{From the source files}

Once the \texttt{ADG} source files are downloaded from the CPC library or the GitHub repository\footnote{\url{https://github.com/adgproject/adg}}, one must enter the project folder and either run
\begin{verbatim}
pip2 install .
\end{verbatim}
or
\begin{verbatim}
python2 setup.py install
\end{verbatim}
With this method, \emph{pip}\footnote{Depending on the system, it might be necessary either to use the "--user" flag to install it only for a specific user or to run the previous command with "sudo -H" to install it system-wide.} also takes care of downloading and installing \texttt{ADG}, \emph{NumPy}, \emph{NetworkX} and \emph{SciPy}.

\subsection{Run the program}

\subsubsection{Batch mode}

The most convenient way to use \texttt{ADG} is to run it in batch mode with the appropriate flags. For example, to run the program and generate BMBPT diagrams at order 4 for example, one can use
\begin{verbatim}
adg -o 4 -t BMBPT -d -c
\end{verbatim}
where the \texttt{-o} flag is for the order, \texttt{-t} for the type of theory, \texttt{-d} indicates that the diagrams must be drawn and \texttt{-c} that \texttt{ADG} must compile the \LaTeX\ output. A complete list of the program's options can be obtained via the program's documentation (see Sec.~\ref{documentation}) or through
\begin{verbatim}
adg -h
\end{verbatim}

Currently, \texttt{ADG} can be run either in relation to HF-MBPT by using \texttt{-t MBPT} or to BMBPT by using \texttt{-t BMBPT}. Though the algorithms described in the previous sections can be used regardless of the diagrams' orders, \texttt{ADG} has been arbitrarily restricted to order 10 or lower to avoid major overloads of the system. Future users are nevertheless advised to first launch calculations at low orders (2, 3 or 4 typically) as the time and memory needed for computations rise rapidly with the perturbative order.

\subsubsection{Interactive mode}

As an alternative to the batch mode, \texttt{ADG} can be run on a terminal by entering the command
\begin{verbatim}
adg -i
\end{verbatim}
A set of questions must be answered using the keyboard to configure and launch the calculation. The interactive mode then proceeds identically to the batch mode.

\subsection{Steps of a program run}

Let us briefly recapitulate the different steps of a typical \texttt{ADG} run
\begin{itemize}
\item Run options are set either by using the command-line flags entered by the user or during the interactive session via keyboard input.
\item \texttt{ADG} creates a list of adjacency matrices for the appropriate theory and perturbative order, and via \emph{NumPy}, feeds them to \emph{NetworkX} that creates \emph{MultiDiGraph} objects.
\item Checks are performed on the list of graphs to remove topologically equivalent or anomalous graphs.
\item The list of topologically unique graphs is used to produce \emph{Diagram} objects that store the graph as well as some of its associated properties depending on the theory (HF status, excitation level, etc.). The expression associated to the graphs are extracted.
\item The program then prints on the terminal the number of diagrams per category and writes the \LaTeX\ output file, the details of which depend on the options selected by the user, as well as a list of adjacency matrices associated to the diagrams. Other output files may be produced, depending on the theory and the user's input.
\item If asked by the user, the program performs the PDF\LaTeX\ compilation.
\item Unnecessary temporary files are removed and the programs exits.
\end{itemize}

\subsection{Documentation}
\label{documentation}

\subsubsection{Local documentation}

Once the source files have been downloaded, a quick start guide is available in the \texttt{README.md} file. Once \texttt{ADG} is installed, it is possible to read its manpages through
\begin{verbatim}
man adg
\end{verbatim}
or a brief description of the program and its options through
\begin{verbatim}
adg -h
\end{verbatim}
A more detailed HTML documentation can be generated directly from the source files by going into the \texttt{docs} directory and run
\begin{verbatim}
make html
\end{verbatim}
The documentation is then stored in \texttt{docs/build/html}, with the main file being \texttt{index.html}. A list of other possible types of documentation format is available by running
\begin{verbatim}
make help
\end{verbatim}

\subsubsection{Online documentation}

The full HTML documentation is available online under \url{https://adg.readthedocs.io/}, and help with eventual bugs of the program can be obtained by opening issues on the GitHub repository at \url{https://github.com/adgproject/adg}.

\section{Conclusions}
\label{conclusions}

The diagrammatic translation of certain quantum many-body methods, e.g. many-body perturbation theory~\cite{goldstone57a,hugenholtz57a,Shavitt_Bartlett_2009,Ti16,Hu16,Ti17}, self-consistent Green's function theory~\cite{Dickhoff:2004xx,CiBa13,Carbone:2013eqa,Soma:2011aj,SoCi13}, coupled cluster theory~\cite{KoDe04,BaRo07,HaPa10,PiGo09,BiLa14,Si15} etc, is used to build an intuition about the systematic contributions to a physical observable and to derive the corresponding algebraic expressions at minimal cost. However, (1) the need in nuclear physics to tackle three-nucleon interactions, i.e., six-leg vertices, (2) the development of novel many-body methods based on generalized diagrammatics~\cite{Duguet:2014jja,Duguet:2015yle} and (3) the implementation of high-order contributions authorized by the rapid progress of computational power, welcome the development of a versatile code capable of both generating and evaluating many-body diagrams automatically. 

In the present publication, we have focused our attention on Bogoliubov many-body perturbation theory (BMBPT) that has been recently formulated~\cite{Duguet:2015yle,Art18b} and first implemented at low orders~\cite{Tichai:2018mll} to tackle (near) degenerate Fermi systems, e.g. open-shell nuclei displaying a superfluid character. This many-body method perturbatively expands the exact solution of the Schrödinger equation around a so-called Bogoliubov reference state, i.e., a general product state breaking $U(1)$ global-gauge symmetry associated with the conservation of good particle number in the system. 

The paper describes the first version (v1.0.0) of the code \texttt{ADG} that generates all valid BMBPT diagrams and  evaluates their algebraic expression to be implemented in a numerical application. This is realized at an arbitrary order $p$ for a Hamiltonian containing both two-body (four-legs) and three-body (six-legs) interactions (vertices). The automated generation of BMBPT diagrams of order $p$ is achieved by  producing all oriented adjacency matrices of size $(p+1) \times (p+1)$ satisfying topological Feynman's rules. The automated evaluation of all BMBPT diagrams of order $p$ relies both on the application of algebraic Feynman's rules and on the design of a systematic method to perform the remaining $p$-tuple time integral. This method provides a novel diagrammatic rule allowing for the straight summation of large classes of time-ordered diagrams at play in the time-independent formulation of BMBPT. The standard resolvent rule employed to compute time-ordered diagrams one by one happens to be a particular case of the general rule presently identified. The code \texttt{ADG} is written in \emph{Python2.7} and uses the graph manipulation package \emph{NetworkX}. It is made flexible enough to be expanded throughout the years to tackle the diagrammatics at play in various many-body formalisms that either already exist or are yet to be formulated.

\section*{Acknowledgments}

This publication is based on work supported in part by the framework of the Espace de Structure et de r\'eactions Nucl\'eaires Th\'eorique (ESNT) at CEA.

The authors thank C.~Drischler, M.~Drissi, A.~Gallo and C.~Wellenhofer for fruitful discussions, B.~Bally and V.~Som\`a for beta-testing the program as well as J.~Ripoche and V.~Som\`a for proofreading the manuscript.

\appendix

\section{Basic elements of graph theory}
\label{s:graph_theory}

Graph theory is a domain of discrete mathematics focusing on the study of graphs and their properties. In this section one introduces basic notation and terminology required for reformulating aspects of many-body theory in terms of graph-theory language. For an extensive discussion see, e.g., the classical textbook~\cite{Bolo98}.
\begin{definition}
A \emph{graph} is a triplet $G=(V, E, \psi)$ consisting of a set $V$ whose elements are called \emph{vertices} and a set $E$ whose elements are called \emph{edges} together with an incidence relation $\psi$. Let further $E^\prime \subset E$, $V^\prime \subset V$ and $\psi\vert_{E^\prime}$ be the restriction of $\psi$ to $E^\prime$ then the triplet $G^\prime=(E^\prime, V^\prime, \psi\vert_{E^\prime} )$ is called a \emph{subgraph} of $G$. We call a graph \emph{oriented} if every edge has a fixed direction.
\end{definition}
Note that the finiteness of either $V$ or $E$ is not assumed. In applications for BMBPT diagrams, however, both sets will always be finite for any perturbative order.

Two nodes $v_1,v_2$ are \emph{adjacent} if there is exists an edge $e$ connecting $v_1$ with $v_2$. An edge that starts and end at the same vertex is called a \emph{loop}. If an edge $e\in E$ starts or ends at a vertex $v\in V$ then $e$ is called \emph{incident} to $e$. The number of incident edges of a vertex is called \emph{degree} and denoted as $\text{deg}(v)$.

\begin{definition}
Let $G=(V, E, \psi)$ be a graph with $v_i\in V$ and $e_i \in E$. The sequence
\begin{align}
v_0 e_1 v_1 e_1 .... e_n v_n \,
\end{align}
is called a \emph{walk}. The walk is \emph{closed} if $v_0=v_n$. Furthermore, the \emph{length} of a walk is the number of edges $\vert \{e_0,...,e_n\}\vert$.
\end{definition}
The first node $v_0$ and last node $v_n$ are called \emph{initial} and \emph{terminal} nodes, respectively. If all nodes in a walk are distinct, it is a \emph{path}. Of particular importance is the case where the initial and terminal node coincide, which is called a \emph{cycle}.

\begin{definition}
A graph $G=(V, E, \psi)$ is \emph{connected} there is a path connecting any pair of nodes.
\end{definition}
The definition of connectedness of graphs is crucial since it directly relates to physical properties of the many-body expansion. A connected graph without a cycle is called \emph{tree}.

For now graphs have been treated as abstract objects. For computational purposes it is convenient to have a representation of graph.

\begin{definition}
Let $G=(V, E, \psi)$ be a graph. The \emph{incidence matrix} $M(g)$ is the $\vert V \vert \times \vert E\vert$ matrix with entries
\begin{align}
m_{ij}= \begin{cases}
	1, \quad \text{if $e_j$ is incident to $v_i$} \\
	0, \quad \text{otherwise}
\end{cases}\, .
\end{align}
\end{definition}
The notation of incidence matrices is an edge-based representation of the graph. However, for the present applications the use of a vertex-based description is more useful.
\begin{definition}
Let $G=(V, E, \psi)$ be a graph. The \emph{adjacency matrix} $A(g)$ is the $\vert V \vert \times \vert V\vert$ matrix with entries
\begin{align}
a_{ij}= \big \vert \{ e_k \in E \, : \,   e_k \, \text{connects} \, v_i, v_j \} \big \vert\, .
\end{align}
\end{definition}
For oriented graphs the definition needs to be slightly extended:
\begin{definition}
Let $G=(V, E, \psi)$ be an oriented graph. The \emph{oriented adjacency matrix} $\tilde A(g)$ is the $\vert V \vert \times \vert V\vert$ matrix with entries
\begin{align}
\tilde{a}_{ij}= \big \vert \{ e_k \in E \, :  \,  e_k \, \text{goes from} \, v_i \text{to} \,  v_j \} \big \vert\, .
\end{align}
\end{definition}
We emphasize that the (oriented) adjacency matrix of a graph $G$ encodes all relevant structural information.

\begin{proposition}
Let $G = (V, E, \psi)$ be a graph with $\vert V\vert= n$ $\vert E\vert = m$ then the following are equivalent
\begin{itemize}
\item[$(i)$] $G$ is connected and and contains no cycle.
\item[$(ii)$] $G$ has no cycle and $m=n-1$.
\item[$(iii)$] $G$ is connected and $m=n-1$.
\item[$(iv)$] $G$ is connected but would not be if any of its edges were suppressed.\footnote{This notion corresponds to one-line irreducible diagrams in physics.}
\item[$(v)$] $G$ contains no cycle and adding a new edge to it creates a unique cycle.
\item[$(vi)$] For any pair of nodes $v_i$ and $v_j$, there exists a single path from $v_i$ to $v_j$.
\end{itemize}
\end{proposition}

\section{Structure of the \texttt{ADG} program}

The previously described methodology has been implemented to build a \emph{Python 2.7} program called \texttt{ADG} for \texttt{Automatic Diagram Generator}. This program uses the external Python packages \emph{NumPy} for matrix-related operations and \emph{NetworkX} for producing and manipulating diagrams. Python allows us to develop an easy-to-use, low-maintenance program without having to tamper directly with low-level concepts such as memory allocation. The wide ecosystem of open-source packages available helps focus on physics-related parts of the code. Furthermore, the possibility to use object-oriented programming has proven useful to design a program that could easily be extended to a various range of many-body diagrammatic theories. For readability and maintainability purposes, the program has been separated into different modules whose functions are detailed below.

\subsection{Main module}

The main function, contained in file \emph{main.py}, organizes the whole program and makes use of the other modules when needed. The function first parses the command-line options enterred by the user, asking they for keyboard input if needed. The calling options comprises the theory being used, the two- or three-body operator character of the operators as well as other features regarding output formatting.

The function is designed to be as theory-agnostic as possible, calling for wrapper functions defined in the \emph{run} module to generate the appropriate adjacency matrices, which are then recasted as \emph{NumPy} matrices and fed to \emph{NetworkX} to produce graphs that are then used to initialize the actual \emph{Diagram} objects that the program uses. A few tests are applied to make sure only appropriate matrices are kept (corresponding to connected graphs, etc.) before MBPT or BMBPT diagrams are produced, which encapsulate the \emph{NetworkX} graph as well as other related properties stored as attributes (two- or three-body character, various tags, degrees of the vertices,...).

Checks for topologically identical diagrams are then performed. As this part of the program scales factorially with the number of diagrams considered, it constitutes the more time-costly part of the program when going to higher orders. Especially, the \texttt{is\_isomorphic} interface of \emph{NetworkX} is itself time-costly as it performs permutations between the graph nodes. Consequently, the algorithm has been optimize to call it as rarely as possible. Diagrams are therefore first selected based on their two- or three-body character and Hartree-Fock or non-Hartree-Fock status, such that comparisons are done within a smaller set of diagrams. Additionally, checks on the set of in- and out-degrees of the vertices of the two graphs are made, leaving the need for a call to \texttt{is\_isomorphic} to the fewest possible cases.

Once only topologically distinct diagrams are kept, the program extracts the expressions associated to the diagrams depending of the formalism involved, and stores them as attributes of the diagram objects.

Finally, output \LaTeX\ files are produced, the exact content and formatting of which depend on user's input. Computer-readable files are available as well for MBPT diagrams. Compilation of the main \LaTeX\ file is then proposed to the user, and useless files produced by the program and the \LaTeX\ compilation are deleted before exiting.

\subsection{Run management module}

The \emph{run.py} file contains routines related to run management, command-line interface and managing the code output.

\subsubsection{Routines}

The routines of the module are:
\begin{itemize}
\item \texttt{parse\_command\_line} sets up the calculation depending on the used command-line flags.
\item \texttt{interactive\_interface} sets up the calculation using keyboard input when \texttt{ADG} has been called with the flag \texttt{-i}.
\item \texttt{attribute\_directory} creates the appropriate folder for the output of the program, depending of the theory, order and other options.
\item \texttt{generate\_diagrams} is used as a wrapper for the different class-dependent diagram creation routines.
\item \texttt{order\_diagrams} is used as a wrapper for the different class-dependent diagram ordering routines.
\item \texttt{print\_diags\_numbers} prints out information about the produced diagram on the terminal.
\item \texttt{prepare\_drawing\_instructions} launches the production of graph-related drawing instructions.
\item \texttt{create\_feynmanmp\_files} then stores the instructions in an appropriately-labelled text file.
\item \texttt{write\_file\_header} takes care of writing the beginning of the \LaTeX\ output file with the appropriate formatting options.
\item \texttt{compile\_manager} takes care of compiling the \LaTeX\ file with PDF\LaTeX\ \hspace{-3pt}.
\item \texttt{clean\_folders} then deletes the auxiliary files that are no longer needed.
\end{itemize}

\subsection{Generic diagram module}

The \emph{diag.py} file contains various routines for diagrams that can be used regardless of the theory of interest, i.e., tests on the degree of vertices in a matrix, tests for topologically identical diagrams, or routines used to label the vertices and propagators of a diagram. It also contains the routines that are used to produce the FeynMP instructions starting from a \emph{NetworkX} graph and various routines used for the production of the \LaTeX\ output files. Finally, it contains the definition of a \texttt{Diagram} abstract class that is inherited by the classes associated to MBPT and BMBPT diagrams.

\subsubsection{Routines}

Let us now describe briefly the different routines of the module:
\begin{itemize}
\item \texttt{no\_trace} takes as input a list of matrices and returns it without any matrix with a non-zero diagonal matrix element.
\item \texttt{check\_vertex\_degree} checks the degree of a specific vertex. It is used during the matrix generation to remove ill-defined matrices.
\item \texttt{topologically\_distinct\_diagrams} checks a list of \texttt{Diagram} objects and removes the topologically equivalent ones.
\item \texttt{label\_vertices} is used to attribute labels to the nodes in a \emph{NetworkX} graph depending on their general operator or grand canonical potential character.
\item \texttt{feynmf\_generator} is the routine used to generate the \emph{FeynMP} drawing instructions starting from a \emph{NetworkX} graph.
\item \texttt{propagator\_style} selects the appropriate drawing instructions for propagators.
\item \texttt{draw\_diagram} recovers the drawing instructions of a given diagram and copies them in the \LaTeX\ file.
\item \texttt{to\_skeleton} returns only the non-redundant links in a diagram, i.e., only the minimal set of edges to infer the time relations. It is mostly used for time-structure diagrams, though its scope could be more general.
\item \texttt{extract\_denom} returns the appropriate denomiantor for a diagram using the subgraph rule.
\item \texttt{print\_adj\_matrices} prints a computer-readable file with the adjacency matrices of the diagrams.
\end{itemize}

\subsubsection{\texttt{Diagram} class}

The \texttt{Diagram} class is used to describe a general diagram and comprises the following attributes:
\begin{itemize}
\item \texttt{graph}, the \emph{NetworkX} graph associated to the diagram.
\item \texttt{unsort\_degrees}, a tuple with the graph vertex degrees.
\item \texttt{degrees}, a sorted tuple with the degrees of the graph's vertices.
\item \texttt{unsort\_io\_degrees}, a tuple with the in- and out-degrees of the vertices.
\item \texttt{io\_degrees}, that correspond to a sorted version of \texttt{unsort\_io\_degrees}.
\item \texttt{max\_degree}, the highest vertex degree in the graph.
\item \texttt{tags}, a list of integers associated with the graph to keep track of topologically identical diagrams.
\item \texttt{adjacency\_mat}, a \emph{NumPy} array with the adjacency matrix of the graph.
\end{itemize}

The \texttt{Diagram} class has two methods described below:
\begin{itemize}
\item \texttt{\_\_init\_\_} takes as input a \emph{NetworkX} graph that it stores in \texttt{graph} and uses to initialize the other attributes.
\item \texttt{write\_graph}, an abstract method for drawing the graph using \emph{FeynMP} instructions.
\end{itemize}

\subsection{MBPT module}

This \emph{mbpt.py} file contains the routines that are related to MBPT diagrams, be it the generation of the adjacency matrices associated to them or the treatment of the MBPT expressions, as well as producing a computer-readable output suitable for automated calculation frameworks \citep{Drischler:2017wtt}. It also contains the \texttt{MbptDiagram} class that inherits from the \texttt{Diagram} class defined in the \emph{diag} module.

\subsubsection{Routines}

Let us now describe briefly the different routines of the module:
\begin{itemize}
\item \texttt{diagrams\_generation} produces all the adjacency matrices associated to MBPT diagrams of a given order.
\item \texttt{write\_diag\_exp} writes the expression associated to a diagram in the \LaTeX\ file.
\item \texttt{write\_header} writes the appropriate header for the \LaTeX\ output file.
\item \texttt{print\_cd\_output} prints a computer-readable output file.
\item \texttt{order\_diagrams} order the diagrams depending on their excitation level.
\item \texttt{attribute\_conjugate} searches for the conjugate partner of a diagram in the list of all diagrams.
\item \texttt{extract\_cd\_denom} extracts the denominator of the graph and writes it in a computer-readable format.
\end{itemize}

\subsubsection{\texttt{MbptDiagram} class}

Additionnally to the attributes defined in the class \texttt{Diagram}, the class \texttt{MbptDiagram} possesses the following attributes:
\begin{itemize}
\item \texttt{incidence}, a \emph{NumPy} array with the incidence matrix of the graph.
\item \texttt{excitation\_level}, an integer coding for the single, double, etc., character of the diagram.
\item \texttt{complex\_conjugate}, the tag of the conjugate partner of the diagram.
\item \texttt{expr} a string that stores the expression associated to a diagram.
\item \texttt{cd\_expr}, the expression associated to the graph in a computer-readable format.
\end{itemize}

The methods of the \texttt{MbptDiagram} class are described below:
\begin{itemize}
\item Its constructor \texttt{\_\_init\_\_} calls the \texttt{Diagram} class constructor and additionnaly inializes the diagram \texttt{tags} before calling \texttt{attribute\_expressions}.
\item \texttt{attribute\_expression} is used to generate the expression associated with the MBPT diagram and stores it in its attribute \texttt{expr}.
\item \texttt{calc\_excitation} returns the integer associated with the excitation level of the diagram.
\item \texttt{count\_hole\_lines} returns the number of hole lines in the graph.
\item \texttt{is\_complex\_conjug\_of} returns True if a diagram is the complex conjugate diagram of the object.
\item \texttt{attribute\_ph\_labels} attributes the appropriate particle or hole label to the lines of the diagram.
\item \texttt{extract\_denominator} returns a string with the denominator associated to the diagram.
\item \texttt{cd\_denominator} returns a string with the denominator associated to the graph in a computer-readable format.
\item \texttt{extract\_numerator} returns the numerator associated to the diagram.
\item \texttt{cd\_numerator} returns the numerator associated to the diagram in a computer-readable format.
\item \texttt{loops\_number} returns the number of loops in the diagram, using a specific convention for reading its representation.
\item \texttt{write\_section} writes the information associated to the graph in the \LaTeX\ output file.
\end{itemize}

\subsection{BMBPT module}

The \texttt{bmbpt.py} file contains the routines related to BMBPT diagrams: generation of the associated adjacency matrices, some on-the-fly tests for this generation, tests used to characterize the BMBPT diagrams with respect to their two- or three-body operator or Hartree-Fock character, different routines for the extraction of the associated numerators, denominators and different symmetry factors, and finally routines used to format the output files. This module also defines a \texttt{BmbptFeynmanDiagram} class, similar to the \texttt{MbptDiagram} one.

\subsubsection{Routines}

Let us now describe briefly the different routines of the module:
\begin{itemize}
\item \texttt{diagrams\_generation} generates all the adjacency matrices associated to BMBPT diagrams of a given order.
\item \texttt{check\_unconnected\_spawn} is used by \texttt{BMBPT\_generation} to avoid producing matrices that would in the end correspond to unconnected diagrams.
\item \texttt{write\_header} takes care of the appropriate formatting of the output \LaTeX\ file in the case \texttt{ADG} has been called for BMBPT.
\item \texttt{produce\_expressions} produces and stores the expressions associated to the BMBPT diagrams.
\item \texttt{order\_diagrams} order the diagrams depending on their use of two- or three-body forces and their Hartree-Fock character, and discard topologically equivalent diagrams.
\end{itemize}

\subsubsection{\texttt{BmbptFeynmanDiagram} class}

Additionnally to the attributes defined in the class \texttt{Diagram}, the class \texttt{BmbptFeynmanDiagram} possesses the following attributes:
\begin{itemize}
\item \texttt{two\_or\_three\_body} that stores as an integer the two-body-only or three-body character of the operators comprised in the diagram.
\item \texttt{time\_tag} is an integer that keep track of the associated time-structure diagram.
\item \texttt{tsd\_is\_tree} is set to True if the associated TSD has a tree structure, False if it has not.
\item \texttt{feynman\_exp}, a string that stores the time-dependent expression associated to a diagram.
\item \texttt{diag\_exp}, a string that stores the time-independent expression associated to a diagram.
\item \texttt{vert\_exp}, a list of strings that stores the expressions associated to each vertex.
\item \texttt{hf\_type}, a string that says if a diagram is of Hartree-Fock character, Hartree-Fock if the generic operator is replaced by the grand canonical potential, or non-Hartree-Fock.
\end{itemize}

The \texttt{BmbptFeynmanDiagram} class has fifteen methods described below:
\begin{itemize}
\item Its constructor \texttt{\_\_init\_\_} calls the \texttt{Diagram} class constructor and initializes the other attributes.
\item \texttt{attribute\_expressions} is used to generate the time-dependent and time-independent expressions associated with the BMBPT diagram and stores it in its attributes \texttt{feynman\_exp} and \texttt{diag\_exp}.
\item \texttt{vertex\_expression} returns the expression associated to a given vertex of the BMBPT diagram.
\item \texttt{write\_graph} writes the graph and its associated TSD to the \LaTeX\ file.
\item \texttt{write\_tsd\_info} writes information relative to the TSD associated to the diagram in the output file.
\item \texttt{write\_section} writes sections and subsections in the output file.
\item \texttt{write\_vertices\_values} writes the quasiparticle energies associated to each vertex of the graph in the output file.
\item \texttt{write\_diag\_exps} writes the expressions associated to a diagram in the \LaTeX\ file.
\item \texttt{vertex\_exchange\_sym\_factor} returns the symmetry factor associated with vertex exchange.
\item \texttt{extract\_integral} returns as a string the integral part of the Feynman expression of the graph.
\item \texttt{attribute\_qp\_labels} is used to attribute the appropriate quasiparticle label to the edges of the \emph{NetworkX} graph.
\item \texttt{extract\_numerator} returns as a string the numerator associated to the graph.
\item \texttt{has\_crossing\_sign} returns True if there is a minus sign associated with crossing propagators in the graph.
\item \texttt{multiplicity\_symmetry\_factor} returns the symmetry factor associated with propagators multiplicity.
\item \texttt{time\_tree\_denominator} returns as a string the time-integrated denominator associated to a BMBPT graph that has a tree time-structure.
\end{itemize}

\subsection{TSD module}

Finally, the \emph{tsd.py} file contains routines related to time-structure diagrams (TSDs). Though designed specifically for TSDs related to BMBPT diagrams, it could be extended to encompass other types of TSDs. The various routines this module contains deal with the production of a TSD diagram out of a BMBPT diagram, different tests on BMBPT diagrams with respect to their associated TSD, extraction of the denominator resulting from the time integration  associated with the TSD, and production of the corresponding section of the output file, including the drawing of the TSD. Finally, it contains a \texttt{TimeStructureDiagram} class that inherits from the \texttt{Diagram} class, with its constructor that generates a TSD starting from a BMBPT diagram.

\subsubsection{Routines}

Let us now describe briefly the different routines of the module:
\begin{itemize}
\item \texttt{time\_structure\_graph} returns the time-structure graph associated to a BMBPT graph.
\item \texttt{tree\_time\_structure\_den} returns the denominator associated to a tree time-structure graph.
\item \texttt{equivalent\_labelled\_TSDs} returns the list of labelled TSDs corresponding to the equivalent tree TSDs of a previously given non-tree TSD.
\item \texttt{write\_section} takes care of the proper formatting of the section devoted to TSDs in the output \LaTeX\ file.
\item \texttt{disentangle\_cycle} is used by \texttt{treat\_cycle} to separate a cycle in a sum of trees.
\item \texttt{find\_cycle} returns the start and end nodes of an elementary cycle and is called by \texttt{disentangle\_cycle}.
\item \texttt{treat\_tsds} orders the TSDs, produces their expressions and returns the number of tree TSDs.
\end{itemize}

\subsubsection{\texttt{TimeStructureDiagram} class}

Additionnally to the attributes defined in the class \texttt{Diagram}, the class \texttt{TimeStructureDiagram} possesses the following attributes:
\begin{itemize}
\item \texttt{perms}, a dictionnary of permutations necessary for the treatment of expressions for topologically equivalent TSDs.
\item \texttt{equivalent\_trees}, a list of integers to keep track of the topologically equivalent TSDs.
\item \texttt{is\_tree}, set to True if the TSD is a tree, False if it not.
\item \texttt{expr}, a string to store the denominator associated with the TSD.
\item \texttt{resum}, the resummation power of the tree TSD stored has an integer.
\end{itemize}

The \texttt{TimeStructureDiagram} class has four methods described below:
\begin{itemize}
\item Its constructor \texttt{\_\_init\_\_} calls the \texttt{Diagram} class constructor and then initializes the other attributes.
\item \texttt{treat\_cycles} finds and treat the cycles in a non-tree TSD.
\item \texttt{draw\_equivalent\_tree\_tsds} draws the equivalent tree TSDs of a given non-tree TSD.
\item \texttt{resummation\_power} returns the resummation power associated to a tree TSD.
\end{itemize}

\section*{References}
\bibliographystyle{apsrev4-1}
\bibliography{Bibliographie}

\end{document}